\documentclass[pre,twocolumn,nobalancelastpage,amsmath,amssymb,showpacs,
nofootinbib,superscriptaddress]{revtex4-1}

\usepackage{graphics}      % standard graphics specifications
\usepackage{graphicx}      % alternative graphics specifications
\usepackage{longtable}     % helps with long table options
\usepackage{url}           % for on-line citations
\usepackage{bm}            % special 'bold-math' package
\usepackage{xcolor}
\usepackage{amsmath}
\usepackage{dcolumn}

\newcommand{\gdot}{\dot\gamma}

\newcommand{\skippa}[1]{}

\begin{document}
\title{Shear banding, discontinuous shear thickening, and rheological phase transitions in athermally sheared frictionless disks}
\author{Daniel V{\aa}gberg}
\affiliation{Laboratoire Charles Coulomb, UMR 5221 CNRS, Universit{\'e} Montpellier, Montpellier, France
}
\author{Peter Olsson}
\affiliation{Department of Physics, Ume{\aa} University, 901 87 Ume{\aa}, Sweden}
\author{S. Teitel}
\affiliation{Department of Physics and Astronomy, University of Rochester, Rochester, NY 14627}
\date{\today}   
\begin{abstract}
We report on numerical simulations of simple models of athermal, bidisperse, soft-core, massive disks in two dimensions, as a function of packing fraction $\phi$, inelasticity of collisions as measured by a parameter $Q$, and applied uniform shear strain rate $\dot\gamma$.  Our particles have contact interactions consisting of normally directed elastic repulsion and viscous dissipation, as well as tangentially directed viscous dissipation, but no inter-particle Coulombic friction.  Mapping the phase diagram in the $(\phi,Q)$ plane for small $\dot\gamma$, we find a sharp first-order rheological phase transition from a region with Bagnoldian rheology to a region with Newtonian rheology, and show that the system is always Newtonian at jamming.  
We consider the rotational motion of particles and demonstrate the crucial importance that the coupling between rotational and translational degrees of freedom has on the phase structure at small $Q$ (strongly inelastic collisions).  
At small $Q$ we show that, upon increasing $\dot\gamma$, the sharp Bagnoldian-to-Newtonian transition becomes a coexistence region of finite width in the $(\phi,\dot\gamma)$ plane, with coexisting Bagnoldian and Newtonian shear bands.  Crossing this coexistence region by increasing $\dot\gamma$ at fixed $\phi$, we find that discontinuous shear thickening can result if $\dot\gamma$ is varied too rapidly for the system to relax to the shear-banded steady state corresponding to the instantaneous value of $\dot\gamma$.
\end{abstract}
\maketitle

\section{Introduction}

Simple models of spherical particles, interacting via soft- or hard-core repulsive contact interactions, have been used to model a wide variety of granular and soft-matter materials, such as dry granular particles, foams, emulsions, non-Brownian suspensions, and colloids.  The rheology of such systems under applied shear strain has been a topic of much active investigation.  For athermal systems ($T=0$), such models generally give a rheology with the following features: (i) At low density and low strain rate the system is in a flowing liquid-like state where the pressure $p$ and shear stress $\sigma$ scale with the shear strain rate $\dot\gamma$ according to either a Newtonian ($p,\sigma\sim\dot\gamma$) or Bagnoldian ($p,\sigma\sim\dot\gamma^2$) rheology, depending on the form of energy dissipation in the system \cite{VOT_rheology}.  
Newtonian rheology is often associated with overdamped  motion due to the Stokes drag on particles that are embedded in a host fluid \cite{OT_2007}.  Bagnoldian rheology is often associated with dry hard granular particles, assuming that the frequency of collisions and the momentum change per collision are both proportional to the strain rate $\dot\gamma$ \cite{LLC}.  However Newtonian rheology may still be observed in models which neglect interactions with a host fluid \cite{VOT_rheology,VOT_RDCD}, and Bagnoldian rheology may still be observed even in the presence of a host fluid, as the strain rate increases \cite{Bagnold,Fall}.
%As $\dot\gamma$ increases away from the small $\dot\gamma$ limit, $p$ and $\sigma$ usually grow less slowly, a phenomenon referred to as ``shear thinning." 
(ii) As the particle packing fraction $\phi$ increases, a shear driven jamming transition occurs at a well defined $\phi_J$, where the macroscopic transport coefficient of the liquid diverges \cite{OT_2007,VOT_RDCD, Barrat,OT_RD,Lerner,DeGiuli,Ikeda,VOT_CDn,VOT_CDnQ}.  (iii) For soft-core particles, above $\phi_J$ the system develops a finite yield stress $\sigma_Y(\phi)$.  For $\sigma < \sigma_Y$, the system is in a static jammed solid state, while for $\sigma>\sigma_Y$ the system  is in a state of plastic flow, generally described by a Herschel-Bulkley rheology, $\sigma = \sigma_Y + C\dot\gamma^b$ \cite{OT_HB}.  When particles are frictionless, the jamming transition is continuous, with $\sigma_Y$ vanishing continuously as $\phi\to\phi_J$ from above.  In this case, critical scaling is found to well describe behavior near the jamming transition \cite{OT_2007,Barrat,OT_RD, Lerner,VOT_RDCD, DeGiuli,Ikeda,VOT_CDn}.  When particles have inter-granular friction, the jamming transition may become discontinuous, with $\sigma_Y$ dropping discontinuously to zero at $\phi_J$;  the rheology may now show hysteretic and re-entrant behavior near $\phi_J$ \cite{OH_friction,Bi_Nature,Heussinger,Ciamarra,Grob,Grob2,Schwarz,Jaeger_Nature}.

However in many materials, the rheology is observed to be more complex.  One such departure from the above scenario is the phenomenon of ``shear banding," in which the shear strain rate $\dot\gamma$ becomes spatially inhomogeneous, sometimes breaking into well-defined distinguishable bands with different values of $\dot\gamma$.  
Shear banding has been found in a wide range of materials, including complex fluids \cite{CatesFielding}, 
non-Brownian suspensions \cite{Coussot,Ovarlez0},
pastes, gels, and emulsions \cite{Ovarlez}, foams \cite{Kabla}, granular matter \cite{Jaeger_RMP, Mueth}, soils and rocks \cite{Rudnicki}, metallic  glasses \cite{Spaepen,Li}, and colloidal glasses \cite{Besseling, Chikkadi}.
In some cases, shear banding is believed to result from 
inhomogeneous stress fields that vary spatially from values above to values below the yield stress of the material \cite{Schall}.  In other cases, shear banding is believed to result from the coupling of the system to a boundary wall \cite{Latzel,Wang}.
However, in yet other cases, shear banding appears to exist as a bulk phenomenon, and in situations of spatially uniform stress \cite{Goddard,Moller}.
In such cases, shear banding is believed to result from some inherent instability in the shear flow
that leads to a non-monotonic stress-strain rate curve; shear bands then result as coexisting phases at a discontinuous transition, as in analogy to the van der Waals picture for the liquid-gas transition.  
In some of these cases, one of the bands is seemingly stationary, while the flowing band exists only above some  critical strain rate $\dot\gamma_c$ \cite{Coussot,Ovarlez0,Moller,Dennin}.  
In other cases, however, bands form which have different finite values of $\dot\gamma$ \cite{CatesFielding, Besseling, Chikkadi}.  
% In some such systems the shear banding has been associated with an ordering phenomenon, such as the orientational ordering of non-spherical particles [dogic].  However shear banding has also been found in colloids of polydisperse spherical particles [schall].
Various theoretical models have been proposed \cite{Fielding,Dhont,Olmsted} for shear banding, and specific microscopic mechanisms have been suggested for specific materials \cite{CatesFielding, Besseling,Dogic,Bonn1,Bonn2,Bonn3}.  For more complete reviews see Refs.~\cite{Ovarlez} and \cite{Schall}.  It remains of interest to see how general the phenomenon of shear banding might be, and whether relatively simple mechanisms can produce shear banding in simple homogeneous systems of repulsive spheres. 

Another phenomenon that has received considerable recent attention is discontinuous shear thickening (DST).  In dense non-Brownian suspensions, the canonical example being cornstarch in water, increasing the shear strain rate can lead to a sudden and dramatic increase of shear stress \cite{Brown}.  There seems to be a growing consensus that inter-particle frictional interactions are key to understanding DST.
Several recent works have sought to explain DST as a transition from a state where hydrodynamic lubrication forces prevent particles from prolonged contact, to a state where such lubrication forces are overcome, particle contacts increase, and elastic Coulomb frictional forces at those contacts become important \cite{Fernandez,Seto,Mari,Cohen}.  Other works \cite{OH_friction,Heussinger,Grob, Grob2, Mari, WyartCates, Zippelius,Hayakawa1} have described DST as a precursor of the discontinuous jump in stress associated with the shear-driven jamming of frictional particles \cite{OH_friction, Bi_Nature}.  While both DST and shear banding have  been associated with discontinuous transitions, it remains unclear if there is a direct connection between these two phenomena.

In this work we report on numerical simulations of simple models of athermal, bidisperse, soft-core  disks in two dimensions.  We consider massive particles and energy dissipation occurs via particle collisions only.
While we take our particles to have no inter-granular elastic frictional force, we do include a tangential viscous dissipative force.  This tangential dissipation is key to the novel behaviors we find.  In one model, denoted as model ``CD", this dissipative force is taken proportional to the center of mass velocity difference between two particles in contact.  In a second model, denoted as ``CD$_\mathrm{rot}$", it is taken proportional to the local velocity difference between the two particles at their point of contact.  Model CD$_\mathrm{rot}$ thus introduces a non-trivial coupling between the translational and rotational motions of the particles.  Exploring the behavior of these models at low shear strain rates $\dot\gamma$, we map out the phase diagram as a function of packing fraction $\phi$, and a parameter $Q$ that measures the elasticity of collisions (see Figs.~\ref{Qc-vs-phi} and \ref{Qc-Qstar-phic}).  We find for both models that the $(\phi, Q)$ phase diagram includes a sharp first-order phase transition separating a region with Newtonian rheology from a region with Bagnoldian rheology.  In both cases, the system transitions to Newtonian behavior as one gets sufficiently close to the jamming transition.  The models differ in their behavior at small $Q$, corresponding to strongly inelastic collisions.  At small $Q$, model CD is Newtonian at all $\phi$, while CD$_\mathrm{rot}$  has a first-order Bagnoldian to Newtonian transition as $\phi$ increases.

For CD$_\mathrm{rot}$ we explore in greater depth the consequences of this rheological transition at small $Q=1$, as one varies the applied shear strain rate $\dot\gamma$.  We find that in the $(\phi,\dot\gamma)$ plane, the first-order rheological transition becomes a region of coexisting shear bands of Bagnoldian and Newtonian rheology.  The width of this coexistence region appears to vanish as $\dot\gamma\to 0$ at a critical $\phi_c$ that lies below the jamming $\phi_J$.  Upon increasing $\dot\gamma$ the coexistence region appears to vanish at a critical end-point.  Thus we find that model CD$_\mathrm{rot}$ displays shear banding within a well defined region of the $(\phi,\dot\gamma)$ phase space, and that this shear banding is a consequence of an underlying Bagnoldian to Newtonian first-order phase transition.

We next consider the behavior of the shear stress $\sigma$ as one crosses through the shear banded coexistence region by varying $\dot\gamma$ at fixed $\phi$.  We show that if $\dot\gamma$ is varied too rapidly for the system to relax to the shear-banded steady state corresponding to the instantaneous value of $\dot\gamma$,
then there can be large jumps in stress, as well as hysteresis upon cycling $\dot\gamma$ up and down through the coexistence region, as the system switches from Bagnoldian to Newtonian rheology or vice versa.  Thus the first-order shear banding transition becomes a possible mechanism for DST.

Our object in this work is not to claim that our model is a realistic description of any particular physical system.  Nor do we argue that friction is not an important element for the observed DST in real non-Brownian suspensions.  Our goal is rather to study a model, built with interactions commonly considered in the literature,  yet still sufficiently simple so that we may systematically investigate wide ranges of the rheological phase diagram.  We show that this phase diagram depends sensitively on the nature of the dissipative interaction, and that, even for frictionless particles, behavior may be richer than previously believed.

The remainder of our paper is organized as follows.  In Sec.~\ref{s2} we define our numerical models, the quantities we measure, and our method of simulation. In Sec~\ref{s3} we present our numerical results.  In Sec.~\ref{s3CD} we give results and map out the $(\phi,Q)$ phase diagram for low strain rates $\dot\gamma$ in model CD.  We characterize the varying nature of particle collisions as one crosses the Bagnoldian to Newtonian transition line.  In Sec.~\ref{s3CDrot} we give results and map out the $(\phi,Q)$ phase diagram at low $\dot\gamma$ for model CD$_\mathrm{rot}$.  We pay particular attention to how the rotational motion of the particles effects dissipation and leads to a different phase structure at small $Q$, as compared to model CD.  In Sec.~\ref{s3SB} we focus on model CD$_\mathrm{rot}$ in the limit of strongly inelastic collisions at $Q=1$, and show how the Bagnoldian to Newtonian rheological transition becomes a coexistence region of Bagnoldian and Newtonian shear bands.  In Sec.~\ref{s3DST} we show how crossing the coexistence region by ramping the strain rate $\dot\gamma$ up and down at fixed volume can lead to stress jumps and hysteresis. Finally in Sec.~\ref{s4} we summarize our conclusions and discuss directions for future investigation.  Appendixes A and B provide some technical details concerning the location of phase boundaries in model CD$_\mathrm{rot}$.

\section{Model and Simulation Method}
\label{s2}

We use a well studied model \cite{OHern} of frictionless, bidisperse, soft-core circular disks in two dimensions, with equal numbers of big and small particles with diameter ratio $d_b/d_s=1.4$.  Particles interact only when they come into contact,  in which case they repel with an elastic potential,
\begin{equation}
{\cal V}_{ij}(r_{ij})=\left\{
\begin{array}{cc}
\frac{1}{\alpha} k_e\left(1-r_{ij}/d_{ij}\right)^\alpha,&r_{ij}<d_{ij}\\
0,&r_{ij}\ge d_{ij}.
\end{array}
\right.
\label{eInteraction}
\end{equation}
Here $r_{ij}\equiv |\mathbf{r}_{ij}|$, where $\mathbf{r}_{ij}\equiv \mathbf{r}_i-\mathbf{r}_j$ is the center to center displacement from particle $j$ at position $\mathbf{r}_j$ to particle $i$ at $\mathbf{r}_i$, and $d_{ij}\equiv (d_i+d_j)/2$ is the average of their diameters.  In this work we will use the value $\alpha=2$, corresponding to a harmonic repulsion.
%We will measure energy in units such that $k_e=1$.
The resulting elastic force on particle $i$ from particle $j$ is,
\begin{equation}
\mathbf{f}_{ij}^\mathrm{el}=-\dfrac{d{\cal V}_{ij}(r_{ij})}{d\mathbf{r}_i} =\frac{k_e}{d_{ij}} \left(1-\frac{r_{ij}}{d_{ij}}\right)^{\alpha-1} \mathbf{\hat r}_{ij},
\label{efel}
\end{equation}
where $\mathbf{\hat r}_{ij}\equiv \mathbf{r}_{ij}/r_{ij}$ is the inward pointing normal direction at the surface of particle $i$.

Because shearing injects energy into the system, we also need to have a mechanism of energy dissipation so as to attain a shear driven steady state.  In the present work we will take dissipation to be due solely to particle collisions. We will consider two different models for this dissipation.  Both models will include a tangential viscous dissipation, but no tangential elastic friction.

\subsection{Model CD}

The first, simpler, model is to take the dissipative force proportional to the difference in the center of mass velocities of the two colliding particles,
\begin{equation}
\mathrm{model\>CD:}\qquad\mathbf{f}_{ij}^\mathrm{dis}=-k_d(\mathbf{v}_i-\mathbf{v}_j),
\label{efdisCD}
\end{equation}
where $\mathbf{v}_i=d\mathbf{r}_i/dt$ is the center of mass velocity of particle $i$.
With this dissipation our model becomes a massive version of the Durian ``bubble model" for foams \cite{Durian}, and we have earlier denoted this model as ``CD" for ``contact dissipation" \cite{VOT_rheology}.  This model has also been used in simulations of convection in granular materials \cite{Taguchi,Luding1,Luding2}. Note that the constant $k_d$ has different physical units from the constant $k_e$.

Given the above forces, particle motion is governed by the deterministic Newton's equation,
\begin{equation}
m_i\dfrac{d^2\mathbf{r}_i}{dt^2}=\sum_j\left[\mathbf{f}_{ij}^\mathrm{el}+\mathbf{f}_{ij}^\mathrm{dis}\right],
\label{eEqMotion}
\end{equation}
where $m_i$ is the mass of particle $i$ and the sum is over all particles $j$ in contact with particle $i$.  In this work we take all particles to have equal mass independent of their size, $m_i= m$.  

To apply a uniform shear strain rate $\dot\gamma$ in the $\mathbf{\hat x}$ direction, we use periodic Lees-Edwards boundary conditions \cite{LeesEdwards}, so that a particle at position $\mathbf{r}=(r_x,r_y)$ has images at positions $(r_x+mL+n\gamma L, r_y+nL)$, with $n$, $m$ integer and $\gamma=\dot\gamma t$ the total shear strain at time $t$. We then numerically integrate the equations of motion (\ref{eEqMotion}) to determine the particle trajectories, and from that the stress tensor of the system.

If we take the collisional forces to act on a particle at the point of contact, then $\mathbf{f}^\mathrm{el}_{ij}$, being normally directed, exerts no torque on the particle.  However $\mathbf{f}^\mathrm{dis}_{ij}$ can have a non-zero component tangent to the particle's surface at the point of contact, and so can exert a torque and cause the particles to rotate.  However, since such rotational motion does not feed back into the force $\mathbf{f}_{ij}^\mathrm{dis}$ of Eq.~(\ref{efdisCD}), in model CD the translational motion remains completely decoupled from the rotational motion and so rotational motion may be ignored when computing stresses.  

\subsection{Model CD$_\mathrm{rot}$}

This leads us to introduce our next model ``CD$_\mathrm{rot}$" in which rotational and translational motion couple in a non-trivial fashion.  In this model we take the dissipation to be proportional to the velocity difference of colliding particles {\em at the point of contact}, 
\begin{equation}
\mathrm{model\>CD}_\mathrm{rot}:\qquad\mathbf{f}_{ij}^\mathrm{dis}=-k_d(\mathbf{v}^{C_j}_i-\mathbf{v}^{C_i}_j),
\label{efdisCDrot}
\end{equation}
where $\mathbf{v}_i^{C_j}$ is the local velocity of particle $i$ at its point of contact with particle $j$. $\mathbf{v}_i^{C_j}$ thus includes a term arising from the rotation of the particle,
\begin{equation}
\mathbf{v}^{C_j}_{i} = \mathbf{v}_i+\dfrac{d\theta_i}{dt}\mathbf{\hat z}\times \mathbf{s}_{ji},\quad
\mathbf{s}_{ji}\equiv\mathbf{r}_{ji}\left(\frac{d_i}{d_i+d_j}\right).
\label{evC}
\end{equation}
Here $\theta_i$ gives the angular orientation of the disk and $\mathbf{s}_{ji}$ is the moment arm pointing from the center of particle $i$ to the point of contact with particle $j$.  

We now need to supplement Eq.~(\ref{eEqMotion}) for the translational motion with an equation for rotational motion,
\begin{equation}
I_i\dfrac{d^2\theta_i}{dt^2} = \mathbf{\hat z}\cdot \sum_j \mathbf{s}_{ji}\times \mathbf{f}^\mathrm{dis}_{ij},
\end{equation}
where $I_i=\tfrac{1}{2}m(d_i/2)^2$ is the moment of inertia of particle $i$, and the sum is over all particles $j$ in contact with $i$.  We note that such a coupling of rotational to translational motion is necessarily a feature of models that include inter-granular friction, although the tangential force in a frictional model is different from what we use here.

To generalize further, we might wish to split the dissipative force of Eq.~(\ref{efdisCDrot}) into components normal and tangential to the particle's surface at the point of contact, with separate dissipative constants $k_{d}^{(n)}$ and $k_{d}^{(t)}$ respectively.  
If we took $k_d^{(n)}>0$ but $k_d^{(t)}=0$, then again the translational motion decouples from rotational motion, there being no tangential force.  We have previously denoted \cite{VOT_rheology} this model as ``CD$_n$" (for normal contact dissipation) and it is well known to have Bagnoldian rheology, with stress $\sim\dot\gamma^2$,  for all values of $\phi$ and $Q$ \cite{VOT_CDn,VOT_CDnQ}.
In the present work, however,  we will keep $k_d^{(n)}=k_d^{(t)}\equiv k_d$ for simplicity.  
%\begin{align}
%\mathbf{f}^\mathrm{dis}_{ij}&=-k_{dn}[(\mathbf{v}_i-\mathbf{v}_j)\cdot\mathbf{\hat r}_{ij}]\mathbf{\hat r}_{ij}\\
%&- k_{dt}\left([(\mathbf{v}_i-\mathbf{v}_j)\cdot\mathbf{\hat t}_{ij}]\mathbf{\hat t}_{ij}\right.\\
%&+\left.\tfrac{1}{2}\mathrm{\hat z}\times \left[\dfrac{d\theta_i}{dt}d_i-\dfrac{d\theta_j}{dt}d_j\right]\right)
%\end{align}

\subsection{Times Scales and Dimensionless Quantities}

The above microscopic dynamics possesses two important microscopic time scales \cite{VOT_rheology}, the elastic and dissipative relaxation times,
\begin{equation}
\tau_e\equiv\sqrt{md_s^2/k_e},\qquad \tau_d\equiv m/k_d.
\label{etaus}
\end{equation}
We will measure time in units of $\tau_e$.
The parameter
\begin{equation}
Q\equiv \tau_d/\tau_e=\sqrt{mk_e/(k_dd_s)^2}
\end{equation}
measures the degree of elasticity of the collisions.
For the harmonic interaction  that we use, if we regarded the elastic potential of Eq.~(\ref{eInteraction}) as a spring which did {\em not} break when particles lose contact, then $2\pi\tau_e$ would give the undamped natural period of oscillation, $2\tau_d$ would be the decay time, and $Q$ would be the quality factor.
We will denote small $Q$, where the head-on collision of two particles is strongly inelastic, as the ``strongly inelastic" region; large $Q$ will then be denoted as the ``weakly inelastic" region.

It is also useful to define the time scale \cite{VOT_rheology},
\begin{equation}
\tau_0\equiv \tau_e^2/\tau_d=d_s^2k_d/k_e,
\end{equation}
which has a well defined value in the limit that the particle mass $m\to 0$.  This time scale will be useful for describing overdamped systems where inertial effects are unimportant.

%and is related to the coefficient of restitution $e$.  
%For the head-on collision of two particles $i$ and $j$, we have,
%\begin{equation}
%e=\mathrm{exp}\left[-\pi\middle/\sqrt{4\left(\dfrac{m_{ij}}{m_0}\right)\left(\dfrac{d_s}{d_{ij}}\right)^2 Q^2-1}\,\,\right],
%\end{equation}
%where $m_{ij}= m_im_j/(m_i+m_j)$ is the reduced mass of the two particles \cite{Schafer}.  When $Q<(d_{ij}/2d_s)\sqrt{m_0/m_{ij}}$, so that the argument of the square root would be negative, then the collision is completely inelastic with $e=0$.

Our system consists of a fixed total number particles $N$ in a square box of fixed length $L$.  $L$ is chosen to set the particle packing fraction $\phi$,
\begin{equation}
\phi=\dfrac{\pi N}{2L^2}\left[\left(\dfrac{d_s}{2}\right)^2+\left(\dfrac{d_b}{2}\right)^2\right].
\end{equation}
The behavior of our system is thus controlled by the three dimensionless parameters, $\phi$, $Q$, and the dimensionless strain rate $\dot\gamma\tau_e$.

To determine the global rheology of the system we measure the pressure tensor of each configuration, which consists of three pieces \cite{LeesEdwards}: the elastic part $\mathbf{p}^\mathrm{el}$, arising from the repulsive elastic forces of Eq.~(\ref{efel}),
\begin{equation}
\mathbf{p}^\mathrm{el}= \frac{1}{L^2}\sum_{i<j}\mathbf{f}^\mathrm{el}_{ij}\otimes\mathbf{r}_{ij},
\label{epel}
\end{equation}
the dissipative part $\mathbf{p}^\mathrm{dis}$, arising from the dissipative forces of Eqs.~(\ref{efdisCD}) or (\ref{efdisCDrot}),
\begin{equation}
\mathbf{p}^\mathrm{dis}= \frac{1}{L^2}\sum_{i<j}\mathbf{f}^\mathrm{dis}_{ij}\otimes\mathbf{r}_{ij},
\label{epdis}
\end{equation}
and the kinetic part $\mathbf{p}^\mathrm{kin}$ (sometimes called the streaming part),
\begin{equation}
\mathbf{p}^\mathrm{kin}=\frac{1}{L^2}\sum_i m_i\delta\mathbf{v}_i\otimes\delta\mathbf{v}_i,
\label{epkin}
\end{equation}
where $\delta\mathbf{v}_i$ is the fluctuation away from the average velocity profile that characterizes the shear flow.
%where $\delta\mathbf{v}_i\equiv \mathbf{v}_i-\dot\gamma y_i\mathbf{\hat x}$ is the fluctuation away from the linear average velocity profile that characterizes the uniform shear strain flow.  
The total pressure tensor is then,
\begin{equation}
\mathbf{p}=\mathbf{p}^\mathrm{el}+\mathbf{p}^\mathrm{dis}+\mathbf{p}^\mathrm{kin}.
\end{equation}
The average pressure $p$ and shear stress $\sigma$ in the system are then,
\begin{equation}
p=\frac{1}{2}\left[\langle p_{xx}\rangle +\langle p_{yy}\rangle\right],\quad
\sigma = -\langle p_{xy}\rangle,
\label{ep}
\end{equation}
where $\langle\dots\rangle$ represents an ensemble average over configurations in the sheared steady state.

It will be convenient to work in terms of dimensionless quantities.  We take the diameter of the small particles $d_s$, the mass $m$, and the time $\tau_e$ as our units of length, mass, and time respectively.  With these choices, stress in two dimensions is measured in units of $m/\tau_e^2$, and so we can define a dimensionless pressure tensor,
\begin{equation}
\mathbf{P}\equiv(\tau_e^2/m)\mathbf{p},
\end{equation}
with similarly,
\begin{equation}
P\equiv(\tau_e^2/m)p,\quad \Sigma\equiv(\tau_e^2/m)\sigma,
\end{equation}
as the dimensionless pressure and shear stress.

For a system with  Bagnoldian rheology (i.e. $p,\sigma\sim\dot\gamma^2$, as $\dot\gamma\to 0$) we define the dimensionless Bagnold transport coefficients,
\begin{equation}
B_p\equiv\frac{P}{(\dot\gamma\tau_e)^2}= \frac{p}{m\dot\gamma^2},\quad B_\sigma\equiv\frac{\Sigma}{(\dot\gamma\tau_e)^2}=\frac{\sigma}{m\dot\gamma^2}.
\label{eB}
\end{equation}
For a system with Newtonian rheology (i.e. $p,\sigma\sim\dot\gamma$, as $\dot\gamma\to 0$) we define the dimensionless transport coefficients in term of the components of $Q\mathbf{P}/(\dot\gamma\tau_e)=\mathbf{P}/(\dot\gamma\tau_0)$,
\begin{equation}
\eta_p\equiv \dfrac{P}{\dot\gamma\tau_0}=\dfrac{1}{k_d}\dfrac{p}{\dot\gamma},\quad\eta_\sigma\equiv\dfrac{\Sigma}{\dot\gamma\tau_0}=\dfrac{1}{k_d}\dfrac{\sigma}{\dot\gamma}.
\label{eeta}
\end{equation}
We define $\eta_p$ and $\eta_\sigma$ in terms of $\mathbf{P}/(\dot\gamma\tau_0)$ and not $\mathbf{P}/(\dot\gamma\tau_e)$ because we have previously shown \cite{VOT_rheology} that, for systems with Newtonian rheology, the first form becomes independent of $Q$ in the hard-core limit of sufficiently small $\dot\gamma\tau_0$.  $\eta_\sigma$ is just the dimensionless shear viscosity of the system, while $\eta_p$ is the pressure analog of viscosity.

In much of this work we will focus specifically on the elastic part of the pressure tensor, $\mathbf{p}^\mathrm{el}$.  In general we find that, at the densities and strain rates we are considering, the elastic part gives the largest contribution to the total $\mathbf{p}$, and it is also the part that is easiest to compute accurately in simulations.
Furthermore, any signature of a transition that we find in the elastic part must necessarily have a corresponding signature in the total stress.  We will use the superscript ``el" when referring to a quantity derived from only the elastic part of the pressure tensor, i.e., $P^\mathrm{el}$ for the elastic part of the pressure and $B_p^\mathrm{el}$ for the elastic part of the Bagnold coefficient.

%The dimensionalizing units in both Eqs.~(\ref{eB}) and (\ref{eeta}) are chosen so that the transport coefficients approach a well defined finite limit as $Q\to 0$ (or equivalently as $m\to 0$).

%We will refer to the limit $\dot\gamma\tau_e\to 0$ as the ``hard-core" limit. For a Newtonian system in the hard-core limit, $\eta_p$ and $\eta_\sigma$ will approach constants depending only on $\phi$ and $Q$; in a Bagnoldian system $B_p$ and $B_\sigma$ will approach constants.

\subsection{Simulation Method}

In our numerical simulations, we choose the diameter of the small particles to be $d_s=1$,
the mass $m=1$, and take the unit of time $\tau_e=1$ (which implies the elastic coupling
$k_e=1$). Since the net force $\mathbf{f}_i$ on particle $i$ depends on particle velocities, we use a modified velocity Verlet
algorithm: With the position and velocity of particle $i$ at times $t$ and $t+\Delta t$ denoted by $(\mathbf{r}_{i0},
\mathbf{v}_{i0})$ and $(\mathbf{r}_{i1}, \mathbf{v}_{i1})$, respectively, and the net force by $\mathbf{f}_i(\{\mathbf{r}_j,\mathbf{v}_j\})$, our equations are
%\begin{subequations}
\begin{eqnarray}
 \mathbf{\tilde v}_{i1} & = & \mathbf{v}_{i0} + \mathbf{f}_i(\{\mathbf{r}_{j0},\mathbf{v}_{j0}\}) \frac{\Delta t }{ m},\label{eq:tildev1}\\
\mathbf{ \tilde r}_{i1} & = & \mathbf{r}_{i0} + \mathbf{v}_{i0} \Delta t,\\
 \mathbf{v}_{i1} & = & \mathbf{v}_{i0} + \left[ \mathbf{f}_i(\{\mathbf{r}_{j0},\mathbf{v}_{j0}\})+ \mathbf{f}_i(\{\mathbf{\tilde r}_{j1},\mathbf{\tilde v}_{j1}\})\right] \frac{\Delta t }{2m},\label{eq:v1}\\
 \mathbf{r}_{i1} & = & \mathbf{r}_{i0} + (\mathbf{v}_{i0} + \mathbf{v}_{i1}) \frac{\Delta t }{ 2}.
\end{eqnarray}
%\end{subequations}

%\begin{subequations}
%\begin{eqnarray}
% \mathbf{\tilde v}_{i1} & = & \mathbf{v}_{i0} + \mathbf{f}_i(\mathbf{r}_{j0},\mathbf{v}_{j0}) \frac{\Delta t }{ m},\label{eq:tildev1}\\
%\mathbf{ \tilde r}_{i1} & = & \mathbf{r}_{i0} + \mathbf{v}_{i0} \Delta t,\\
% \mathbf{v}_{i1} & = & \mathbf{v}_{10} + [ \mathbf{f}_i(\mathbf{r}_{j0},\mathbf{v}_{j0}) + \mathbf{f}_i(\mathbf{\tilde r}_{j1},\mathbf{\tilde v}_{j1})] \frac{\Delta t }{2m},\label{eq:v1}\\
% \mathbf{r}_{i1} & = & \mathbf{r}_{i0} + (\mathbf{v}_{i0} + \mathbf{v}_{i1}) \frac{\Delta t }{ 2}.
%\end{eqnarray}
%\end{subequations}

The equivalent of Eqs.~(\ref{eq:tildev1}) and (\ref{eq:v1}) are used for the rotational
velocity $d\theta_i/dt\equiv\dot\theta_i$.  In general we use an integration time step $\Delta
t/\tau_e=0.05$. The exception is for $Q<0.5$ (large $k_d$) where too large a time step may
cause particles to unphysically reverse direction rather than just slow down; we then use
time steps down to $\Delta t/\tau_e=0.01$.

We simulate with two different system sizes. For the smaller size we have $N=1024$ particles, and most of our
presented results are for $\gdot\tau_e=10^{-5}$ and $10^{-6}$.  In these cases we simulate to a total shear strain
$\gamma=\gdot t$ of typically 100.
%, so as to be sure to have reached steady state. 
For our larger system of $N=65536$ particles we simulate for a range of strain rates
from $\gdot\tau_e=2\times 10^{-7}$ up to $10^{-3}$.  The length of our runs, as measured by the total shear  strain $\gamma$, varies substantially for different shear rates. However, even for our lowest
shear strain rate the run length is always $\gamma>1$.

\section{Results}
\label{s3}

\subsection{Model CD in the $\dot\gamma\tau_e\to 0$ limit}
\label{s3CD}

We first discuss our results for model CD.
In this section (as well as the next section for model CD$_\mathrm{rot}$) we are concerned with inferring the qualitative behavior of our system in the limit of small strain rate $\dot\gamma\tau_e\to 0$.  Because we are simulating over a wide range of parameters $(\phi, Q)$, we use a relatively small system with $N=1024$ particles.  Most of our results will be for the strain rate $\dot\gamma\tau_e=10^{-5}$, although at some $\phi$ we will also consider strain rates of $10^{-6}$ and $10^{-7}$ so as to infer the $\dot\gamma\tau_e\to 0$ limit.  But we caution that the phase diagram we find (see Fig.~\ref{Qc-vs-phi} for CD and Fig.~\ref{Qc-Qstar-phic} for CD$_\mathrm{rot}$) may be only approximate, with possible effects due to the finite values of $N$ and $\dot\gamma\tau_e$, especially as we get closer to critical points.

\subsubsection{Rheological transitions}

\begin{figure}
\includegraphics[width=3.2in]{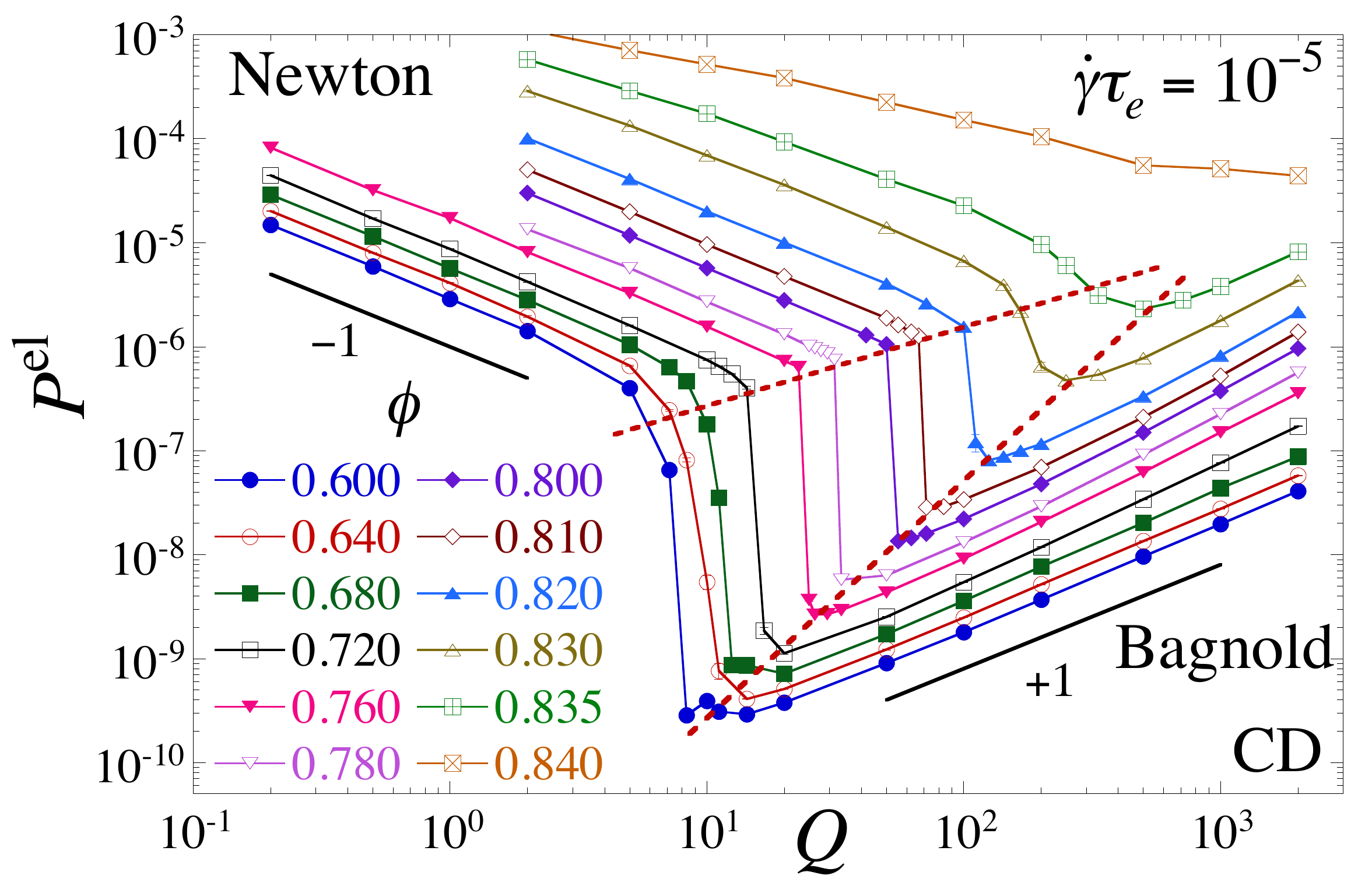} 
\caption{(Color online) Dimensionless pressure $P^\mathrm{el}$ vs collision elasticity parameter $Q$ in model CD, for different packing fractions $\phi=0.60$ to 0.84 as curves go from bottom to top.  Dashed lines indicate discontinuous jump in pressure between region of Newtonian rheology (low $Q$) and region of Bagnoldian rheology (high $Q$).  Strain rate is $\dot\gamma\tau_e=10^{-5}$ and system has $N=1024$ particles.  Short solid lines indicate the power law behaviors as denoted.
}
\label{pe-vs-Q}
\end{figure}

In Fig.~\ref{pe-vs-Q} we plot results for the elastic part of the dimensionless pressure $P^\mathrm{el}$
vs the collision elasticity parameter $Q$, for several different packing fractions $\phi$.  We show results for the strain rate $\dot\gamma\tau_e=10^{-5}$.  We have previously found for this model that the shear driven jamming transition takes place at $\phi_J\approx 0.8433$ \cite{VOT_RDCD}.  As $Q$ increases we see that $P^\mathrm{el}$ decreases roughly as $1/Q$, but then at a certain $Q_c(\phi)$ takes a sharp discontinuous drop.  Above $Q_c(\phi)$, $P^\mathrm{el}$ then increases roughly linearly in $Q$.  As $\phi$ increases, the magnitude of this drop decreases, and the width in $Q$ over which the drop occurs broadens somewhat. As will be shown below, this broadening as $\phi$ increases is a consequence of the finite strain rate $\dot\gamma\tau_e$; the drop will sharpen if one goes to lower values of $\dot\gamma\tau_e$.  Very similar behavior is seen for the shear stress $\sigma^\mathrm{el}$, with a discontinuous drop at the same $Q_c(\phi)$ as found for pressure.

\begin{figure}
\includegraphics[width=3.2in]{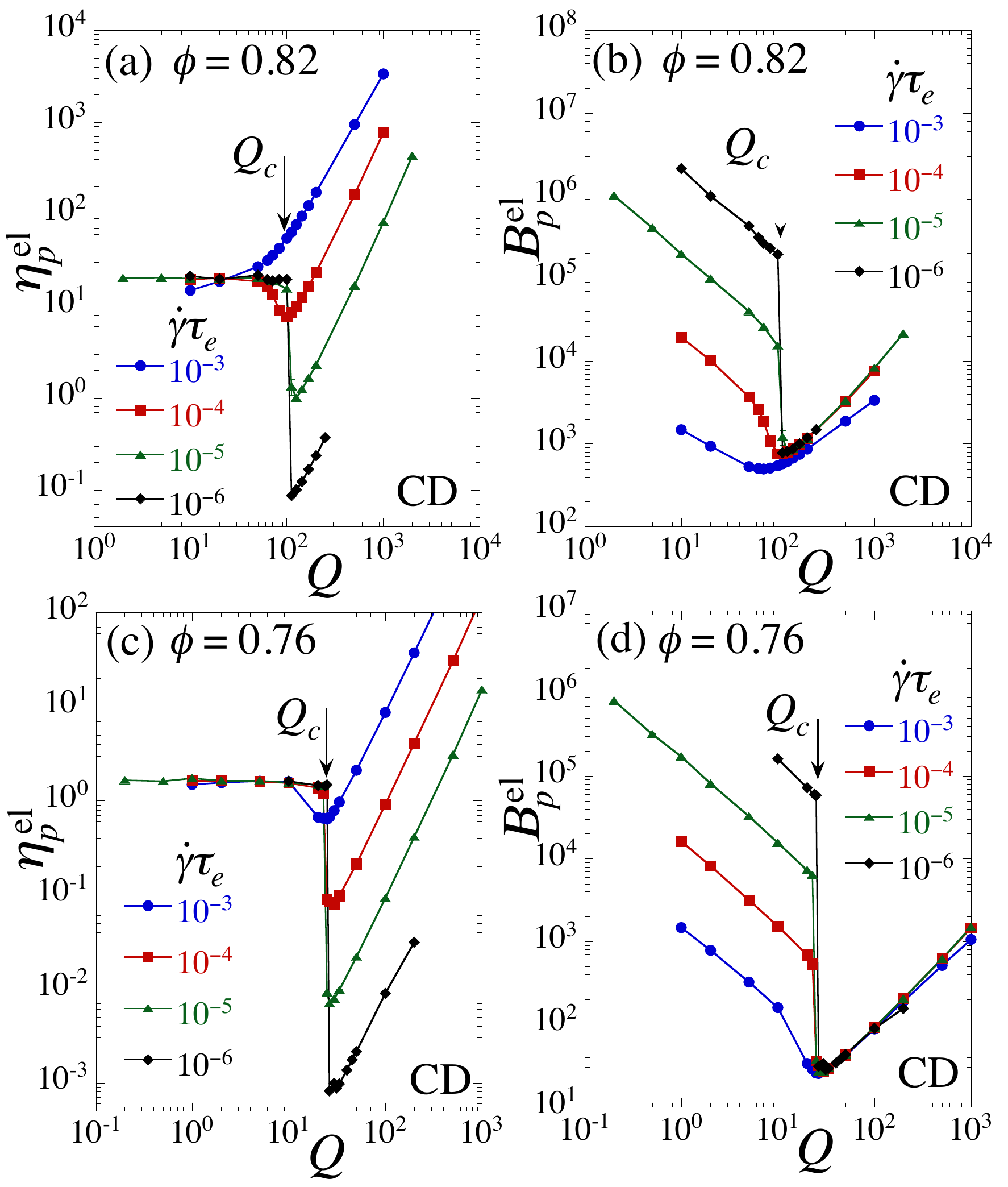} 
\caption{(Color online) (a) Pressure analog of viscosity, $\eta_p^\mathrm{el}=P^\mathrm{el}/(\dot\gamma\tau_0)=p^\mathrm{el}/(k_d\dot\gamma)$, and (b) Bagnold coefficient for pressure, $B_p^\mathrm{el}=P^\mathrm{el}/(\dot\gamma\tau_e)^2=p^\mathrm{el}/(m\dot\gamma^2)$, vs $Q$ for several different strain rates $\dot\gamma\tau_e$ at $\phi=0.82$ in model CD.  Panels (c) and (d) show similar results at $\phi=0.76$.  System has $N=1024$ particles.  In (a) and (c) $\dot\gamma\tau_e$ decreases as curves go from top to bottom; in (b) and (d) $\dot\gamma\tau_e$ decreases as curves go from bottom to top.
}
\label{etap-Bp-vs-Q}
\end{figure}

We now show that for $Q<Q_c(\phi)$  the system has Newtonian rheology, while for $Q>Q_c(\phi)$ the rheology is Bagnoldian.  In Figs.~\ref{etap-Bp-vs-Q}(a) and (b) we show the pressure analog of viscosity,
$\eta_p^\mathrm{el}=P^\mathrm{el}/(\dot\gamma\tau_0)$, and the Bagnold coefficient for pressure,  $B_p^\mathrm{el}=P^\mathrm{el}/(\dot\gamma\tau_e)^2$, vs $Q$ for several different strain rates $\dot\gamma\tau_e$ at the packing fraction $\phi=0.82$. 
For $Q<Q_c$ we see that $\eta_p^\mathrm{el}$ becomes independent of $\dot\gamma\tau_e$ for $\dot\gamma\tau_e$ sufficiently small, thus confirming Newtonian rheology.  Moreover, below $Q_c$ we find $\eta_p^\mathrm{el}$ to be independent of $Q$, as we have found previously \cite{VOT_rheology}.  This independence of $\eta_p^\mathrm{el}$ on $Q$ thus explains the $1/Q$ dependence of $P^\mathrm{el}$ seen in Fig.~\ref{pe-vs-Q} for $Q<Q_c$ (since $P^\mathrm{el}=\eta_p^\mathrm{el}\dot\gamma\tau_e/Q$).  For $Q>Q_c$ we see that $B_p^\mathrm{el}$ becomes independent of $\dot\gamma\tau_e$, thus confirming Bagnoldian rheology.  The roughly linear rise in $B_p^\mathrm{el}$ as $Q$ increases above $Q_c$ has been observed by us in other Bagnoldian systems at large $Q$ \cite{VOT_CDnQ}, and explains the roughly linear increase in $P^\mathrm{el}$ seen in Fig.~\ref{pe-vs-Q} for $Q>Q_c$.  In both cases we see that the drop at $Q_c$ becomes sharper as $\dot\gamma\tau_e$ decreases.  Similar results are shown for the lower packing fraction $\phi=0.76$ in Figs.~\ref{etap-Bp-vs-Q}(c) and (d).

\subsubsection{Phase diagram}

\begin{figure}[h!]
\includegraphics[width=3.2in]{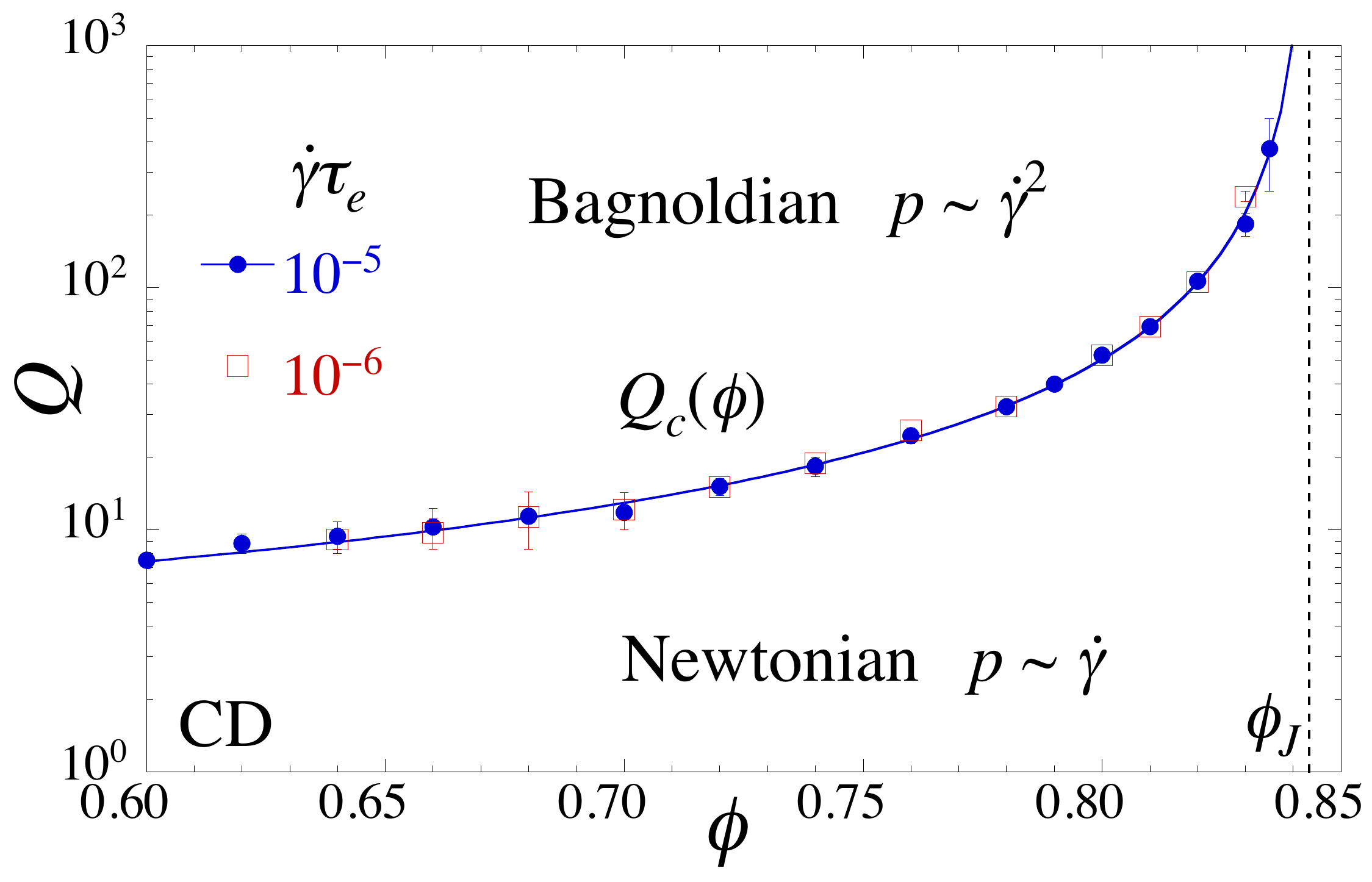} 
\caption{(Color online) Phase diagram for model CD in the $(\phi, Q)$ plane showing the location of the discontinuous transition $Q_c(\phi)$ between regions of Newtonian and Bagnoldian rheology. Solid circles are from simulations at $\dot\gamma\tau_e=10^{-5}$ while open squares are from $\dot\gamma\tau_e=10^{-6}$; we see complete agreement.  Solid line is a fit to $Q_c=Q_0 + c(\phi_J-\phi)^{-x}$ for the $\dot\gamma\tau_e=10^{-5}$ data, with fixed $\phi_J=0.8433$, and yields $x\approx 1.1\pm0.1$.  Vertical dashed line locates the jamming packing fraction $\phi_J$.
}
\label{Qc-vs-phi}
\end{figure}

We thus conclude that $Q_c(\phi)$ marks a sharp discontinuous transition from a region of Newtonian rheology ($Q<Q_c$) to a region of Bagnoldian rheology ($Q>Q_c$), in the $\dot\gamma\tau_e\to 0$ limit.  Estimating $Q_c$ from the midpoint of the sharp drop in $P^\mathrm{el}$ of Fig.~\ref{pe-vs-Q}, in Fig.~\ref{Qc-vs-phi} we plot $Q_c$ vs $\phi$, showing results from two different strain rates, $\dot\gamma\tau_e=10^{-5}$ and $10^{-6}$.  We find complete agreement between the two different strain rates, suggesting that our results reasonably give the $\dot\gamma\tau_e\to 0$ limit.  We see that $Q_c(\phi)$ appears to diverge as $\phi$ approaches the jamming $\phi_J$ from below.  Thus in model CD, at any fixed $Q$, one eventually enters the Newtonian region sufficiently close to $\phi_J$, and so the jamming criticality is always that of the Newtonian system.  Moreover, since we have seen in Figs.~\ref{etap-Bp-vs-Q}(a) and (c) that $\eta_p^\mathrm{el}$ is independent of $Q$ in the Newtonian region, the divergence of $\eta_p^\mathrm{el}$ as $\phi\to\phi_J$ must be the same for all values of $Q$.  Thus the value of $\phi_J$ and the exponent characterizing the divergence of $\eta_p^\mathrm{el}$ should be independent of $Q$; the only thing that depends on $Q$ is how close one must get to $\phi_J$ in order to enter the $Q$-independent Newtonian region.

Making a power law fit, $Q_c(\phi)=Q_0+c(\phi_J-\phi)^{-x}$, we find good agreement with $x\approx 1.1\pm0.1$ and $Q_0\approx 1.2$.  However our data is not sufficiently close to the jamming $\phi_J\approx 0.8433$ for us to know whether this exponent represents the true asymptotic behavior as $\phi\to \phi_J$, or is only an effective value for the range of $\phi$ we have considered.

\begin{figure}[h!]
\includegraphics[width=3.2in]{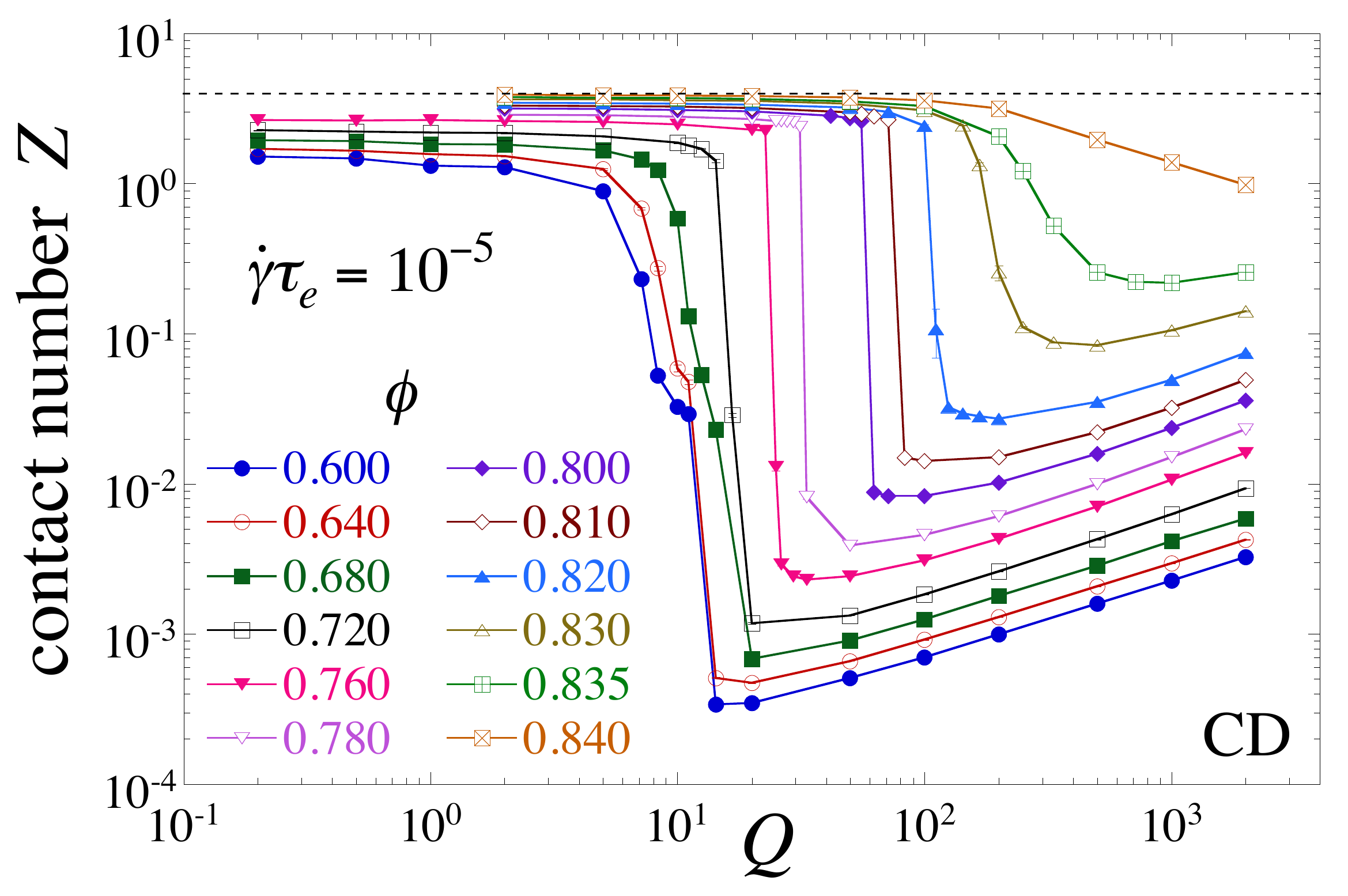} 
\caption{(Color online) Average instantaneous contact number per particle $Z$ vs $Q$ in model CD, for different packing fractions $\phi=0.60$ to 0.84 as curves go from bottom to top. Strain rate $\dot\gamma\tau_e=10^{-5}$ and system has $N=1024$ particles.  The sharp drop in $Z$ as $Q$ increases signals the transition from Newtonian to Bagnoldian rheology. For reference, the horizontal dashed line gives the isostatic value  $Z_\mathrm{iso}=2d=4$ for frictionless disks in two dimensions.
}
\label{z-vs-Q}
\end{figure}

We next ask if there is any obvious physical signature that distinguishes the Newtonian phase of our system from the Bagnoldian phase.  We find that in the Newtonian phase the particles tend to cluster together, with force chains that percolate throughout the system, while in the Bagnoldian phase particles have few instantaneous contacts \cite{VOT_rheology}.  This is seen by computing the average instantaneous number of contacts per particle $Z$.  In Fig.~\ref{z-vs-Q} we plot $Z$ vs $Q$ for several different packing fractions $\phi$,  for the shear strain rate $\dot\gamma\tau_e=10^{-5}$.  We see that $Z$ takes a sharp drop from values of $O(1)$ to values several orders of magnitude smaller, at the same $Q_c$ where $P^\mathrm{el}$ has it's drop.

%Moreover, the finite values of $Z$ found in Fig.~\ref{z-vs-Q} above $Q_c$ are a consequence of the finite strain rate.  We have previously shown that $Z\to 0$ as $\dot\gamma\tau_e\to 0$ for systems with Bagnoldian rheology.  

%We illustrate the differing particle connectivity of the two rheologies in Fig.~\ref{configs} where we show two instantaneous configurations of particles for the packing fraction $\phi=0.70$ at $\dot\gamma\tau_e=10^{-5}$.  Straight line segments connecting particles represent particles in contact, $r_{ij}\le d_{ij}$. For $Q<Q_c$ in Fig.~\ref{configs}a we see many contacts, forming a system spanning cluster; for $Q>Q_c$ in Fig.~\ref{configs}b we see few such contacts. 

\begin{figure}[h!]
\includegraphics[width=3.2in]{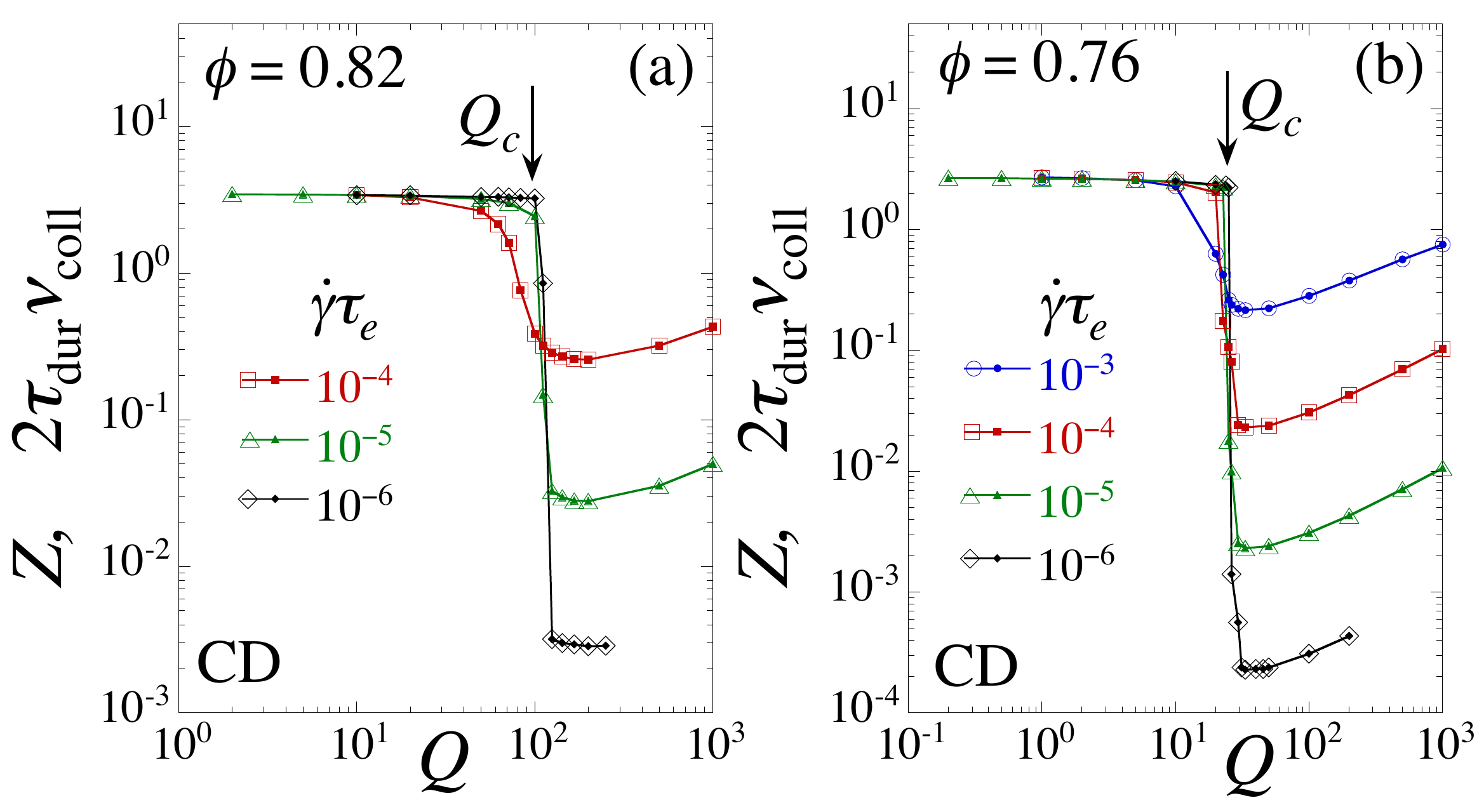} 
\caption{(Color online) Average instantaneous contact number per particle $Z$, and twice the product of the average collision duration time and collision rate $2\tau_\mathrm{dur}\nu_\mathrm{coll}$, vs $Q$ for several different strain rates $\dot\gamma\tau_e$ at (a) $\phi=0.82$ and (b) $\phi=0.76$  in model CD.  System has $N=1024$ particles.  Solid symbols are for $Z$, while open symbols are for $2\tau_\mathrm{dur}\nu_\mathrm{coll}$.  Strain rate $\dot\gamma\tau_e$ decreases as curves go from top to bottom. 
}
\label{Z-vs-Q-gdot}
\end{figure}

In Fig.~\ref{Z-vs-Q-gdot} we plot $Z$ vs $Q$ for several different strain rates $\dot\gamma\tau_e$, at $\phi=0.82$ and 0.76.  We see that in the Newtonian phase $Q<Q_c$, as $\dot\gamma\tau_e$ decreases $Z$ becomes independent of $\dot\gamma\tau_e$ and independent of $Q$.  Thus the Newtonian phase is characterized by a finite  particle connectivity even as $\dot\gamma\tau_e\to 0$.  In the Bagnoldian phase $Q>Q_c$, we see that $Z$ steadily decreases as $\dot\gamma\tau_e$ decreases.  Below we will show that in the Bagnoldian phase the average contact number vanishes linearly, $Z\sim\dot\gamma\tau_e$, as $\dot\gamma\tau_e\to 0$.

To understand better the behavior of $Z$, we consider two time scales associated with particle collisions.  We define the collision duration time $\tau_\mathrm{dur}$ as the average time from the initiation of a particle contact until the breaking of that contact.  We define the collision rate $\nu_\mathrm{coll}$ as the average number of collisions per unit time divided by the number of particles.  As has been argued in Ref.~\cite{OHL}, and as we have shown previously for model CD$_n$ \cite{VOT_CDnQ}, $\tau_\mathrm{dur}$ and $\nu_\mathrm{coll}$ are related to the average contact number $Z$ by,
\begin{equation}
Z=2\tau_\mathrm{dur}\nu_\mathrm{coll}.
\label{Ztaunu}
\end{equation}
The product $\tau_\mathrm{dur}\nu_\mathrm{coll}$ measures the fraction of the time between two collisions that a contact on average persists, while the factor 2 is because each contact is shared by two particles.
In Fig.~\ref{Z-vs-Q-gdot} we show $2\tau_\mathrm{dur}\nu_\mathrm{coll}$ vs $Q$ on the same plot as $Z$.  Although these two quantities are measured by completely different methods within the simulation, we find perfect agreement, thus providing a consistency check on our calculations.

\begin{figure}[h!]
\includegraphics[width=3.2in]{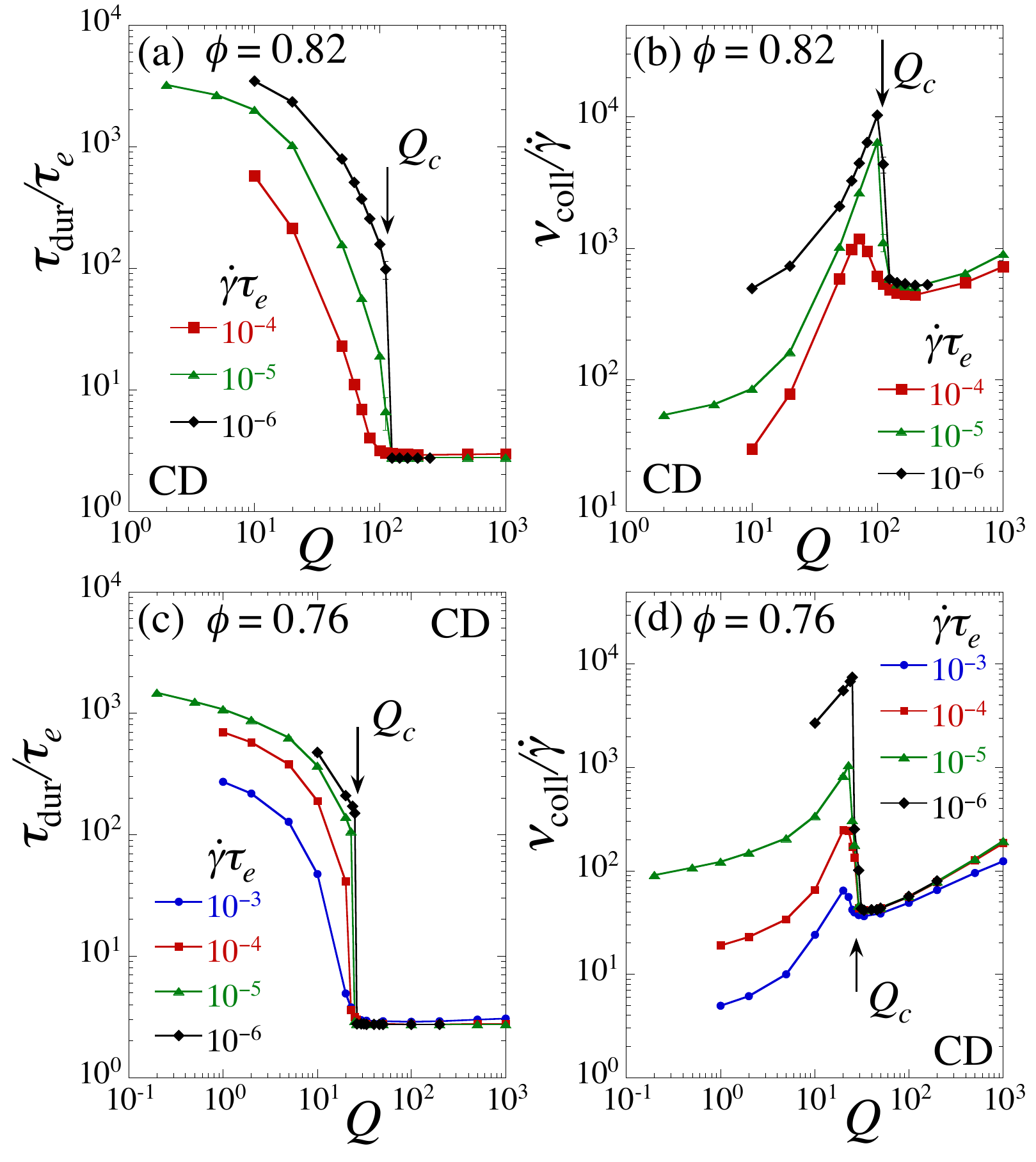} 
\caption{(Color online) (a) Average contact duration time $\tau_\mathrm{dur}/\tau_e$ and (b) average collision rate normalized by the shear strain rate $\nu_\mathrm{coll}/\dot\gamma$, vs $Q$ for several different strain rates $\dot\gamma\tau_e$ at $\phi=0.82$  in model CD.  Panels (c) and (d) show similar results at $\phi=0.76$.  System has $N=1024$ particles. 
Strain rate $\dot\gamma\tau_e$ decreases as curves go from bottom to top. 
}
\label{tau-nu-vs-Q}
\end{figure}

We now consider $\tau_\mathrm{dur}$ and $\nu_\mathrm{coll}$ separately.  In Figs.~\ref{tau-nu-vs-Q}(a) and (b) we plot $\tau_\mathrm{dur}/\tau_e$ and $\nu_\mathrm{coll}/\dot\gamma$ vs $Q$ for several different strain rates $\dot\gamma\tau_e$ at $\phi=0.82$.  We see that for $Q>Q_c$ in the Bagnoldian region, $\tau_\mathrm{dur}/\tau_e$ and $\nu_\mathrm{coll}/\dot\gamma$ become independent of $\dot\gamma\tau_e$ at the smallest $\dot\gamma\tau_e$, thus implying that $Z=2\tau_\mathrm{dur}\nu_\mathrm{coll}\sim\dot\gamma\tau_e$ vanishes as $\dot\gamma\tau_e\to 0$.
In the Bagnoldian region, $\tau_\mathrm{dur}/\tau_e\sim O(1)$, as would be the case for an isolated binary collision.
In the Newtonian region $Q<Q_c$, $\tau_\mathrm{dur}/\tau_e$ rapidly increases by several orders of magnitude as $Q$ decreases, while $\nu_\mathrm{coll}$ decreases less rapidly than linearly in $\dot\gamma$ (so that $\nu_\mathrm{coll}/\dot\gamma$ increases as $\dot\gamma$ decreases),  so as to keep $Z=2\tau_\mathrm{dur}\nu_\mathrm{coll}$ constant.  Similar results are shown in Figs.~\ref{tau-nu-vs-Q}(c) and (d) for the lower $\phi=0.76$.

%\begin{figure}[h!]
%\includegraphics[width=3.2in]{configs} 
%\caption{Instantaneous configurations at packing fraction $\phi=0.70$ and strain rate $\dot\gamma\tau_e=10^{-5}$ for (a) $Q=5<Q_c$ in the region of Newtonian rheology, and (b) $Q=10>Q_c$ in the region of Bagnoldian rheology, for model CD.  The straight line segments connecting particles represent particles in contact.
%}
%\label{configs}
%\end{figure}

%We can give a qualitative explanation for this transition from clustered particles and Newtonian rheology, to isolated particles and Bagnoldian rheology, as follows.
%Since the dissipative force of Eq.~\ref{efdisCD} is proportional to the center of mass velocity difference between particles, collisions serve to damp out relative motion between a pair of colliding particles.  If $Q$ is large (collisions weakly inelastic) particles will exit a collision with a reduced but still finite relative velocity, and so separate.  But if  $Q$ is sufficiently small (i.e. $k_d$ sufficiently large, and so collisions strongly inelastic), all relative motion between colliding particles will be damped out after the collision, so that $\mathbf{v}_i=\mathbf{v}_j$ and the particles stick together.  Increasing $\phi$ reduces the available free volume in which particles can move and so greatly restricts normal relative motion between particles at high densities.  This serves to renormalize the critical $Q_c$ upwards as $\phi$ increases. 

\subsubsection{Geometry of collisions}

\begin{figure}[h!]
\includegraphics[width=3.2in]{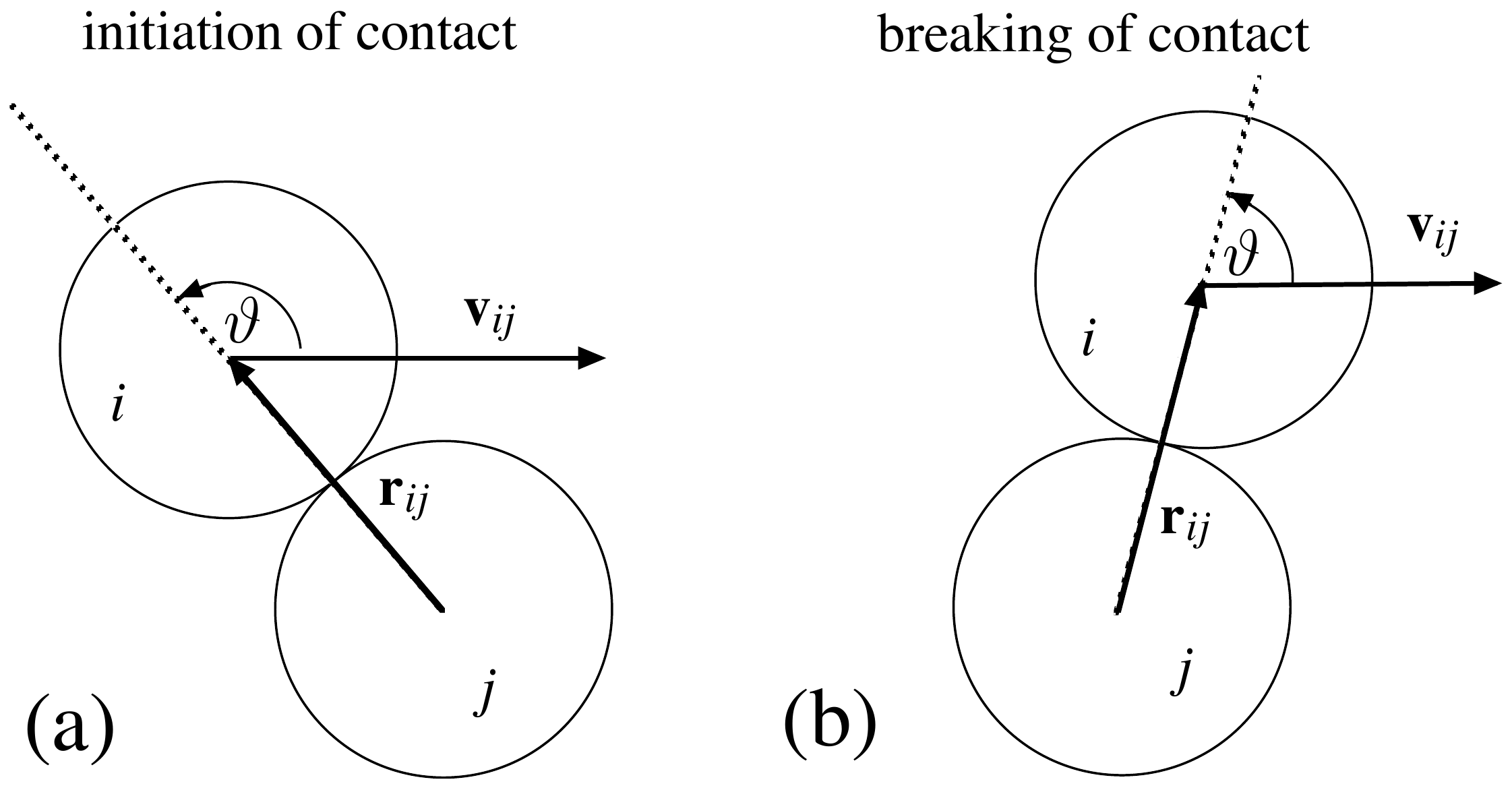} 
\caption{Schematic of the collision of two particles $i$ and $j$. (a) Initiation of contact, where $|\vartheta|>90^\circ$, and (b) breaking of contact, where $|\vartheta|<90^\circ$.  Here $\mathbf{r}_{ij}\equiv \mathbf{r}_i-\mathbf{r}_j$ and $\mathbf{v}_{ij}\equiv \mathbf{v}_i-\mathbf{v}_j$. 
}
\label{collisions}
\end{figure}

Next, we look more directly at the geometry of individual collisions in the phases above and below $Q_c$.
%As a final comment on the differing nature of configurations above and below $Q_c$, we consider the geometry of the collisions.   
To measure this, let us define,
\begin{equation}
\mathbf{r}_{ij}\equiv \mathbf{r}_i-\mathbf{r}_j,\quad \mathbf{v}_{ij}\equiv\mathbf{v}_i-\mathbf{v}_j,
\end{equation}
as the position and velocity of particle $i$ with respect to particle $j$.  We then define the angle $\vartheta$ as the angle by which one must rotate $\mathbf{v}_{ij}$ to align it parallel with $\mathbf{r}_{ij}$,
\begin{equation}
\mathbf{\hat v}_{ij}\cdot\mathbf{\hat r}_{ij} =\cos\vartheta,\quad -180^\circ<\vartheta\le 180^\circ.
\end{equation}
For two particles just initiating a contact, we must have $\cos\vartheta< 0$, so that the particles are driven into each other, as illustrated in Fig.~\ref{collisions}(a).  In this case we must have $|\vartheta|>90^\circ$.  For two particles just breaking a contact, we must have $\cos\vartheta> 0$, so that the particles are driven away from each other, as illustrated in Fig.~\ref{collisions}(b).  In this case we must have $|\vartheta|<90^\circ$.

\begin{figure}[h!]
\includegraphics[width=3.2in]{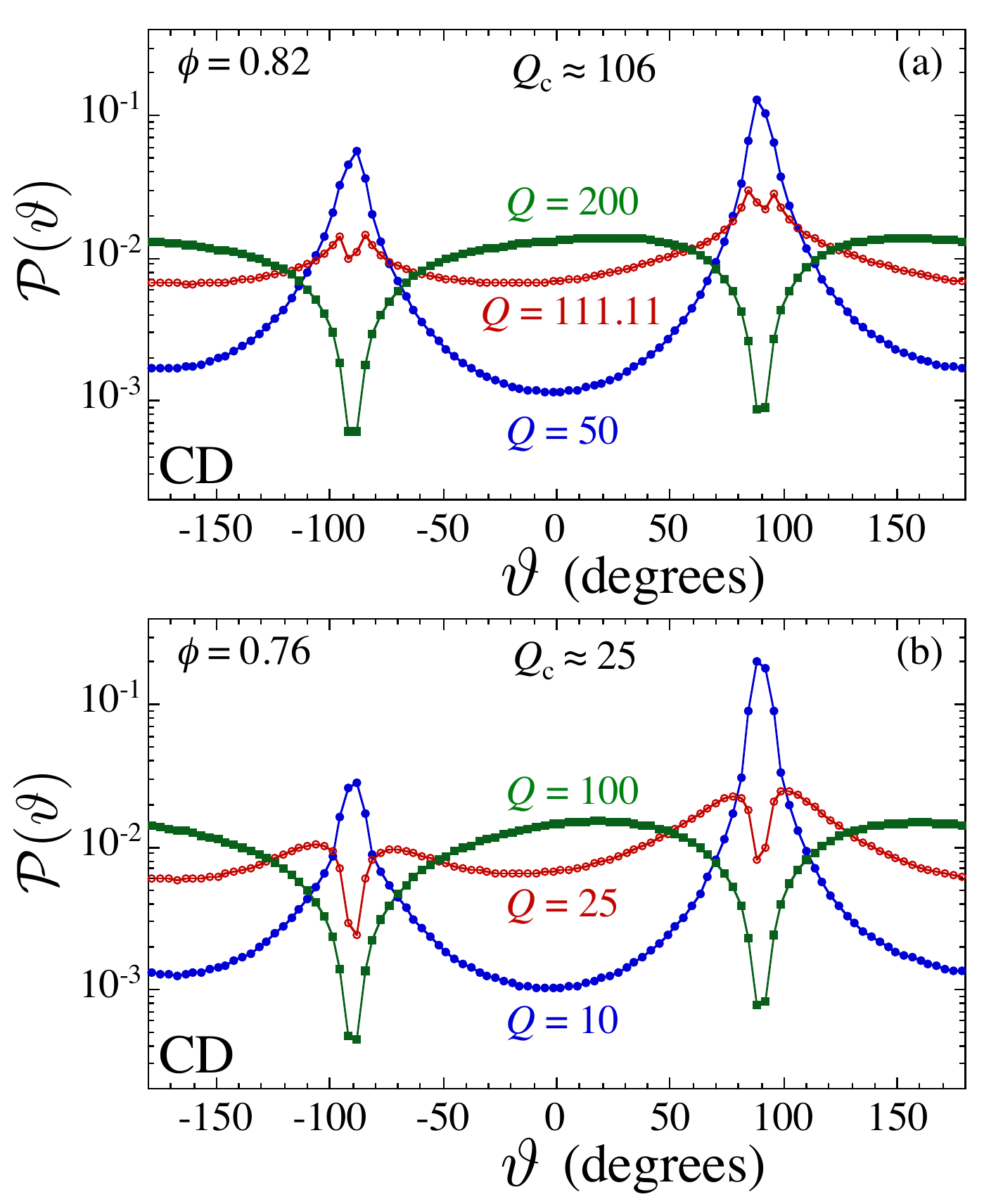} 
\caption{(Color online) Histograms ${\cal P}(\vartheta)$ vs collision angle $\vartheta$, at initiation and breaking of particle contacts (see Fig.~\ref{collisions} for definition of $\vartheta$), at (a) $\phi=0.82$ and (b) $\phi=0.76$ in model CD.  In each case we show results for one value $Q>Q_c$, one value $Q<Q_c$, and for $Q\approx Q_c$.  The value of $Q_c$ is as indicated in each panel.
Results are for the shear strain rate $\dot\gamma\tau_e=10^{-5}$ in a system with $N=1024$ particles. 
}
\label{Ptheta}
\end{figure}

\begin{figure}[h!]
\includegraphics[width=3.2in]{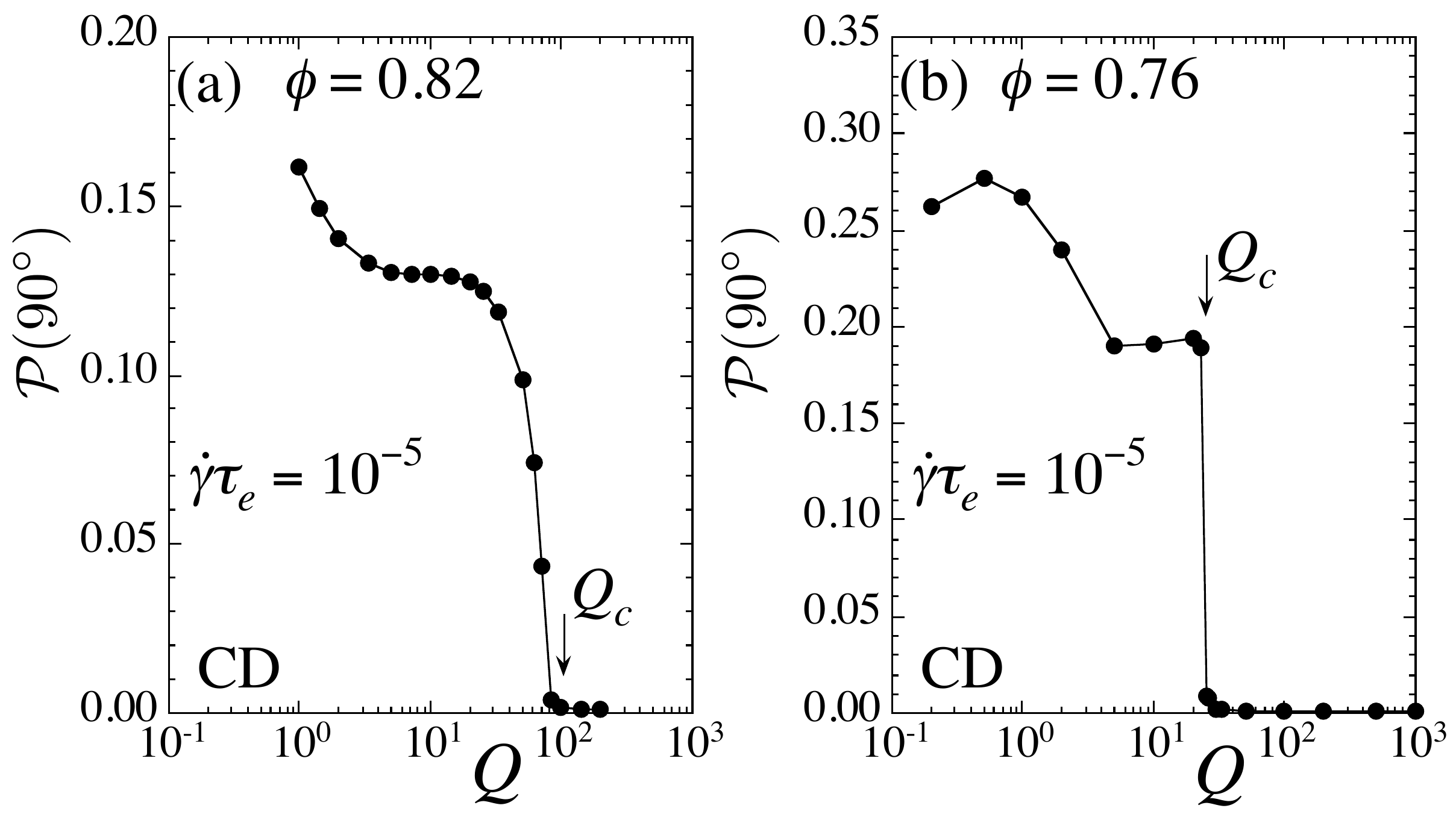} 
\caption{Histogram ${\cal P}(\vartheta=90^\circ)$ vs $Q$ for (a) $\phi=0.82$ and (b) $\phi=0.76$.  The propensity to have tangential collisions, i.e. $\vartheta=90^\circ$, increases sharply as one crosses from the Bagnoldian ($Q>Q_c$) to the Newtonian ($Q<Q_c$) region.
Results are for the shear strain rate $\dot\gamma\tau_e=10^{-5}$ in a system with $N=1024$ particles. 
}
\label{P90}
\end{figure}

Measuring the value of $\vartheta$ each time a contact is initiated and each time a contact is broken, we construct a histogram ${\cal P}(\vartheta)$ which combines both contact initiation and contact breaking events.  In Fig.~\ref{Ptheta} we plot ${\cal P}(\vartheta)$ vs $\vartheta$ for the two packing fractions $\phi=0.82$ and 0.76.  We use a bin width of $\Delta\vartheta=3.6^\circ$ in constructing this histogram.  In each case we show ${\cal P}(\vartheta)$ for three different values of $Q$: one well above $Q_c$, one well below $Q_c$, and one approximately at $Q_c$.  We see that for $Q>Q_c$, the histogram ${\cal P}(\vartheta)$ has minima at $\vartheta=\pm 90^\circ$ and is broadly distributed between; collisions take place with a broad range of impact parameter,
$b=|\mathbf{r}_{ij}\times\mathbf{\hat v}_{ij}|$,  including many normally oriented (i.e. head-on)  collisions.  For $Q<Q_c$, however, ${\cal P}(\vartheta)$ is strongly peaked at $\vartheta=\pm 90^\circ$, indicating that collision are primarily glancing, with particles approaching each other close to tangentially.
In Fig.~\ref{P90} we plot the value of the histogram ${\cal P}(\vartheta=90^\circ)$ vs $Q$, as a measure of the propensity for tangential collisions.  We show results at $\phi=0.82$ and $0.76$ at $\dot\gamma\tau_e=10^{-5}$.  We see that ${\cal P}(90^\circ)$ is essentially zero for $Q>Q_c$ in the Bagnoldian region, and then takes a sharp jump upwards at $Q=Q_c$ with the transition to the Newtonian region.

Looking at behavior more closely at the level of individual collisions, we find the following.  For $Q>Q_c$ in the Bagnoldian region, collisions are mostly isolated binary collisions, where a pair of particles collide with a wide range of impact parameter, and then separate before coming into contact with another particle. If we define $v_{ij,N}$ and $v_{ij,T}$ as the root mean square average of the normal and tangential components of $\mathbf{v}_{ij}$ respectively, then for $Q>Q_c$ we find $v_{ij,N}\approx v_{ij,T}$, with both decreasing as $Q$ decreases.  As $Q$ decreases below $Q_c$, however, $v_{ij,N}$ takes a sharp drop to small values, while $v_{ij,T}$ plateaus to a constant independent of $Q$.  The drop in $v_{ij,N}$ results from the increasing inelasticity of the collisions as $Q$ gets small, combined with the steric interactions of excluded volume at large particle densities; the decrease in free volume as $\phi$ increases, results in the increase of $Q_c(\phi)$. The constant value of $v_{ij,T}$ results from the driving shear strain rate $\dot\gamma$, which forces particles to move relative to one another in shear flow, even for the smallest $Q$.  Thus below $Q_c$ the relative motion between particles is largely tangential.  For such tangential motion, the tangential component of the dissipative forces resists relative particle motion, while the applied $\dot\gamma$ requires such motion.  This results in the build up of extended force chains of elastic force, that drive the particles forward with the required $\dot\gamma$.  Such force chains result in a sharp increase in the average particle contact number $Z$ and the elastic pressure $P^\mathrm{el}$.
We leave further investigation of the detailed connection between particle collisions and the transition from Newtonian to Bagnoldian rheology to future work.

%Looking at behavior more closely in individual configurations, we find the following picture suggested.  For $Q>Q_c$ in the Bagnoldian region, collisions are mostly isolated binary collisions, where a pair of particles collide with a wide range of impact parameter, and then separate before coming into contact with another particle.  For $Q<Q_c$ in the Newtonian region, normally directed relative motion between particles becomes strongly damped and the relative motion between particles is constrained to be mostly tangential.  This tangential motion leads to large dissipative forces due to the tangential component of $\mathbf{f}^\mathrm{dis}_{ij}$ of Eq.~(\ref{efdisCD}).  These dissipative forces would act to slow the motion of the particles, but since the average speed of the particles is fixed by the applied strain rate, there must be a compensating elastic force generated to keep the particles moving according to the fixed strain rate.  This leads to larger particle overlaps and so larger pressure, 
%which in turn leads to larger collision duration times $\tau_\mathrm{dur}$, and so large connected clusters of particles, as compared to the Bagnoldian region.  We may think of $Q_c(\phi)$ as the density renormalized critical damping parameter, below which collisions become completely inelastic and particles stick together after colliding.  

\subsection{Model CD$_\mathrm{rot}$ in the $\dot\gamma\tau_e\to 0$ limit}
\label{s3CDrot}

We now consider behavior in model CD$_\mathrm{rot}$, in which the rotational and translational degrees of freedom are coupled.  Our results in this section are again for a model with $N=1024$ particles.  Our main results are for a strain rate $\dot\gamma\tau_e=10^{-5}$, although we also use $10^{-6}$ and $10^{-7}$ at several values of $\phi$ so as to better illustrate and locate the rheological phase transitions.

\subsubsection{Rheological transitions}

\begin{figure}[h!]
\includegraphics[width=3.2in]{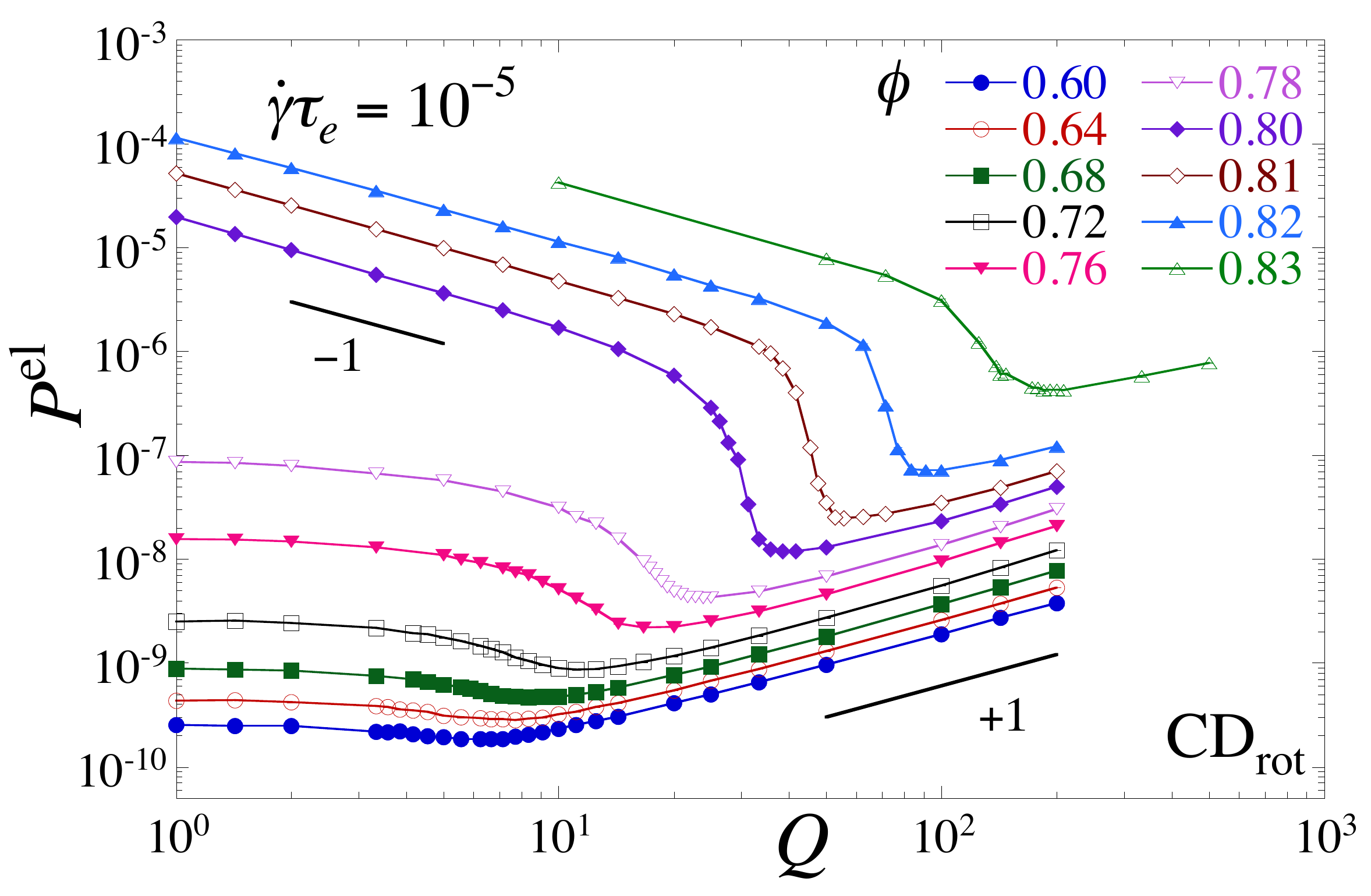} 
\caption{(Color online) Elastic part of the dimensionless pressure $P^\mathrm{el}$ vs collision elasticity parameter $Q$ in model CD$_\mathrm{rot}$, for different packing fractions $\phi=0.60$ to 0.83 as curves go from bottom to top.   Strain rate is $\dot\gamma\tau_e=10^{-5}$ and system has $N=1024$ particles.
}
\label{pe-vs-Q-rot}
\end{figure}

In Fig.~\ref{pe-vs-Q-rot} we plot results for the elastic part of the dimensionless pressure $P^\mathrm{el}$ vs the collision elasticity parameter $Q$, for different packing fractions $\phi$ and strain rate $\dot\gamma\tau_e=10^{-5}$.  For the higher packing fractions, $\phi\gtrsim 0.80$, our results for CD$_\mathrm{rot}$ appear qualitatively the same as found in Fig.~\ref{pe-vs-Q} for model CD; as $Q$ increases, $P^\mathrm{el}$ initially decreases roughly as $\sim1/Q$, until a $Q_c$ is reached at which $P^\mathrm{el}$ takes a sharp drop, and then $P^\mathrm{el}$ increases roughly linearly as $Q$ increases further.  We will see below that this drop at $Q_c$ corresponds to a sharp transition from Newtonian to Bagnoldian rheology, as in model CD.
For smaller $\phi < 0.80$, however, the drop becomes broadened over a wider range of $\phi$, and $P^\mathrm{el}$ appears to be approaching a constant as $Q$ further decreases.  For the smallest $\phi$, the jump almost disappears, leaving only a kink between the small $Q$ and large $Q$ behaviors.  We will see below that this drop/kink corresponds to a non-singular crossover $Q^*$ between two different regions of Bagnoldian rheology.

\begin{figure}[h!]
\includegraphics[width=3.2in]{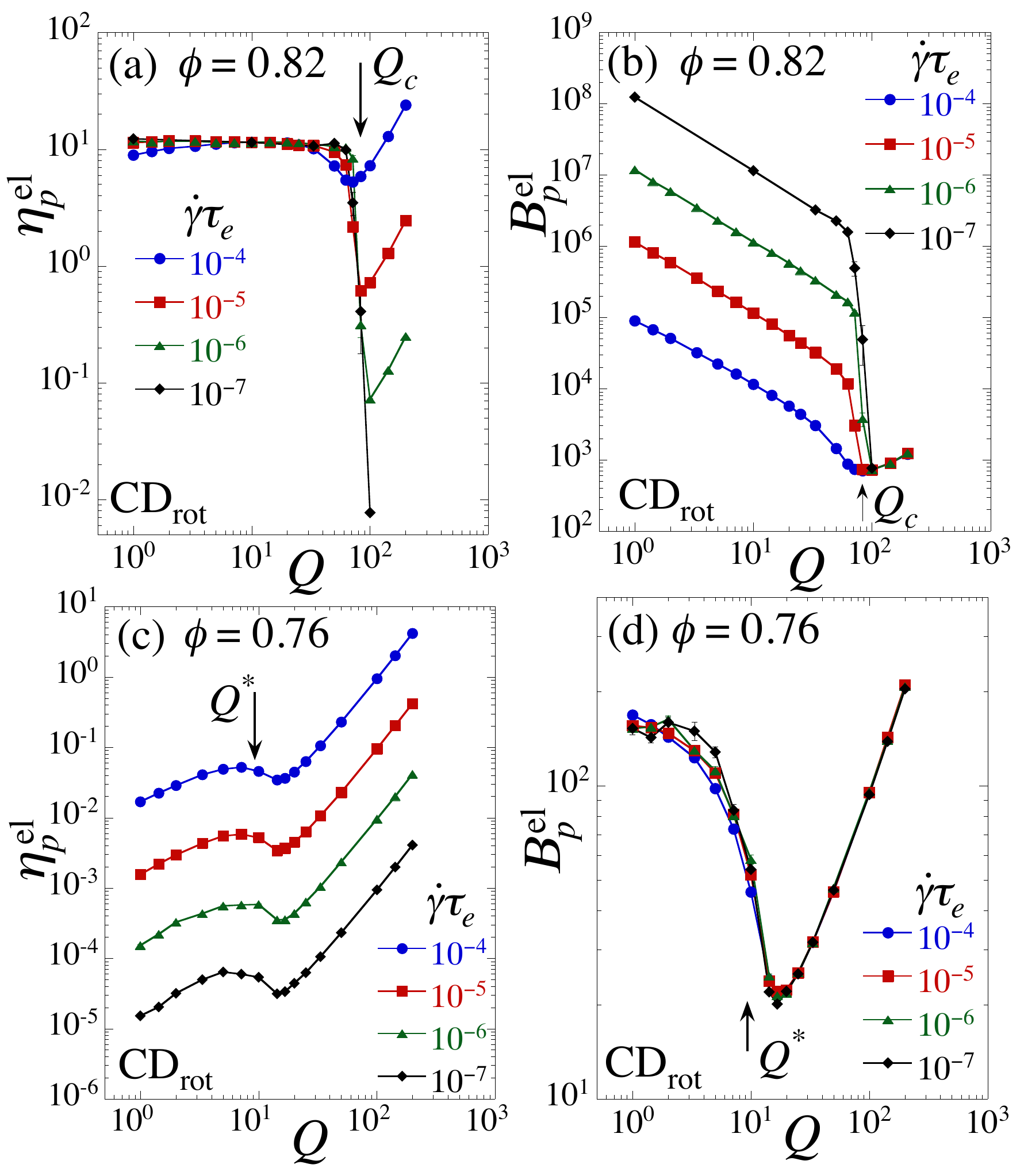} 
\caption{(Color online) (a) Pressure analog of viscosity, $\eta_p^\mathrm{el}=P^\mathrm{el}/(\dot\gamma\tau_0)=p^\mathrm{el}/(k_d\dot\gamma)$, and (b) Bagnold coefficient for pressure, $B_p^\mathrm{el}=P^\mathrm{el}/(\dot\gamma\tau_e)^2=p^\mathrm{el}/(m\dot\gamma^2)$, vs $Q$ for several different strain rates $\dot\gamma\tau_e$ at $\phi=0.82$   in model CD$_\mathrm{rot}$.  Panels (c) and (d) show similar results for $\phi=0.76$.  System has $N=1024$ particles.  In (a) and (c) $\dot\gamma\tau_e$ decreases as curves go from top to bottom; in (b) $\dot\gamma\tau_e$ decreases as curves go from bottom to top.
}
\label{etap-Bp-vs-Q-rot}
\end{figure}

In Figs.~\ref{etap-Bp-vs-Q-rot}(a) and (b) we plot $\eta_p^\mathrm{el}$ and $B_p^\mathrm{el}$ for strain rates $\dot\gamma\tau_e=10^{-4}$ to $10^{-7}$, at $\phi=0.82$.  We see that the behavior is qualitatively the same as in Figs.~\ref{etap-Bp-vs-Q}(a) and (b) for model CD; $\eta_p^\mathrm{el}$ is independent of $\dot\gamma\tau_e$ for $Q<Q_c$, while $B_p^\mathrm{el}$ is independent of $\dot\gamma\tau_e$ for $Q>Q_c$.  We thus conclude that the rheology is Newtonian for $Q<Q_c$, but Bagnoldian for $Q>Q_c$.  In Figs.~\ref{etap-Bp-vs-Q-rot}(c) and (d) we similarly plot $\eta_p^\mathrm{el}$ and $B_p^\mathrm{el}$ for the lower packing fraction $\phi=0.76$.  Here we see rather different behavior; $\eta_p^\mathrm{el}$ keeps decreasing as $\dot\gamma\tau_e$ decreases, while $B_p^\mathrm{el}$ is essentially independent of $\dot\gamma\tau_e$ at all $Q$, indicating that the rheology is Bagnoldian both above and below $Q^*$.

\begin{figure}[h!]
\includegraphics[width=3.2in]{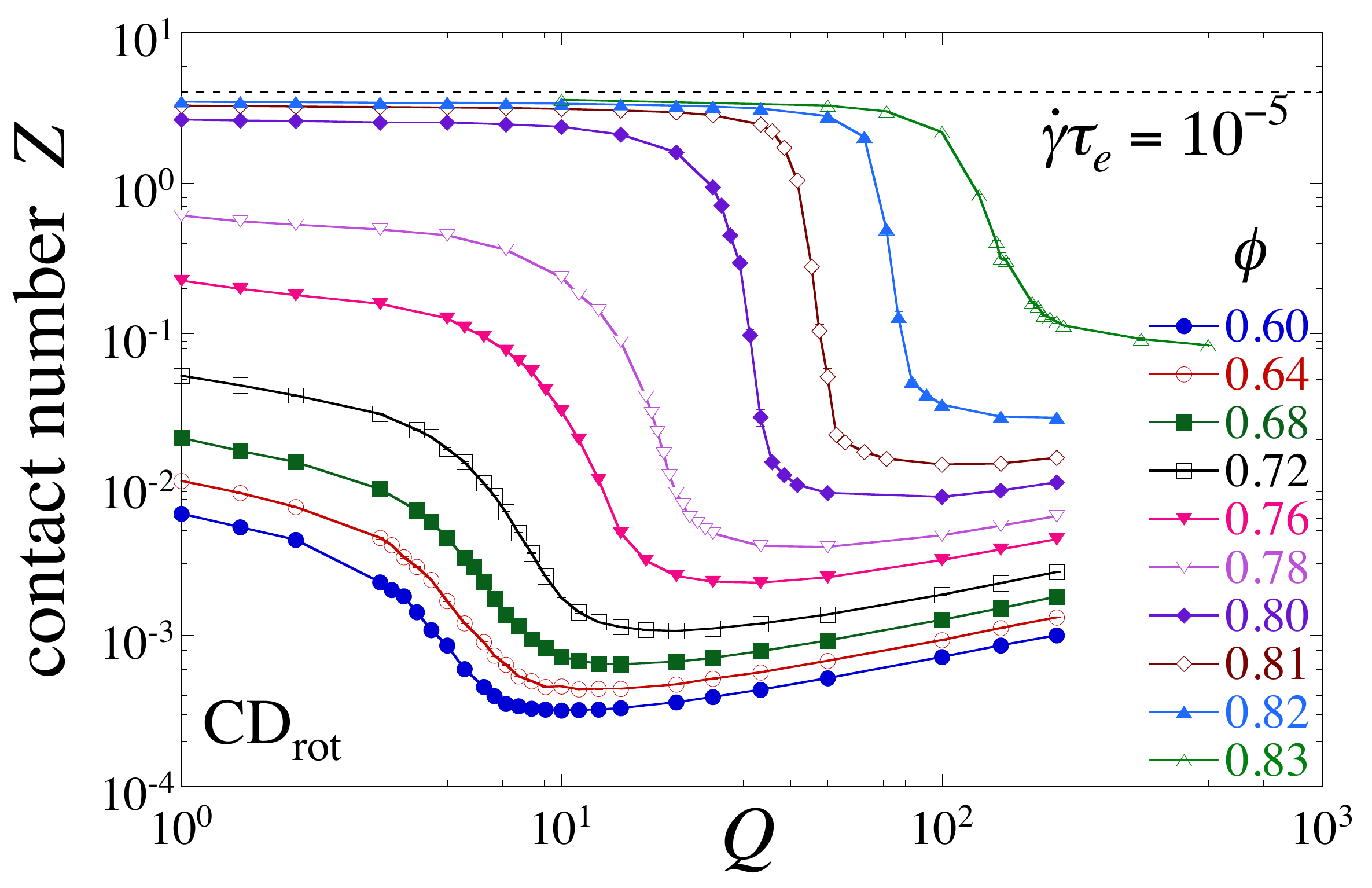} 
\caption{(Color online) Average instantaneous contact number $Z$ vs $Q$ in model CD$_\mathrm{rot}$, for different packing fractions $\phi=0.60$ to 0.83 as curves go from bottom to top. Strain rate $\dot\gamma\tau_e=10^{-5}$ and system has $N=1024$ particles.  For reference, the horizontal dashed line gives the isostatic value  $Z_\mathrm{iso}=2d=4$ for frictionless disks in two dimensions.
}
\label{z-vs-Q-rot}
\end{figure}

\begin{figure}[h!]
\includegraphics[width=3.2in]{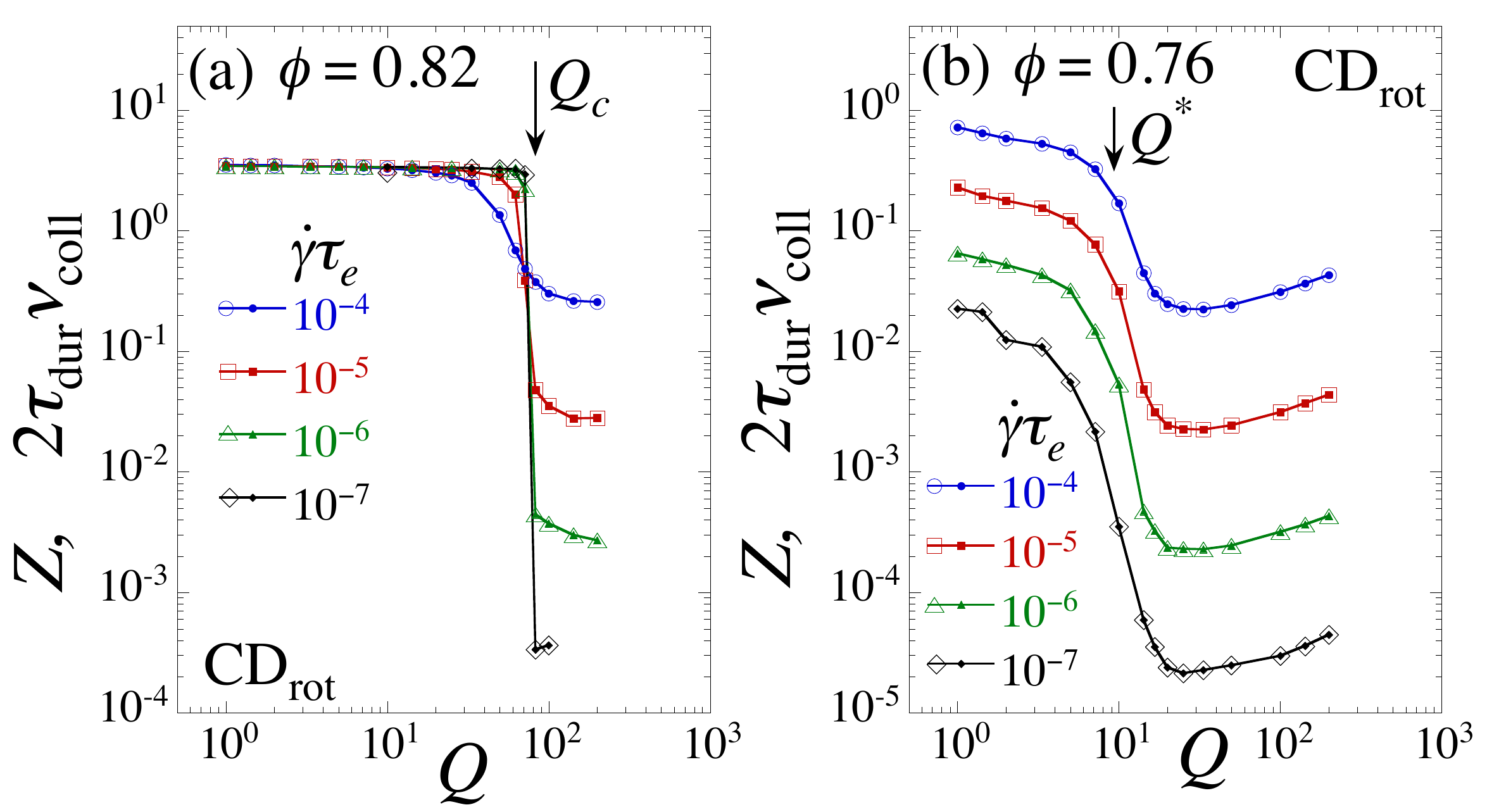} 
\caption{(Color online) Average instantaneous contact number $Z$, and twice the product of the average collision duration time and collision rate $2\tau_\mathrm{dur}\nu_\mathrm{coll}$, vs $Q$ for several different strain rates $\dot\gamma\tau_e$ at (a) $\phi=0.82$ and (b) $\phi=0.76$  in model CD$_\mathrm{rot}$.  System has $N=1024$ particles.  Solid symbols are for $Z$, while open symbols are for $2\tau_\mathrm{dur}\nu_\mathrm{coll}$. Strain rate $\dot\gamma\tau_e$ decreases as curves go from top to bottom. 
}
\label{Z-vs-Q-rotg}
\end{figure}

The same conclusion is obtained by looking at the average instantaneous contact number $Z$, which we plot vs $Q$ in Fig.~\ref{z-vs-Q-rot} for several different $\phi$ at $\dot\gamma\tau_e=10^{-5}$.  For $\phi\gtrsim 0.80$ we see $Z\sim O(1)$ at low $Q$, characteristic of the Newtonian phase; as $Q$ increases, $Z$ sharply drops roughly two orders of magnitude, characteristic of the Bagnoldian phase.  For $\phi < 0.80$, we see $Z<1$ at small $Q$ for all $\phi$, decreasing rapidly as $\phi$ decreases, consistent with a Bagnoldian phase at all $Q$.  This becomes clearer in Fig.~\ref{Z-vs-Q-rotg} where we show $Z$ vs $Q$ for strain rates $\dot\gamma\tau_e=10^{-4}$ to $10^{-7}$ at $\phi=0.82$ and $0.76$.  For $\phi=0.82$ in panel (a) we see that, for $Q<Q_c$, $Z$ is a constant $O(1)$ independent of $\dot\gamma\tau_e$, while for $Q>Q_c$, $Z$ decreases to zero as $\dot\gamma\tau_e$ decreases.  This is the same behavior seen  in Fig.~\ref{Z-vs-Q-gdot}(a) for model CD.  For $\phi=0.76$ in panel (b), however, we see that $Z$ decreases with decreasing $\dot\gamma\tau_e$ for all $Q$.  We also show in Fig.~\ref{Z-vs-Q-rotg} the quantity $2\tau_\mathrm{dur}\nu_\mathrm{coll}$ and find everywhere excellent agreement with $Z$, as expected according to Eq.~(\ref{Ztaunu}).

\begin{figure}[h!]
\includegraphics[width=3.2in]{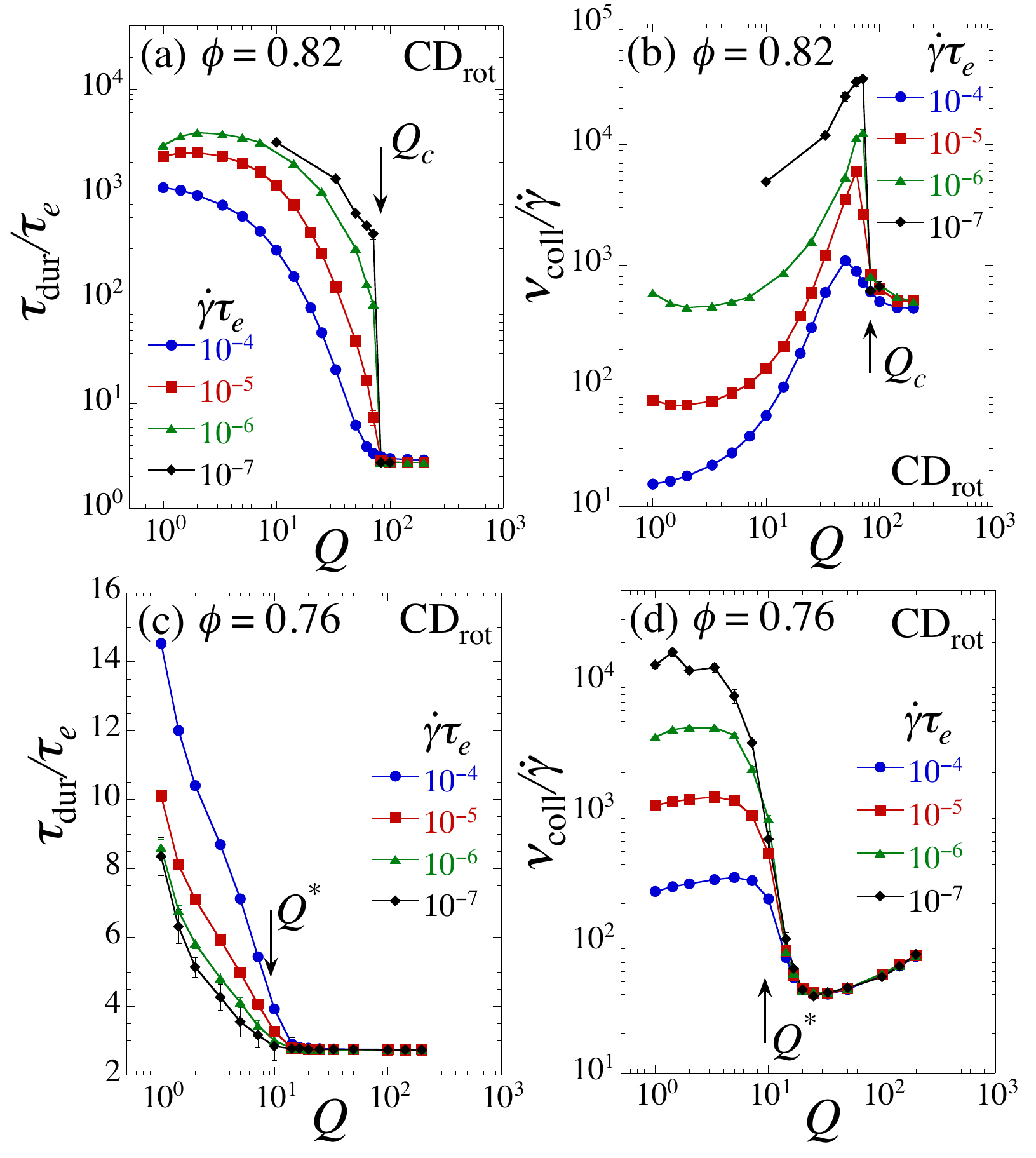} 
\caption{(Color online) (a) Average contact duration time $\tau_\mathrm{dur}/\tau_e$ and (b) average collision rate normalized by the shear strain rate $\nu_\mathrm{coll}/\dot\gamma$, vs $Q$ for several different strain rates $\dot\gamma\tau_e$ at $\phi=0.82$  in model CD$_\mathrm{rot}$.  Panels (c) and (d) show similar results at $\phi=0.76$.  System has $N=1024$ particles. 
In (a), (b) and (d), the strain rate $\dot\gamma\tau_e$ decreases as curves go from bottom to top; in (c) $\dot\gamma\tau_e$ decreases as curves go from top to bottom.
}
\label{tau-nu-vs-Q-rot}
\end{figure}

In Figs.~\ref{tau-nu-vs-Q-rot}(a) and (b) we plot $\tau_\mathrm{dur}/\tau_e$ and $\nu_\mathrm{coll}/\dot\gamma$ at $\phi=0.82$ for strain rates $\dot\gamma\tau_e=10^{-4}$ to $10^{-7}$.  We see behavior similar to that of model CD in Figs.~\ref{tau-nu-vs-Q}(a) and (b).  For $Q>Q_c$ we see that both $\tau_\mathrm{dur}/\tau_e$ and $\nu_\mathrm{coll}/\dot\gamma$ are independent of $\dot\gamma\tau_e$, demonstrating that $Z=2\tau_\mathrm{dur}\nu_\mathrm{coll}\sim \dot\gamma\tau_e$ vanishes linearly in $\dot\gamma\tau_e$ as $\dot\gamma\tau_e\to 0$.  As $Q$ decreases below $Q_c$, $\tau_\mathrm{dur}$ increases several orders of magnitude while $\nu_\mathrm{coll}/\dot\gamma$ decreases correspondingly, so that 
$Z=2\tau_\mathrm{dur}\nu_\mathrm{coll}$ remains constant, as explicitly seen in Fig.~\ref{Z-vs-Q-rotg}(a) for $Q<Q_c$.  

In Figs.~\ref{tau-nu-vs-Q-rot}(c) and (d) we plot these same quantities, but at the lower packing fraction $\phi=0.76$.  Again $\tau_\mathrm{dur}/\tau_e$ and $\nu_\mathrm{coll}/\dot\gamma$ are independent of $\dot\gamma\tau_e$ for $Q>Q^*$, demonstrating that in this large $Q$ region $Z\sim \dot\gamma\tau_e$ vanishes as $\dot\gamma\tau_e\to 0$.  However, for $Q<Q^*$, behavior is now different from what is seen in Figs.~\ref{tau-nu-vs-Q-rot}(a) and (b) at the higher $\phi=0.82$.  We still find that  $\tau_\mathrm{dur}/\tau_e$ increases as $Q$ decreases below $Q^*$, but now that increase is much less dramatic (note the linear vertical scale).  We see that  $\nu_\mathrm{coll}/\dot\gamma$ increases with decreasing $\dot\gamma\tau_e$ for $Q<Q^*$, but since from Fig.~\ref{Z-vs-Q-rotg}(b) we see that $Z$ decreases with decreasing $\dot\gamma\tau_e$, and $\tau_\mathrm{dur}/\tau_e$ appears to be approaching a constant as $\dot\gamma\tau_e$ decreases,  we conclude from Eq.~(\ref{Ztaunu}) that $\nu_\mathrm{coll}$ (and hence $Z$) decreases with decreasing $\dot\gamma\tau_e$, but more slowly than linearly in $\dot\gamma\tau_e$.
This behavior at $\phi=0.76$ is similar to what we have seen previously in model CD$_n$ \cite{VOT_CDnQ}, where dissipation depends only on the normal component of the velocity difference of the colliding particles.

\subsubsection{Phase diagram}

From the results in the preceding section we thus conclude that, for $\phi=0.82$, upon increasing $Q$ there is a sharp transition at $Q_c$ from a phase where the $\dot\gamma\tau_e\to 0$ limiting rheology is Newtonian, to a phase where the rheology is Bagnoldian.  For $\phi=0.76$, as $Q$ increases above $Q^*$, there is a non-singular crossover from a strongly inelastic Bagnoldian rheology to a weakly inelastic Bagnoldian rheology; these two different Bagnoldian regions differ primarily in how $P^\mathrm{el}$ varies with $Q$ and how $Z$ varies with $\dot\gamma\tau_e$.  We have previously found similar behavior in model CD$_n$ \cite{VOT_CDnQ,Lois}. Since we find that the small $Q$ behavior is Newtonian at large $\phi$, and Bagnoldian at small $\phi$, it therefore follows that at small $Q$ there must be a sharp transition from Bagnoldian to Newtonian rheology as $\phi$ increases.

\begin{figure}[h!] 
\includegraphics[width=3.2in]{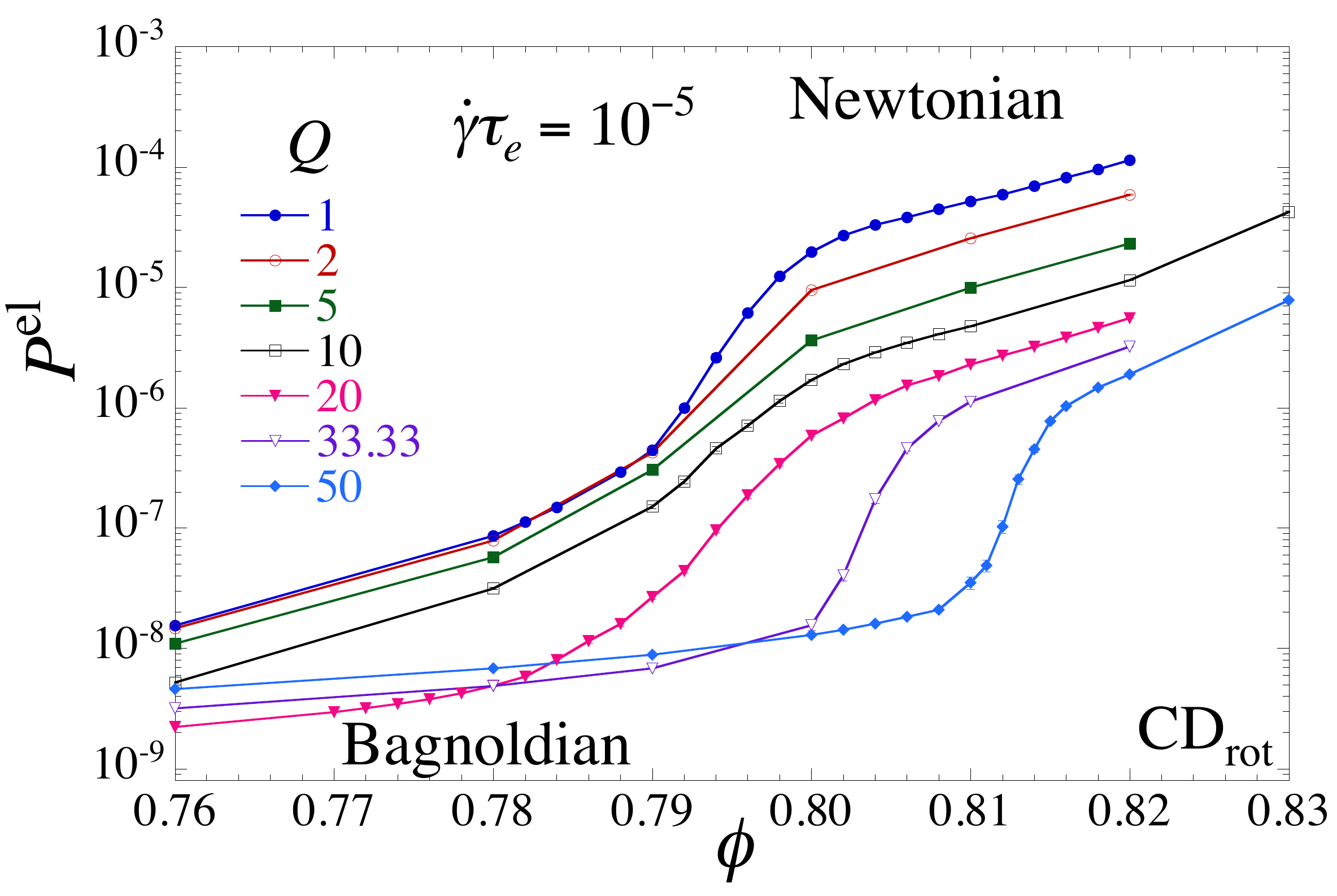} 
\caption{(Color online) Elastic part of the dimensionless pressure $P^\mathrm{el}$ vs packing fraction $\phi$ in model CD$_\mathrm{rot}$, for different values of the collision elasticity parameter $Q$.  The strain rate $\dot\gamma\tau_e=10^{-5}$ and the system has $N=1024$ particles.  $Q$ increases as curves go from top to bottom.
}
\label{pe-vs-phi-rot}
\end{figure}

\begin{figure}[h!] 
\includegraphics[width=3.2in]{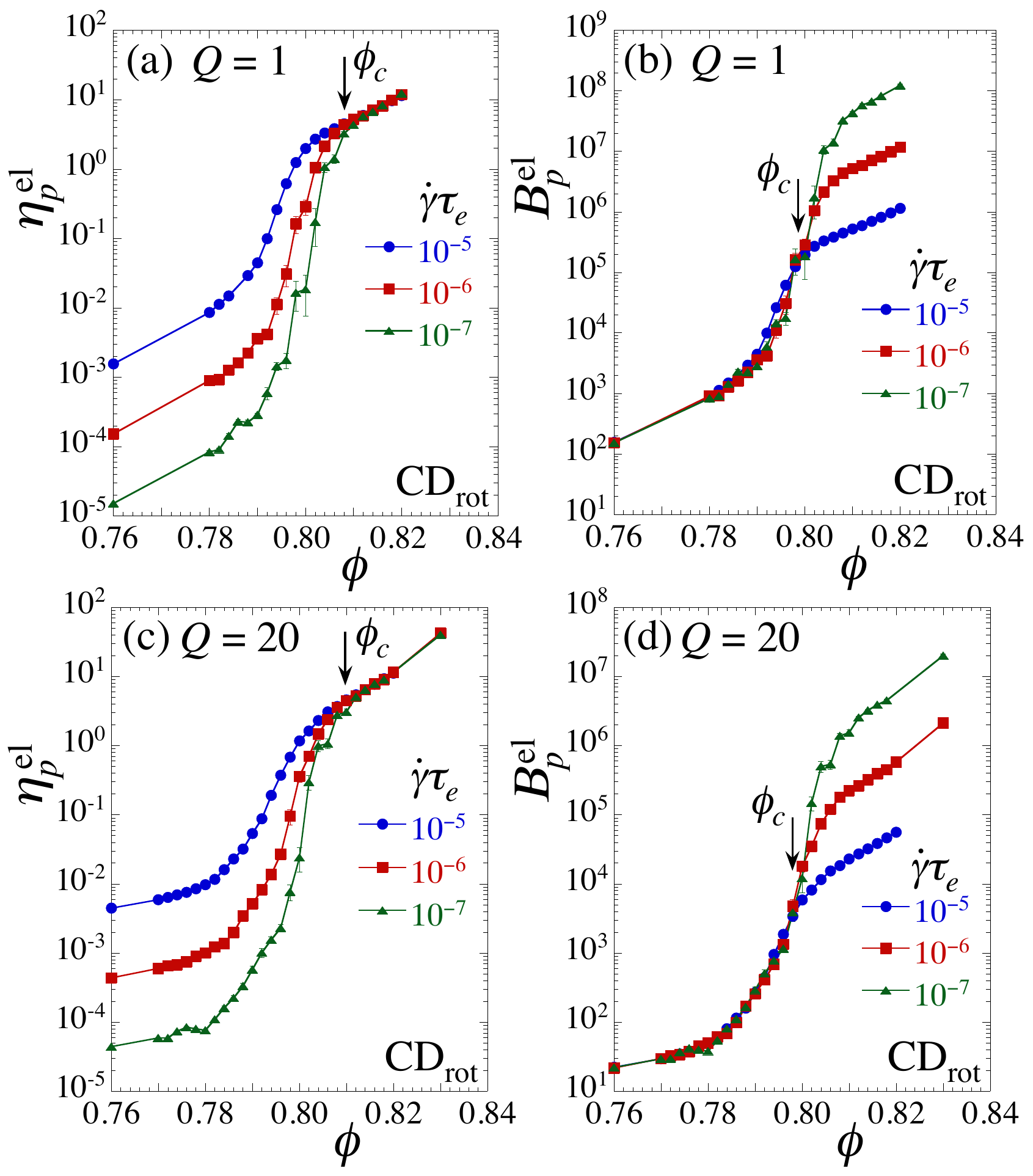} 
\caption{(Color online) (a) Pressure analog of viscosity, $\eta_p^\mathrm{el}=P^\mathrm{el}/(\dot\gamma\tau_0)=p^\mathrm{el}/(k_d\dot\gamma)$, and (b) Bagnold coefficient for pressure, $B_p^\mathrm{el}=P^\mathrm{el}/(\dot\gamma\tau_e)^2=p^\mathrm{el}/(m\dot\gamma^2)$, vs $\phi$ for several different strain rates $\dot\gamma\tau_e$ at $Q=1$   in model CD$_\mathrm{rot}$.  Panels (c) and (d) show similar results for $Q=20$.  System has $N=1024$ particles.  In (a) and (c) $\dot\gamma\tau_e$ decreases as curves go from top to bottom; in (b) and (d) $\dot\gamma\tau_e$ decreases as curves go from bottom to top.
}
\label{etap-Bp-vs-phi}
\end{figure}

To search for this small $Q$ transition from Bagnoldian to Newtonian rheology, in Fig.~\ref{pe-vs-phi-rot} we plot the elastic part of the dimensionless pressure $P^\mathrm{el}$ vs $\phi$ for several different values of $Q$ at fixed strain rate $\dot\gamma\tau_e=10^{-5}$.  We see a clear step upwards in the vicinity of $\phi\approx 0.79$ -- 0.81, depending on the value of $Q$.  
To confirm that these steps do indeed represent a sharp Bagnoldian-to-Newtonian rheological transition, in Figs.~\ref{etap-Bp-vs-phi}(a) and (b) we plot $\eta_p^\mathrm{el}$ and $B_p^\mathrm{el}$ vs $\phi$ at several different strain rates $\dot\gamma\tau_e$, for the value $Q=1$.  We see that there is a $\phi_c$ such that for $\phi>\phi_c$ the curves of $\eta_p^\mathrm{el}$ become independent of $\dot\gamma\tau_e$, thus demonstrating Newtonian rheology.  For $\phi<\phi_c$ the curves of $B_p^\mathrm{el}$ become independent of $\dot\gamma\tau_e$, thus demonstrating Bagnoldian rheology.  In Figs.~\ref{etap-Bp-vs-phi}(c) and (d) we show similar results, with similar conclusions, for $Q=20$.  
We find that this $\phi_c$ tends to lie near the upper shoulder of the steps in Fig.~\ref{pe-vs-phi-rot}.
From Fig.~\ref{etap-Bp-vs-phi} we note that there appears to be a small interval in $\phi$ between the point where the system ceases to be Bagnoldian and the point where it starts to be Newtonian.  We will address this in greater detail in the next section.

\begin{figure}[h!] 
\includegraphics[width=3.2in]{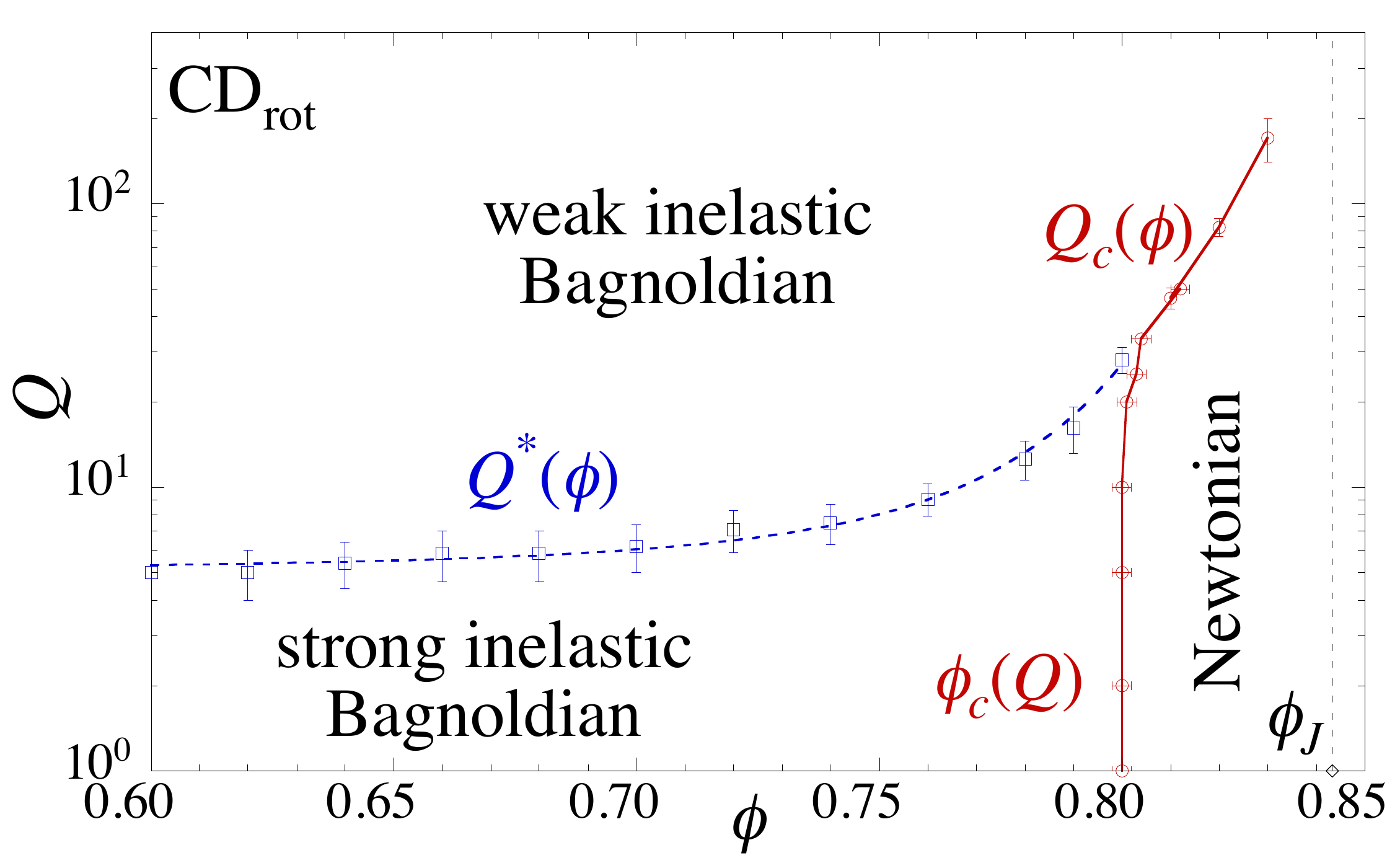}
\caption{(Color online) Phase diagram for model CD$_\mathrm{rot}$ in the $(\phi,Q)$ plane. A region of Newtonian rheology, just below the jamming transition, is separated from the region of Bagnoldian rheology by the continuous solid curve $Q_c(\phi)$ (or equivalently its inverse $\phi_c(Q))$.  The dashed curve $Q^*(\phi)$ represents a non-singular crossover between two different regions of Bagnoldian rheology.  In the weak inelastic Bagnoldian region, the average particle contact number $Z\sim \dot\gamma\tau_e$, while in the strong inelastic Bagnoldian region, $Z$ decreases more slowly than linearly with $\dot\gamma\tau_e$.  The locations of the lines $Q_c$, $Q^*$ and $\phi_c$ are determined by the methods discussed in Appendix A.  The vertical dashed line locates the jamming packing fraction $\phi_J$.
}
\label{Qc-Qstar-phic}
\end{figure}

Using the specific methods detailed in Appendix A, we proceed to determine the location of the crossover line $Q^*(\phi)$ between the regions of weak inelastic and strong inelastic Bagnoldian rheology at smaller $Q$ (as indicated by Fig.~\ref{etap-Bp-vs-Q-rot}(c) and (d)), the transition line $Q_c(\phi)$ between the regions of Bagnoldian and Newtonian rheology at larger $Q$ (as indicated by Fig.~\ref{etap-Bp-vs-Q-rot}(a) and (b)), and the transition line $\phi_c(Q)$ between regions of Bagnoldian and Newtonian rheology at small $Q$ (as indicated by Fig.~\ref{etap-Bp-vs-phi}).  The resulting phase diagram is shown in Fig.~\ref{Qc-Qstar-phic}.  We see that the transition lines $Q_c(\phi)$ and $\phi_c(Q)$ form one continuous curve that separates the Newtonian phase from the Bagnoldian phase.  As in our earlier model CD, our results suggest that the system is always Newtonian at the jamming transition $\phi_J$; the independence of $\eta_p^\mathrm{el}$ on $Q$ in this region, as seen in Fig.~\ref{etap-Bp-vs-Q-rot}(a), suggests that the critical parameters of the jamming transition must also be independent of $Q$.  Unlike model CD, however, we see that the Newtonian region is confined to relatively large packing fractions, $\phi>\phi_c\approx 0.80$.  For  $\phi<\phi_c$, even at small $Q$ where the collisions are strongly inelastic, the rheology remains Bagnoldian; upon increasing $Q$ at fixed $\phi<\phi_c$ there is a non-singular crossover $Q^*$ between two different regions of Bagnoldian rheology, as we have previously found in model CD$_n$ \cite{VOT_CDnQ}.

\subsubsection{Geometry of collisions}

\begin{figure}[h]
\includegraphics[width=3.2in]{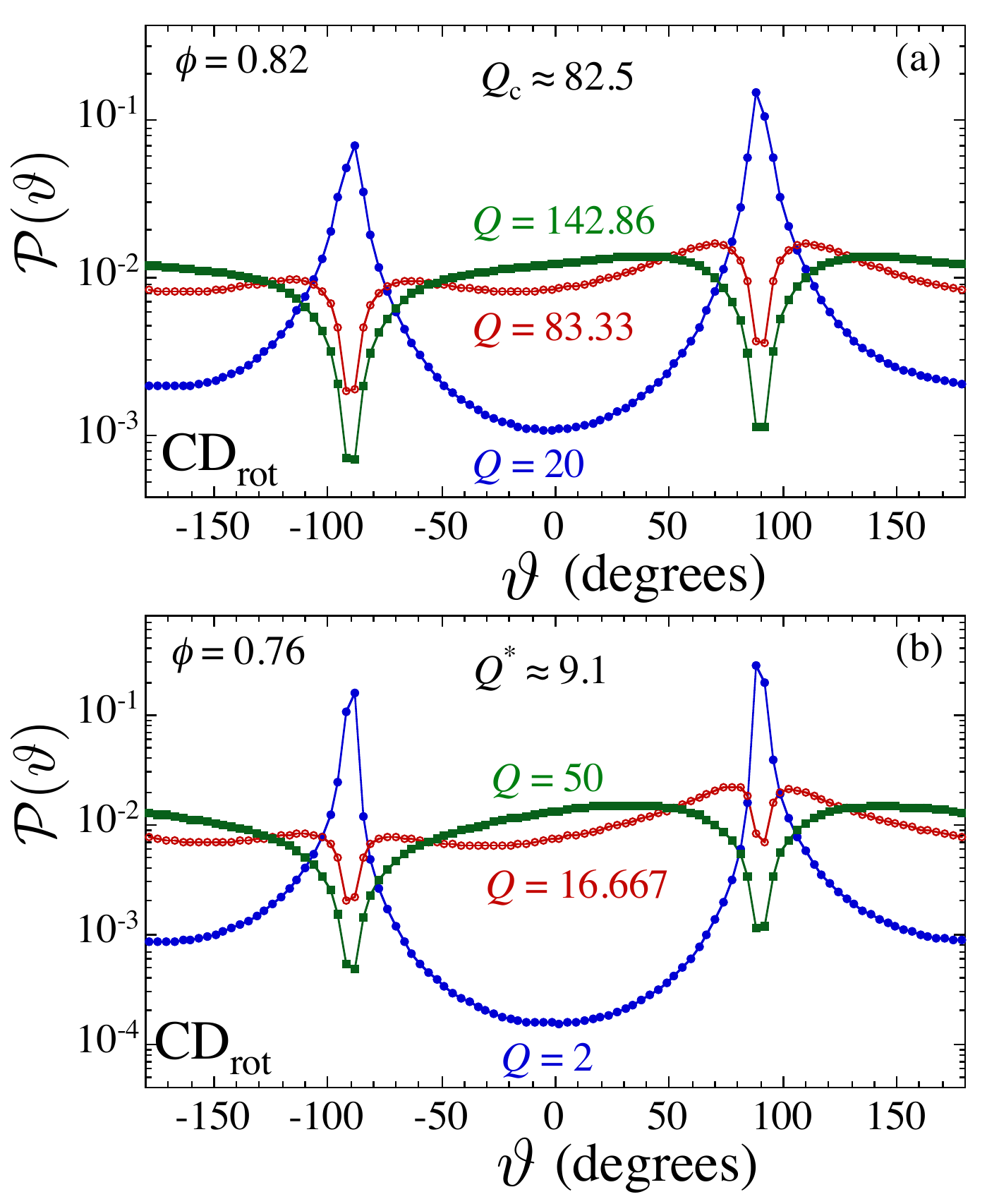} 
\caption{(Color online) Histograms ${\cal P}(\vartheta)$ vs collision angle $\vartheta$ (defined in Fig~\ref{collisions}), at initiation and breaking of particle contacts, at (a) $\phi=0.82$ and (b) $\phi=0.76$  in model CD$_\mathrm{rot}$.  In (a) we we show results for one value $Q>Q_c$, one value $Q<Q_c$, and for $Q\approx Q_c$, with the value of the transition $Q_c$ as indicated in the panel; in (b) we do similarly with respect to the crossover value $Q^*$.
Results are for the shear strain rate $\dot\gamma\tau_e=10^{-5}$ in a system with $N=1024$ particles. 
}
\label{Ptheta-rot}
\end{figure}

\begin{figure}[h!]
\includegraphics[width=3.2in]{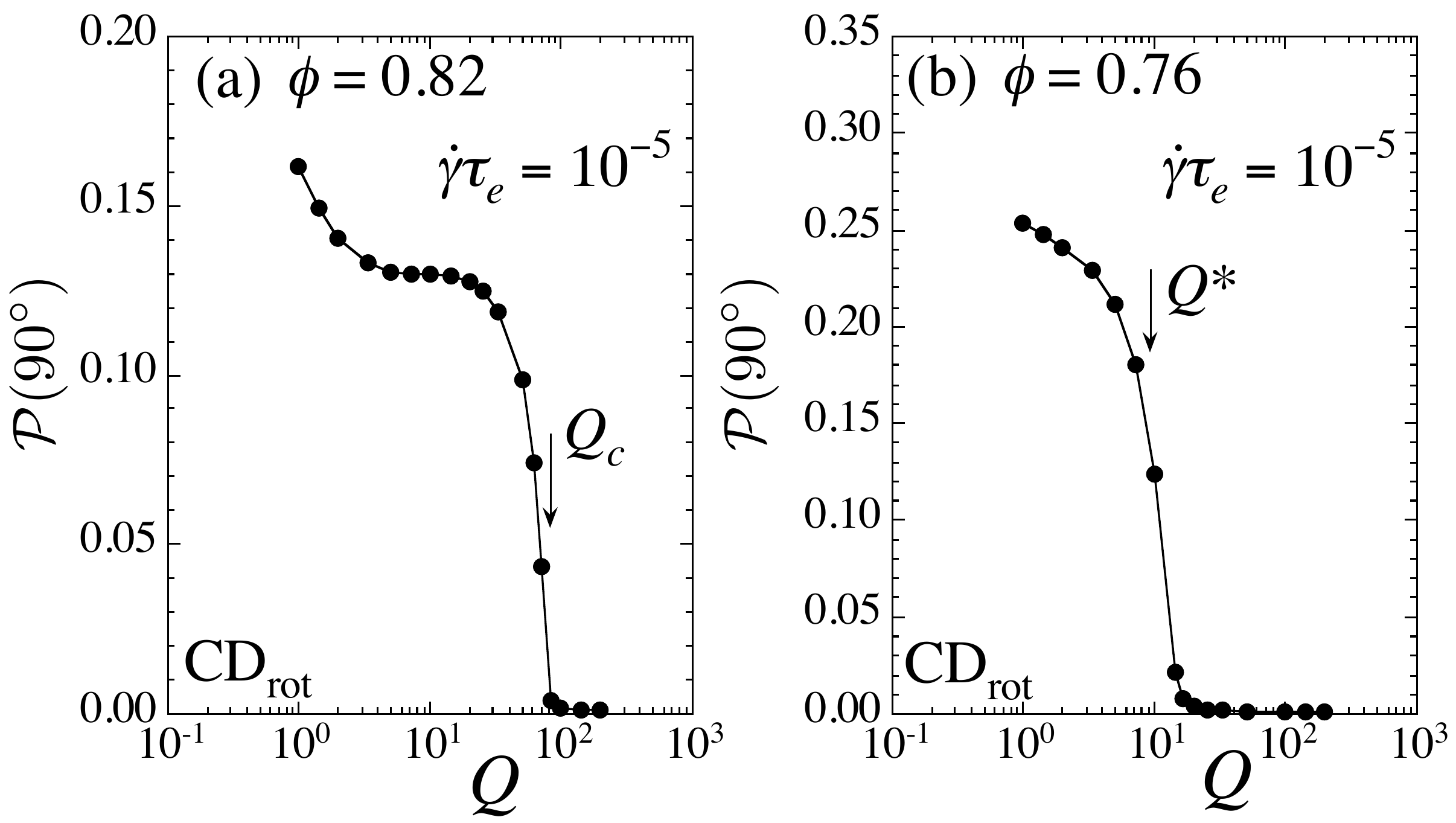} 
\caption{Histogram ${\cal P}(\vartheta=90^\circ)$ vs $Q$ for (a) $\phi=0.82$ and (b) $\phi=0.76$.  The propensity to have tangential collisions, i.e. $\vartheta=90^\circ$, increases sharply as one crosses from the Bagnoldian ($Q>Q_c$) to the Newtonian ($Q<Q_c$) region in (a), or from the weakly inelastic Bagnoldian ($Q>Q^*$) to the strongly inelastic Bagnoldian ($Q<Q^*$) region in (b).
Results are for the shear strain rate $\dot\gamma\tau_e=10^{-5}$ in a system with $N=1024$ particles. 
}
\label{P90-rot}
\end{figure}

We next look at the geometry of collisions, computing the histogram ${\cal P}(\vartheta)$ of the collision angle $\vartheta$ at the time of initiation of a contact and the time of breaking of a contact (see Fig.~\ref{collisions} for the definition of $\vartheta$).
In Fig.~\ref{Ptheta-rot}(a) we plot ${\cal P}(\vartheta)$ vs $\vartheta$ at  $\phi=0.82>\phi_c$,
for three values of $Q$: well above, well below, and near $Q_c$.  In Fig.~\ref{Ptheta-rot}(b) we plot similarly at $\phi=0.76<\phi_c$, for $Q$ well above, well below, and near $Q^*$.  In Fig.~\ref{P90-rot} we plot ${\cal P}(\vartheta=90^\circ)$ vs $Q$ for $\phi=0.82$ and $0.76$.
Both above and below $\phi_c$ we see the same behavior as was found previously in model CD (compare with Figs.~\ref{Ptheta} and \ref{P90}).  For $Q>Q_c$ (at $\phi>\phi_c$) or $Q>Q^*$ (at $\phi<\phi_c$), we see that collisions take place over a broad range of $\vartheta$;  tangentially directed collisions at $\vartheta=90^\circ$ occur a negligible fraction of the time.  For $Q<Q_c$ or $Q<Q^*$, normally directed collisions are greatly suppressed, and the relative motion of particles is primarily tangential.

\subsubsection{Effect of particle rotations}

It therefore remains to explain, at low $Q$ where relative particle motion is primarily tangential at any of the $\phi$ values we have considered here, what is the cause of the different rheology for $\phi<\phi_c$ (Bagnoldian) compared to $\phi>\phi_c$ (Newtonian).  Since CD$_\mathrm{rot}$ couples rotational and translational motion while CD does not, it is natural to look to the rotational motion of the particles for the answer.  We therefore consider the rotational velocity of the particles, $\dot\theta_i\equiv d\theta_i/dt$.

\begin{figure}[h!] 
\includegraphics[width=3.2in]{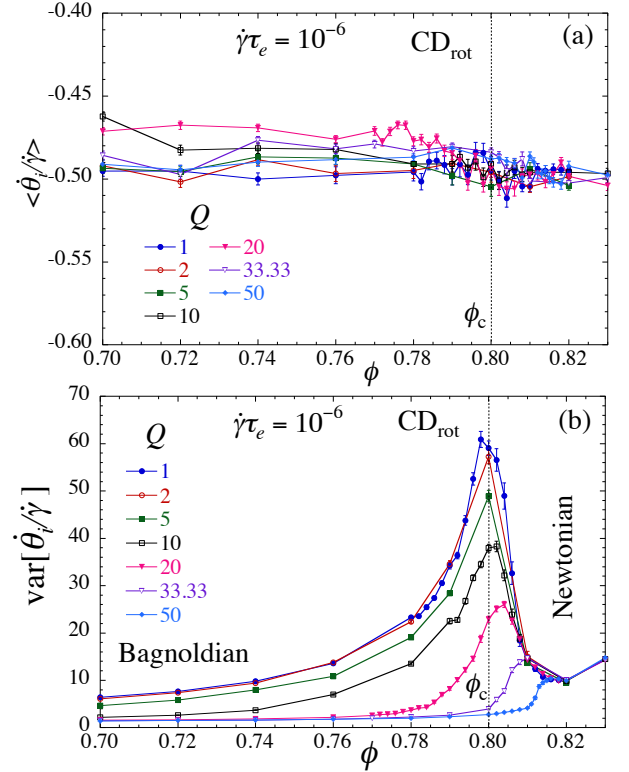} 
\caption{(Color online) (a) $\langle\dot\theta_i/\dot\gamma\rangle$ and (b) $\mathrm{var}[\dot\theta_i/\dot\gamma]$ vs $\phi$, for different values of $Q$  in model CD$_\mathrm{rot}$.  The strain rate is $\dot\gamma\tau_e=10^{-6}$ and the system has $N=1024$ particles.  The vertical dotted line indicates the small $Q$ value of $\phi_c\approx 0.80$.
}
\label{omegagdot-vs-phi-rot}
\end{figure}

In Fig.~\ref{omegagdot-vs-phi-rot}(a) we plot the average $\langle\dot\theta_i/\dot\gamma\rangle$ vs $\phi$ for several of the smaller values of $Q\le 50$, at $\dot\gamma\tau_e=10^{-6}$.  We see that $\langle\dot\theta_i/\dot\gamma\rangle\approx -1/2$ is roughly independent of $\phi$, as would be expected for circular disks in a uniform shear flow (the minus sign indicates a clockwise rotation).  There is no visible feature as $\phi$ passes through $\phi_c$. Next we consider the variance of the single particle rotational velocity, 
\begin{equation}
\mathrm{var}[\dot\theta_i/\dot\gamma]\equiv\left\langle (\dot\theta_i/\dot\gamma)^2\right\rangle-\left\langle\dot\theta_i/\dot\gamma\right\rangle^2,
\end{equation} where $\langle\dots\rangle$ denotes the time average over the steady state, and
\begin{equation}
\left\langle (\dot\theta_i/\dot\gamma)^2\right\rangle\equiv\left\langle\frac{1}{N}\sum_i (\dot\theta_i/\dot\gamma)^2\right\rangle.
\end{equation}
In Fig.~\ref{omegagdot-vs-phi-rot}(b) we plot $\mathrm{var}[\dot\theta_i/\dot\gamma]$ vs $\phi$ for different $Q$, at $\dot\gamma\tau_e=10^{-6}$.
We now see a clear signature of the transition at $\phi_c$. As $\phi$ increases within the Bagnoldian phase, $\mathrm{var}[\dot\theta_i/\dot\gamma]$ steadily increases, reaches a peak at $\phi\approx\phi_c(Q)$, and then drops abruptly as the Newtonian phase is entered, suggesting that fluctuations of rotation are more constrained in the Newtonian phase.  Note, although $\mathrm{var}[\dot\theta_i/\dot\gamma]$ drops in the Newtonian phase, fluctuations remain large compared to the average, $\langle\dot\theta_i/\dot\gamma\rangle\approx - 1/2$.

For a clearer understanding of the effects of the rotational motion on the rheology, we now consider behavior at the level of individual collisions.  Defining the dimensionless velocity $\mathbf{V}\equiv\mathbf{v}/(d_s\dot\gamma)$, we can write the difference in velocities at the point of contact of particle $i$ colliding with particle $j$ as,
\begin{equation}
\mathbf{V}^{C}_{ij}=\mathbf{V}_{ij}+\mathbf{V}^R_{ij},
\label{VC}
\end{equation}
where $\mathbf{V}_{ij}$ is the difference in center of mass velocities, and from Eq.~(\ref{evC}),
\begin{equation}
\mathbf{V}^R_{ij}\equiv \left[ \dfrac{\dot\theta_i}{\dot\gamma}\dfrac{d_i}{d_{s}}+\dfrac{\dot\theta_j}{\dot\gamma}\dfrac{d_j}{d_{s}}\right]
\mathbf{\hat z}\times\mathbf{\hat r}_{ji}
\end{equation}
is the contribution from the rotational motion of the particles (we have used $\mathbf{r}_{ij}=-\mathbf{r}_{ji}$).  With $\mathbf{\hat r}_{ji}$ as the outward unit normal vector at the point of contact on particle $i$, and $\mathbf{\hat t}_{ji}\equiv\mathbf{\hat z}\times\mathbf{\hat r}_{ji}$ as the unit tangent vector at the point of contact, we see that $\mathbf{V}^R_{ij}$ is purely tangential.  We define,
\begin{equation}
V^R_{ij}\equiv\mathbf{V}^R_{ij}\cdot\mathbf{\hat t}_{ji}
\end{equation}
as the rotational part of the contact velocity difference,
\begin{equation}
V^C_{ij,T}\equiv \mathbf{V}^C_{ij}\cdot\mathbf{\hat t}_{ji},\quad
V^C_{ij,N}\equiv \mathbf{V}^C_{ij}\cdot\mathbf{\hat r}_{ji},
\end{equation}
as the tangential and normal components of the contact velocity difference, and
\begin{equation}
V_{ij,T}\equiv \mathbf{V}_{ij}\cdot\mathbf{\hat t}_{ji},\quad
V_{ij,N}\equiv \mathbf{V}_{ij}\cdot\mathbf{\hat r}_{ji},
\end{equation}
as the tangential and normal components of the center of mass velocity difference.
Since $\mathbf{V}^R_{ij}$ is purely tangential, $V^C_{ij,N}=V_{ij,N}$.

\begin{figure}[h!] 
\includegraphics[width=3.2in]{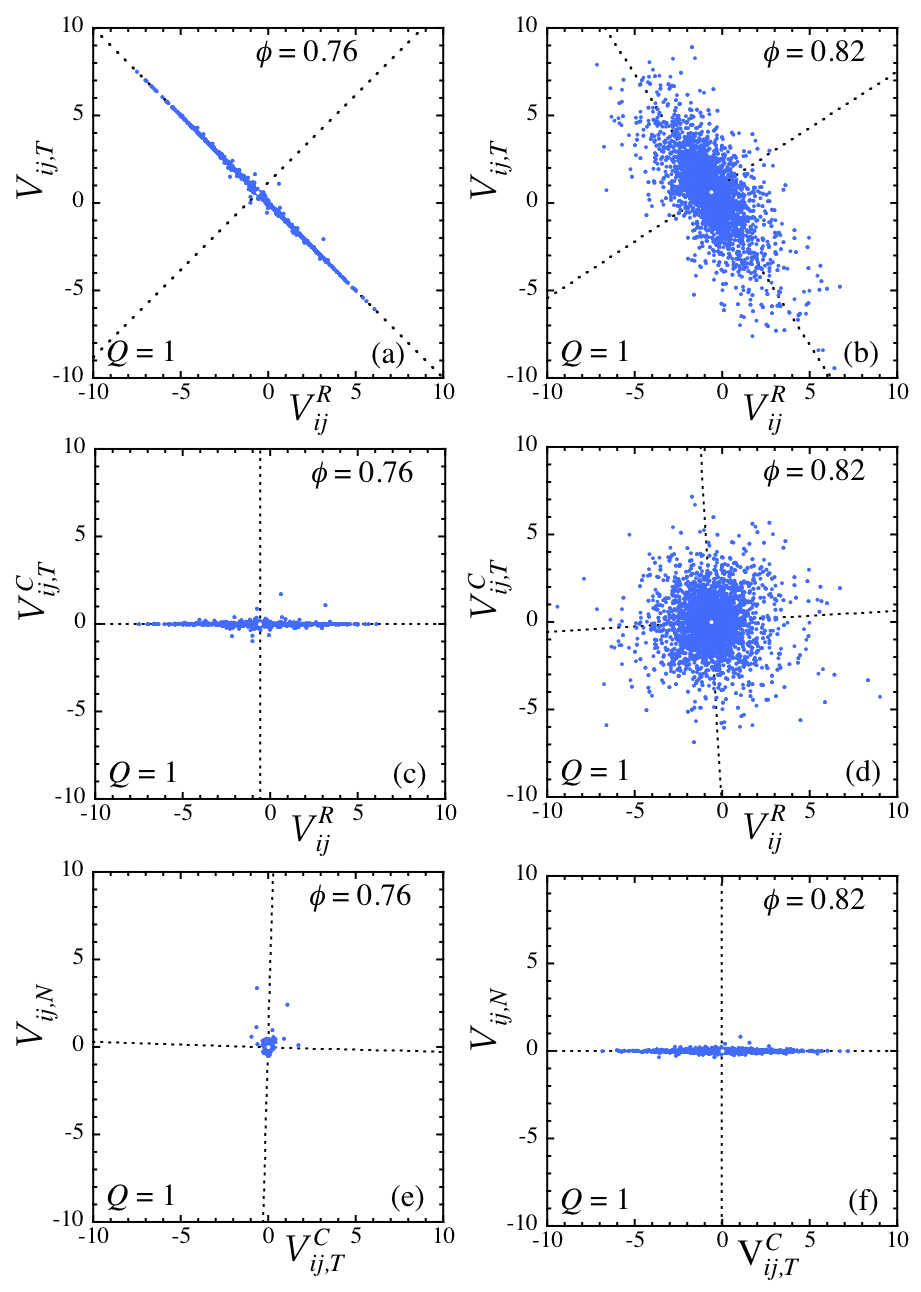} 
\caption{(Color online) Scatter plots looking at correlations of different components of the dimensionless velocity difference between two particles in contact.  The left column is for $\phi=0.76<\phi_c$, the right column is for $\phi=0.82 > \phi_c$.  The data points represent $5000$ independent collisions sampled at random times during the duration of each collision.  In each panel the white dot in the center of the data locates the average of the distribution, while the black dotted lines give the directions of the principal axes of the covariance matrix.
The top row, panels (a) and (b), show the correlations between the tangential component of the center of mass velocity difference, $V_{ij,T}$, and the component of the contact velocity difference due to particle rotations, $V^R_{ij}$.  The middle row, panels (c) and (d), show the correlations between the tangential component of the contact velocity difference, $V^C_{ij,T}$ and $V^R_{ij}$.  The bottom row, panels (e) and (f), show the correlations between the normal component of the velocity difference, $V_{ij,N}$, and $V^C_{ij,T}$.
}
\label{vcorel}
\end{figure}

In Fig.~\ref{vcorel} we show correlation scatter plots of different components of the contact velocity difference at $\phi=0.76<\phi_c$ (left column) and $\phi=0.82>\phi_c$ (right column) for $Q=1$ and $\dot\gamma\tau_e=10^{-5}$.  The data points represent $5000$ independent collisions sampled at random times during the duration of each collision.  In each panel the white dot in the center of the data locates the average of the distribution, while the black dotted lines give the directions of the principal axes of the covariance matrix.

We consider first behavior at $\phi=0.76<\phi_c$.  In panel (a) we show the correlation of $V_{ij,T}$ with $V^R_{ij}$.  We see almost a perfect anticorrellation, with the result from Eq.~(\ref{VC}) that the tangential component of the contact velocity difference $V^C_{ij,T}\approx 0$.  This is confirmed in panel (c), where we show the correlation of $V^C_{ij,T}$ with $V^R_{ij}$.  Despite a wide spread in values for $V^R_{ij}$, we find $V^C_{ij,T}$ is always close to zero.  In panel (e) we show the correlation of the normal component of the contact velocity difference, $V_{ij,N}=V^C_{ij,N}$, with the tangential component $V^C_{ij,T}$. We see that both components are comparably small. Thus at small $Q$ in the Bagnoldian phase below $\phi_c$, the normally directed relative motion between particles is suppressed, relative motion is primarily tangential, but the particles are able to adjust their rotation to approximately cancel out the difference in center of mass motion, so that the tangential component of the contact velocity difference is comparably as small as the normal component.  Thus both normal and tangential components of the dissipative force remain small.  Behavior here is thus similar to what we have previously found at low $Q$ in model CD$_n$ \cite{VOT_CDnQ}, where the tangential component of the dissipative force is explicitly set to zero.

We next consider the behavior at $\phi=0.82>\phi_c$.  In panel (b) we show $V_{ij,T}$ vs $V^R_{ij}$.  Compared to panel (a) we see that there is still correlation, however the scatter of the data is broader and the slope of the major axis of the covariance matrix axis is no longer $-1$, showing that the sum of these two, i.e. $V^C_{ij,T}$, remains finite.  This is confirmed in panel (d) where we show $V^C_{ij,T}$ vs $V^R_{ij}$.  We see that the spread in both quantities remains comparable, and the nearly vertical and horizontal orientations of the covariance matrix axes show that these quantities are at best very weakly correlated.  Thus the tangential component of the contact velocity difference is now largely independent of the rotational motion.  
This observation is sufficient to explain the correlations seen in panel (b).
In panel (f) we show $V_{ij,N}$ vs $V^C_{ij,T}$.  We see that the spread in the normal component of the contact velocity difference remains small, comparable to that of panel (e), but that the tangential component is comparatively broadly distributed.  Thus at small $Q$ in the Newtonian phase above $\phi_c$, the normally directed relative motion between particles is suppressed, relative motion is primarily tangential, the particles fluctuate in their rotational motion, but these fluctuations are largely uncorrelated with the 
%translational motion so that they do not much effect the 
tangential component of the contact velocity difference.  The normal component of the dissipative force is small, but the tangential component is comparatively large; both are largely unaffected by the rotational motion of the particles.  Behavior here is thus similar to what we find in model CD.

We thus see that the transition from Bagnoldian to Newtonian rheology, upon increasing $\phi$ at small $Q$, is due to the changing nature of correlations between rotational and translational motion as the system gets denser.  Below $\phi_c$ the rotations are almost perfectly anticorrelated with the differences in center of mass motion, so  that the tangential part of the contact velocity difference $V^C_{ij,T}$ becomes essentially zero, and dissipative forces are small (see Figs~\ref{vcorel}(a) and (c)). Above $\phi_c$, rotational motion becomes less correlated with the translational motion, so that $V^C_{ij,T}$ has sizable fluctuations which are uncorrelated with the particle rotations; rotational motion thus has little effect on the dissipative forces, which become sizable (see Figs.~\ref{vcorel}(b) and (d)).
We find that, as $\phi$ increases, the number of particles in mutual contact during any particular collision is increasing.  We therefore speculate that this change in the correlation between rotational and translational motion is due to the increasing constraint on rotational motion that arises when the number of contacts per collision increases. 
We leave further exploration of this effect to future work.

%%%%%%%%%%%%%%%%%%%%%%%%%%%%%%%%

\subsection{Model CD$_\mathrm{rot}$: Shear banding at finite $\dot\gamma\tau_e$}
\label{s3SB}

In the previous sections we focused on sketching out the behavior in the $(\phi,Q)$ plane as $\dot\gamma\tau_e\to 0$.  In this section we focus more carefully on behavior as a function of $\dot\gamma\tau_e$.  We consider here the specific case of crossing the Bagnoldian to Newtonian transition line at $\phi_c$ for small $Q=1$ in model CD$_\mathrm{rot}$.  We will find that upon increasing $\dot\gamma\tau_e$, the sharp discontinuous transition becomes a coexistence region of finite width, where a shear band of Bagnoldian rheology coexists with a shear band of Newtonian rheology.  Preliminary results suggest that similar behavior  exists when crossing the Bagnoldian to Newtonian transition elsewhere at $Q_c(\phi)$ in both the models CD and CD$_\mathrm{rot}$. To investigate the spatially inhomogeneous configurations associated with this shear banding, it is necessary to consider a much larger system than in the previous sections.  Here we use a system with $N=65536$ particles.

\subsubsection{Indicators for a coexistence region}

\begin{figure}[h!] 
\includegraphics[width=3.2in]{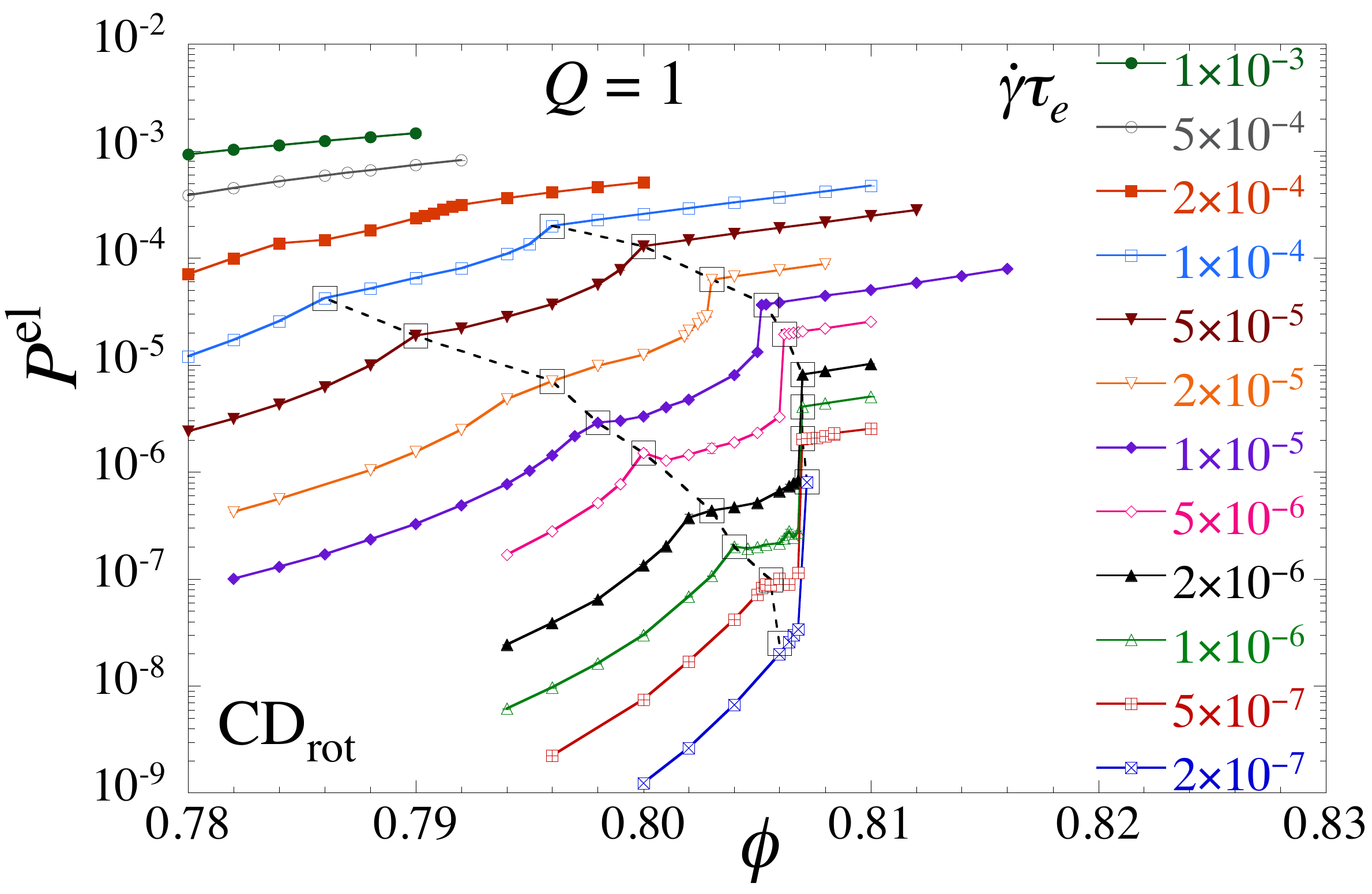} 
\caption{(Color online)  Elastic part of the dimensionless pressure $P^\mathrm{el}$ vs packing fraction $\phi$ for $Q=1$ in model CD$_\mathrm{rot}$.  Results are shown for different strain rates, ranging from  $\dot\gamma\tau_e=10^{-3}$ to $2\times 10^{-7}$, as curves go from top to bottom.  Large open squares denote ``kinks" in the pressure curves; dashed lines connect these kinks.  The system has $N=65536$ particles. 
}
\label{pe-vs-phi-65536}
\end{figure}

In Fig.~\ref{pe-vs-phi-65536} we plot results for the elastic part of the dimensionless pressure $P^\mathrm{el}$ vs packing fraction $\phi$ for different fixed values of the strain rate $\dot\gamma\tau_e$ at $Q=1$.  At the smallest $\dot\gamma\tau_e$ we see a sharp jump in $P^\mathrm{el}$ that signals the transition at $\phi_c$ from Bagnoldian to Newtonian rheology.  The jump here is indeed sharp, compared to the smoother behavior seen previously in Fig.~\ref{pe-vs-phi-rot}, because the strain rate is much lower and our system is much larger; this $\dot\gamma\tau_e\to 0$ limiting value $\phi_c\approx 0.807$ is also slightly larger than shown in Fig.~\ref{Qc-Qstar-phic} for the same reason.
However as $\dot\gamma\tau_e$ increases, we see that this jump broadens and the $P^\mathrm{el}$ vs $\phi$ curve develops two noticeable kinks rather than a single sharp jump.  These kinks are denoted in the figure by the large open squares.  At sufficiently large $\dot\gamma\tau_e$ these kinks disappear, and the $P^\mathrm{el}$ vs $\phi$ curve looks smooth.  For a given $\dot\gamma\tau_e$, we denote the kink at the smaller $\phi$ by $\phi_B(\dot\gamma\tau_e)$, and the kink at the larger $\phi$ by $\phi_N(\dot\gamma\tau_e)$.  The dashed lines in Fig.~\ref{pe-vs-phi-65536} connect these kinks so as to delimit a region in the $(\phi,P^\mathrm{el})$ plane.  We will soon show that this region denotes a coexistence region where a band of Bagnoldian rheology coexists in mechanical equilibrium with a band of Newtonian rheology.

We next consider the macroscopic friction $\mu\equiv\sigma/p$.  Since coexistence between bands of Bagnoldian and Newtonian rheology must result from mechanical equilibrium between states of equal stress, we will consider here the total pressure $p$ and total shear stress $\sigma$, rather than just the elastic parts.  We note however, that for the parameters considered here, the dissipative and kinetic parts of the pressure are negligible, as is the kinetic part of the shear stress; the dissipative $\sigma^\mathrm{dis}$ contributes non-negligibly, roughly 10\% -- 20\%, to the total $\sigma$.  The quantity $\mu$ is of interest to consider since, unlike $p$ or $\sigma$, $\mu$ approaches a finite value as $\dot\gamma\tau_e\to 0$ in both Bagnoldian and Newtonian systems.  Thus for sufficiently small $\dot\gamma\tau_e$ we might expect  that, to a large extent, the curves of $\mu$ vs $\phi$ for different $\dot\gamma\tau_e$ will overlap.

\begin{figure}[h!] 
\includegraphics[width=3.2in]{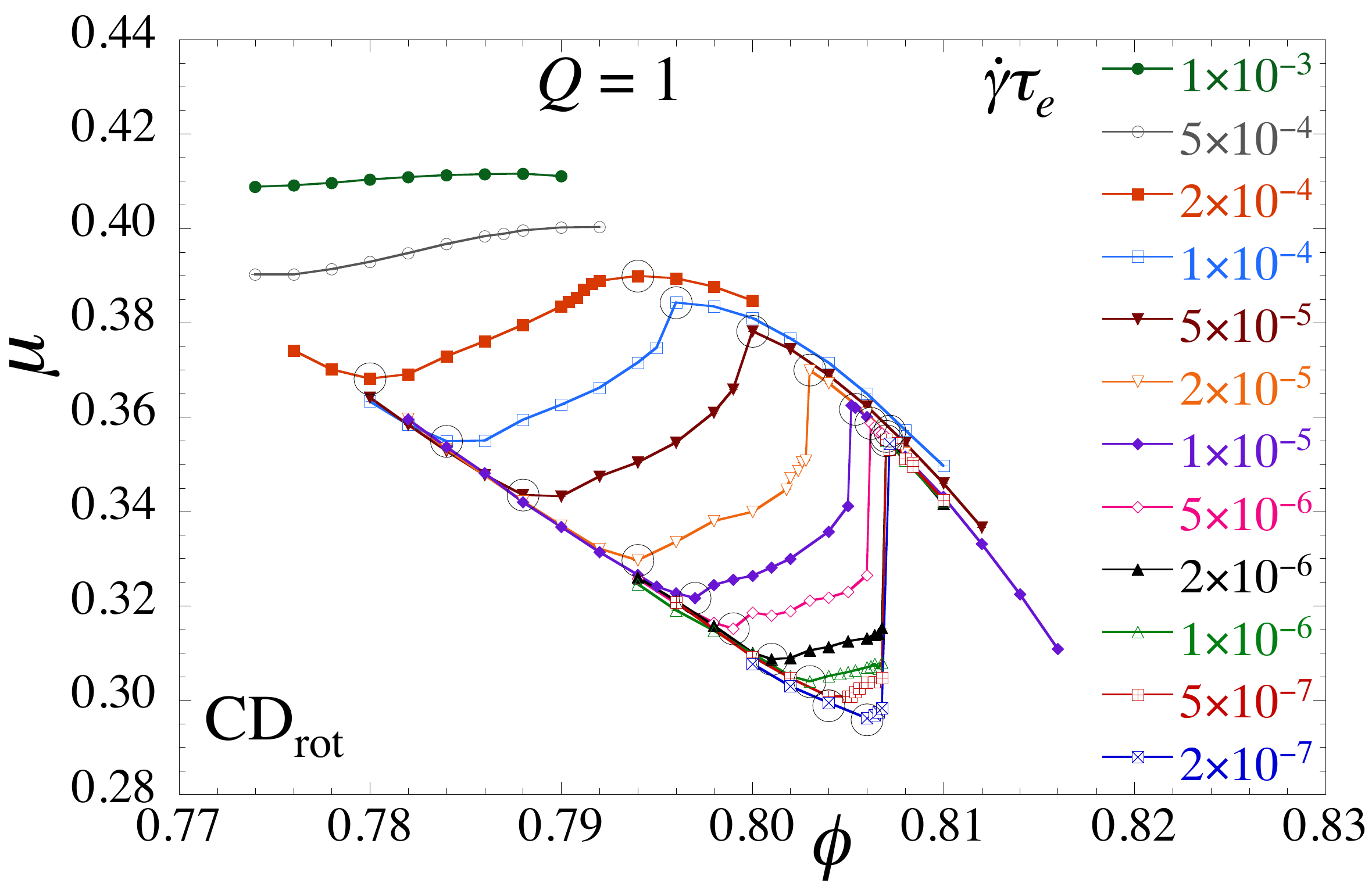} 
\caption{(Color online)  Macroscopic friction $\mu\equiv\sigma/p$ vs packing fraction $\phi$ for $Q=1$ in model CD$_\mathrm{rot}$.  Results are shown for different strain rates, ranging from  $\dot\gamma\tau_e=10^{-3}$ to $2\times 10^{-7}$, as curves go from top to bottom.  Large open circles denote the points where there curves of $\mu$ switch between the two different branches.  The system has $N=65536$ particles. 
}
\label{mu-vs-phi-65536}
\end{figure}

In Fig.~\ref{mu-vs-phi-65536} we plot $\mu$ vs $\phi$ for different values of $\dot\gamma\tau_e$ at $Q=1$.  We see that the data for different $\dot\gamma\tau_e$ do indeed seem to collapse to a common curve, but that this curve has two distinct branches.  For a given $\dot\gamma\tau_e$, the curve switches between the two branches at packing fractions $\phi_B(\dot\gamma\tau_e)$ and $\phi_N(\dot\gamma\tau_e)$, with $\phi_B(\dot\gamma\tau_e)<\phi_N(\dot\gamma\tau_e)$.  These switching points, as judged by eyeball, are denoted in the figure by the large open circles.  As $\dot\gamma\tau_e$ decreases, $\phi_B(\dot\gamma\tau_e)$ and $\phi_N(\dot\gamma\tau_e)$ increase towards a common $\phi_c$.  The two branches we see in $\mu$ suggest a coexistence region, where $\phi_B(\dot\gamma\tau_e)$ locates the Bagnoldian side of the coexistence region while $\phi_N(\dot\gamma\tau_e)$ locates the Newtonian side.

We now consider a more direct global measure of shear banding, by looking how the velocity profile compares to the linear profile $\langle v_x(\mathbf{r})\rangle=\dot\gamma \mathbf{r}\cdot\mathbf{\hat y}$ , $\langle v_y(\mathbf{r})\rangle=0$,
%$\langle\mathbf{v}(\mathbf{r})\rangle = \dot\gamma (\mathbf{r}\cdot\mathbf{\hat y})\mathbf{\hat x}$ 
that is expected for a uniform homogeneous shear flow.
We measure the fluctuation away from this presumed average by, 
\begin{align}
\langle\delta v_x^2\rangle &=\left\langle\dfrac{1}{N}\sum_i\left(v_{ix}-\gdot r_{iy}\right)^2\right\rangle,\label{vx2}\\ 
\langle v_y^2\rangle&=\left\langle\dfrac{1}{N}\sum_i  v_{iy}^2\right\rangle.\label{vy2}
\end{align}
%\begin{align}
%\langle\delta v_x^2\rangle &=\dfrac{1}{L^2}\int d^2r\, \left\langle\left(v_x(\mathbf{r})-\gdot\mathbf{r}\cdot\mathbf{\hat y}\right)^2\right\rangle,\label{vx2}\\ 
%\langle v_y^2\rangle&=\dfrac{1}{L^2}\int d^2r\, \left\langle v_y^2(\mathbf{r})\right\rangle.\label{vy2}
%\end{align}
For a homogeneously sheared state we expect $\langle\delta v_x^2\rangle$ to be roughly the same order of magnitude as $\langle v_y^2\rangle$.  However in a shear banded state, where the local strain rate is nonuniform, we would expect $\langle\delta v_x^2\rangle$, as defined above with respect to the fixed  average $\dot\gamma$, to become anomalously large; this follows since the velocity profile is no longer linear over the entire width of the system, but only piecewise linear, with different slopes, in each shear band (as we will soon see in Fig.~\ref{vx-gdot-phi-vs-y}).  

\begin{figure}
\includegraphics[width=3.2in]{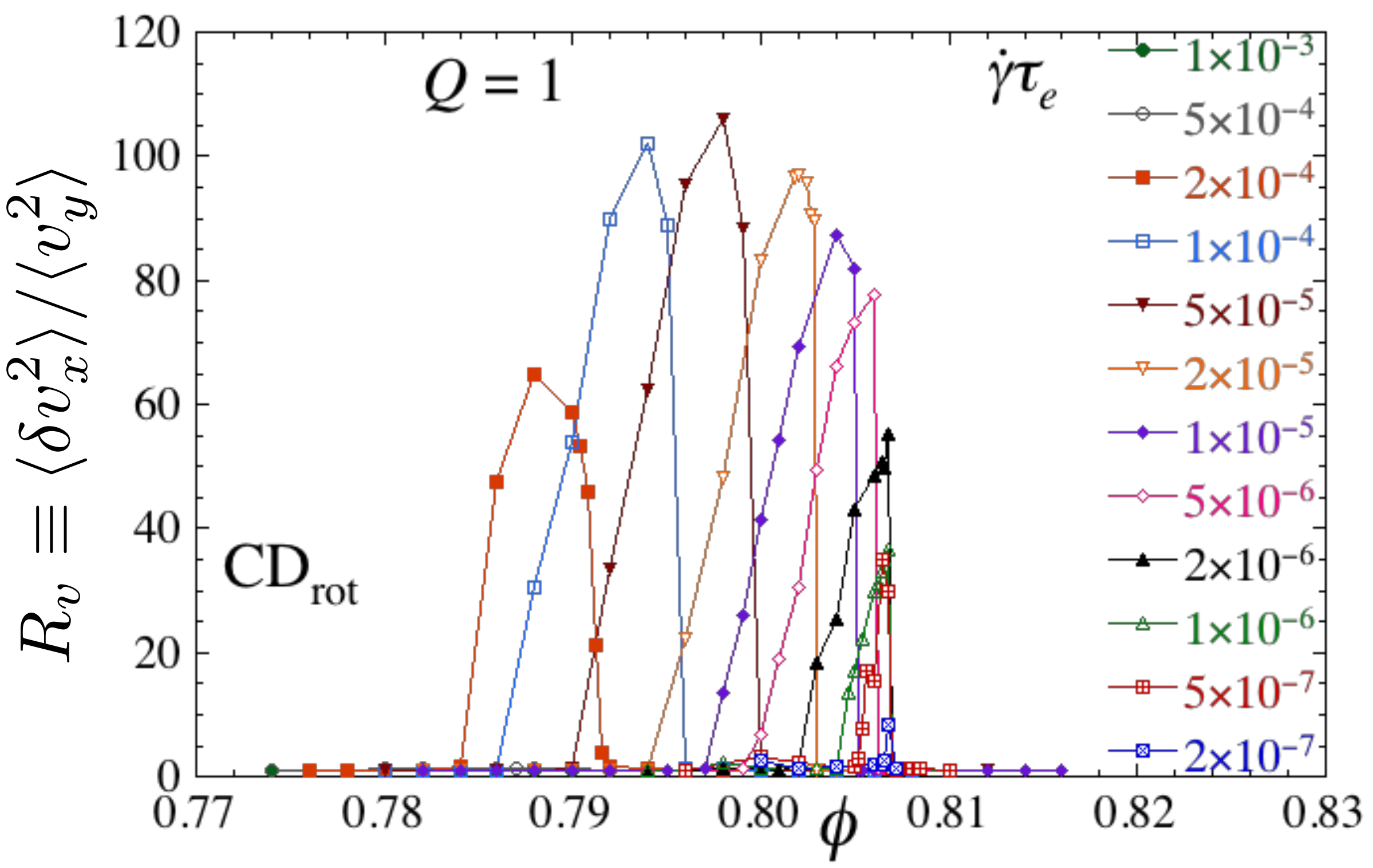} 
\caption{(Color online)  Shear banding paramter $R_v\equiv\langle\delta v_x^2\rangle/\langle v_y^2\rangle$ vs packing fraction $\phi$ for $Q=1$ in model CD$_\mathrm{rot}$.  Results are shown for different strain rates, ranging from  $\dot\gamma\tau_e=10^{-3}$ to $2\times 10^{-7}$, as curves go from left to right.  Within the shear-banded region, $R_v$ is large;  outside the shear-banded region, $R_v\approx 1$.  The system has $N=65536$ particles. 
}
\label{shband-vs-phi-65536}
\end{figure}

Defining the ratio of velocity fluctuations defined in Eqs.~(\ref{vx2}) and (\ref{vy2}) as, 
\begin{equation}
R_v\equiv \langle\delta v_x^2\rangle/\langle v_y^2\rangle,
\label{eRv}
\end{equation}
in Fig.~\ref{shband-vs-phi-65536} we plot $R_v$ vs $\phi$ for different fixed values of $\dot\gamma\tau_e$.  We see that at low and high $\phi$, $R_v$ is indeed of $O(1)$ as in a homogeneously sheared state.  But for intermediate $\phi$ near $\phi_c$, we see a large jump in $R_v$ as expected for a shear banded state.  For a given $\dot\gamma\tau_e$, we will denote as $\phi_B(\dot\gamma\tau_e)$ the value of $\phi$ at which $R_v$ takes the sharp jump upwards from $O(1)$, and the higher $\phi_N(\dot\gamma\tau_e)$ as the value of $\phi$ at which $R_v$ takes the sharp jump back down to $O(1)$.

\begin{figure}[h!] 
\includegraphics[width=3.2in]{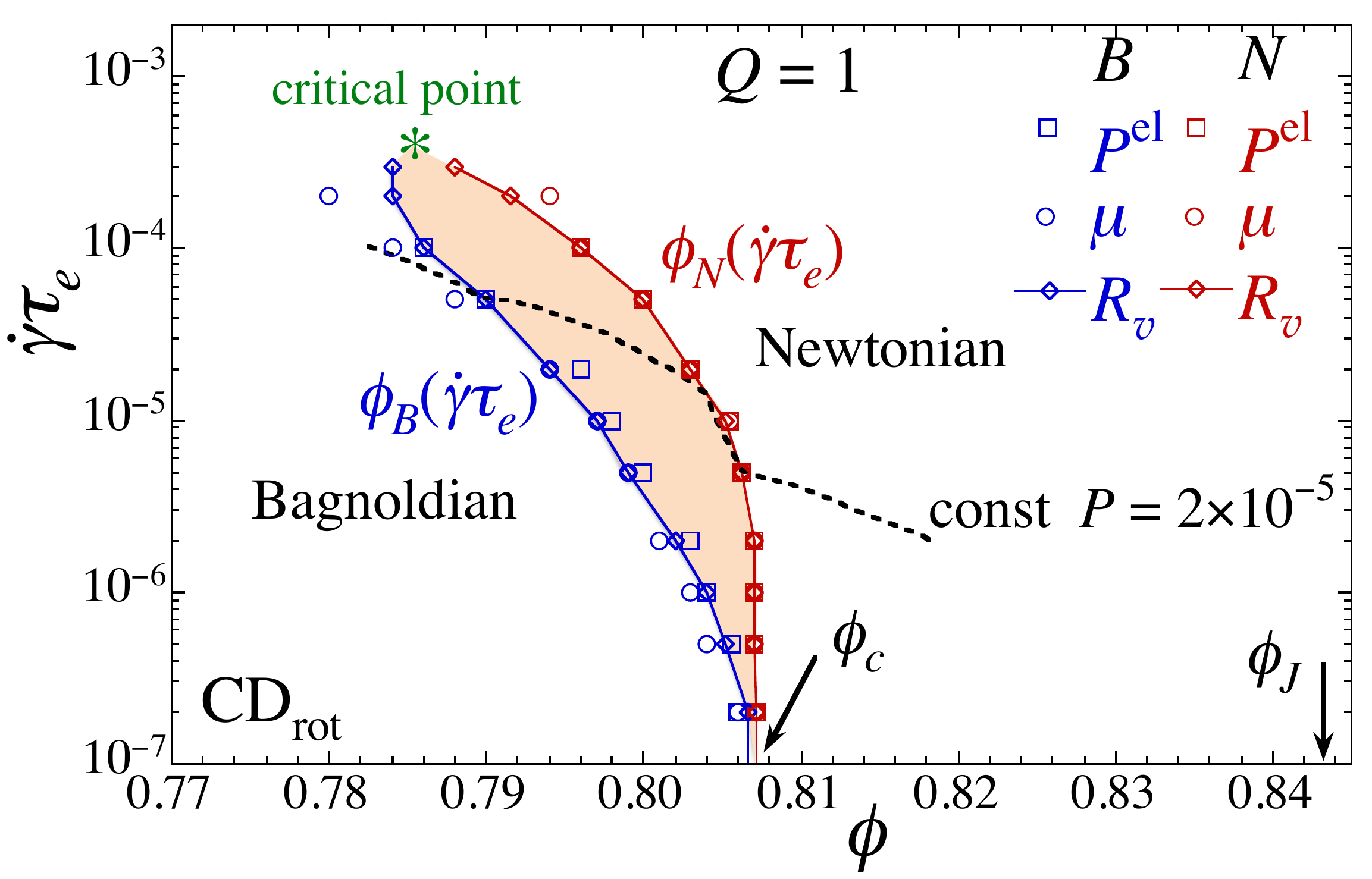} 
\caption{(Color online)  Phase diagram in the $(\phi,\dot\gamma\tau_e)$ plane for a system with $Q=1$.  The shaded region is a coexistence region where shear bands of Bagnoldian rheology and Newtonian rheology coexist in mechanical equilibrium.  The boundary curves $\phi_B(\dot\gamma\tau_e)$ and $\phi_N(\dot\gamma\tau_e)$, on the Bagnoldian and Newtonian sides of the coexistence region respectively, are determined from three different measurements: kinks in the $P^\mathrm{el}$ vs $\phi$ curves of Fig.~\ref{pe-vs-phi-65536} (open squares), switching points in the $\mu$ vs $\phi$ curves of Fig.~\ref{mu-vs-phi-65536} (open circles), jumps in the velocity fluctuation ratio $R_v$ vs $\phi$ curves of Fig.~\ref{shband-vs-phi-65536} (open diamonds).  Solid lines connect the data points from $R_v$, which we regard as the most reliable of the three measures of shear banding.  As $\dot\gamma\tau_e$ increases, $\phi_B(\dot\gamma\tau_e)$ and $\phi_N(\dot\gamma\tau_e)$ merge together at a critical end-point, denoted by the ``$*$''.  The dashed curve cutting through the coexistence region is a contour of constant pressure $P=2\times 10^{-5}$. The jamming point at $\phi_J\approx 0.8433$, above the coexistence region, is indicated.
}
\label{coexist}
\end{figure}

\subsubsection{Phase diagram and shear banding}

We now have from $P^\mathrm{el}$, $\mu$ and $R_v$, three different determinations of the curves $\phi_B(\dot\gamma\tau_e)$ and $\phi_N(\dot\gamma\tau_e)$, which we interpret as the Bagnoldian and Newtonian boundaries of a coexistence region in the $(\phi,\dot\gamma\tau_e)$ plane.  In Fig.~\ref{coexist} we plot these data and see that the results from $P^\mathrm{el}$, $\mu$ and $R_v$ reasonably agree with one another. Since we regard $R_v$ as the most objective of our measurements of these boundaries, we connect the data points from $R_v$ in the figure by solid lines, and denote the enclosed shaded region as the coexistence region of a first-order Bagnoldian-to-Newtonian phase transition. As $\dot\gamma\tau_e$ decreases, we see that $\phi_B$ and $\phi_N$ appear to come together at a common $\phi_c\approx 0.807$.  As $\dot\gamma\tau_e$ increases, the boundary curves $\phi_B(\dot\gamma\tau_e)$ and $\phi_N(\dot\gamma\tau_e)$ merge together at a critical end-point (denoted schematically in the figure by the ``$*$'').  While we have not determined the precise location of this critical end-point, we know that it must occur at a $\dot\gamma^*\tau_e< 5\times 10^{-4}$, since the curve of $R_v$ vs $\phi$ at $\dot\gamma\tau_e=5\times 10^{-4}$ gives no evidence of shear banding.

\begin{figure}[h!] 
\includegraphics[width=3.2in]{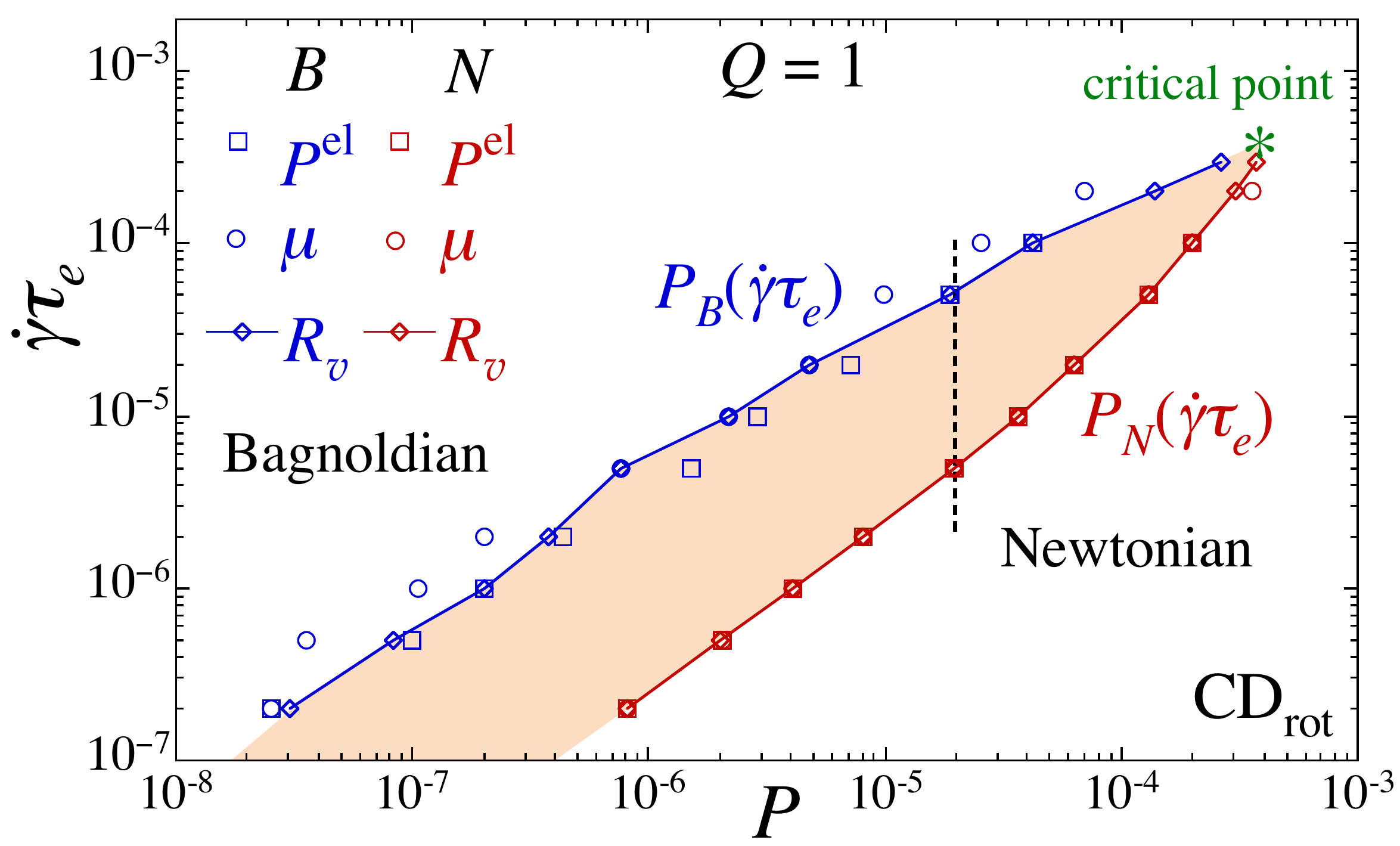} 
\caption{(Color online)  Phase diagram in the $(P,\dot\gamma\tau_e)$ plane for a system with $Q=1$.  The shaded region is a coexistence region where shear bands of Bagnoldian rheology and Newtonian rheology coexist in mechanical equilibrium.  The boundary curves $P_B(\dot\gamma\tau_e)$ and $P_N(\dot\gamma\tau_e)$, on the Bagnoldian and Newtonian sides of the coexistence region respectively, are determined from three different measurements: kinks in the $P^\mathrm{el}$ vs $\phi$ curves of Fig.~\ref{pe-vs-phi-65536} (open squares), switching points in the $\mu$ vs $\phi$ curves of Fig.~\ref{mu-vs-phi-65536} (open circles), jumps in the velocity fluctuation ratio $R_v$ vs $\phi$ curves of Fig.~\ref{shband-vs-phi-65536} (open diamonds).  Solid lines connect the data points from $R_v$, which we regard as the most reliable of the three measures of shear banding.  As $\dot\gamma\tau_e$ increases, $P_B(\dot\gamma\tau_e)$ and $P_N(\dot\gamma\tau_e)$ merge together at a critical end-point, denoted by the ``$*$''.  The vertical dashed curve cutting through the coexistence region is a contour of constant pressure $P=2\times 10^{-5}$.
}
\label{P-gdot-pd}
\end{figure}

\begin{figure}[t]
\includegraphics[width=3.2in]{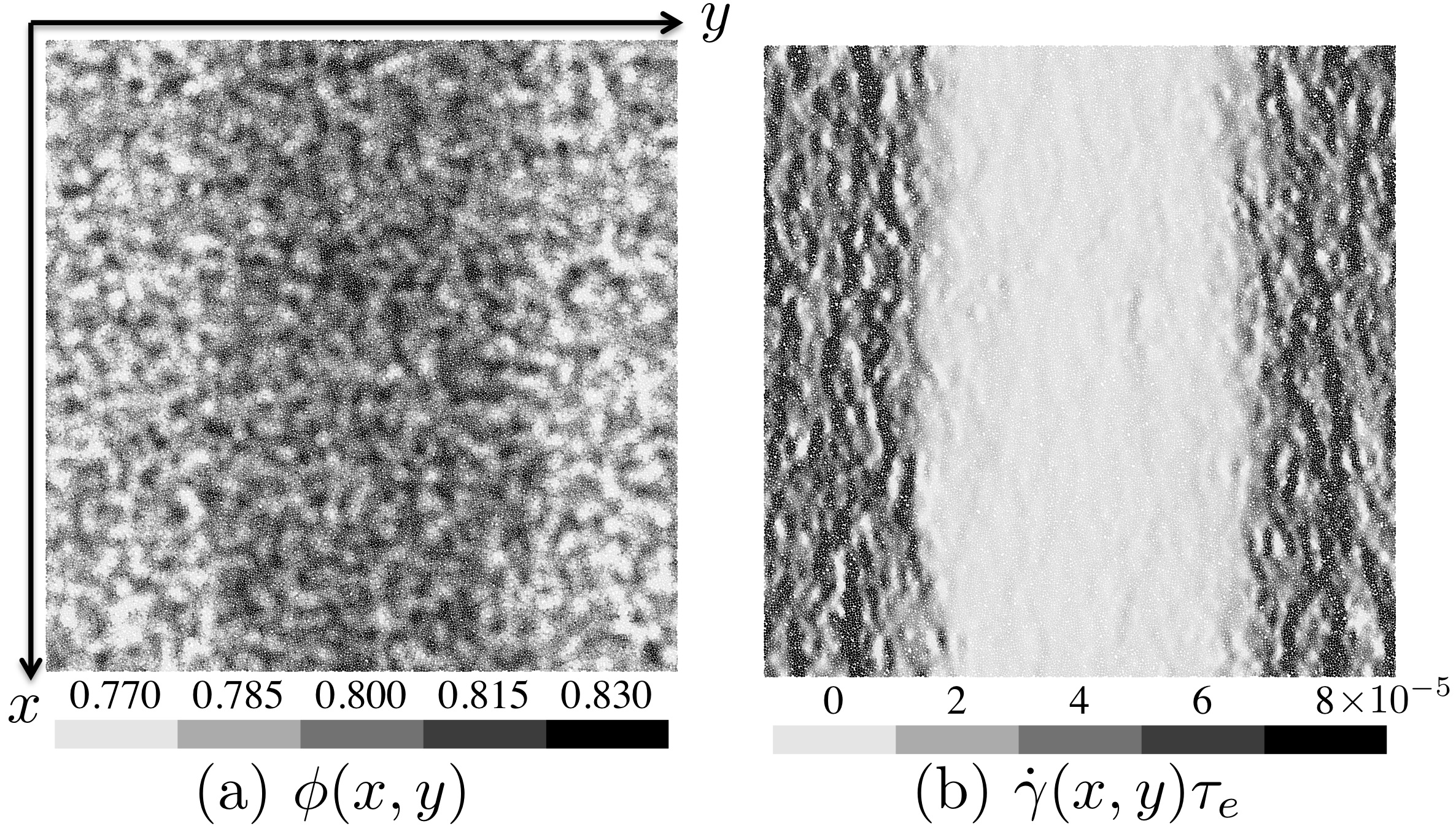}
\caption{Gray-scale encoded snapshot of a typical configuration at $(\phi,\dot\gamma\tau_e)=(0.7984, 3\times 10^{-5})$, within the coexistence region. (a) Local packing fraction $\phi(x,y)$ and (b) local shear strain rate $\dot\gamma(x,y)\tau_e$.  Shear banding is clearly observed. The system has $N=65536$ particles and $Q=1$.
}
\label{snapshot}
\end{figure}

\begin{figure*}[t]
\includegraphics[width=7in]{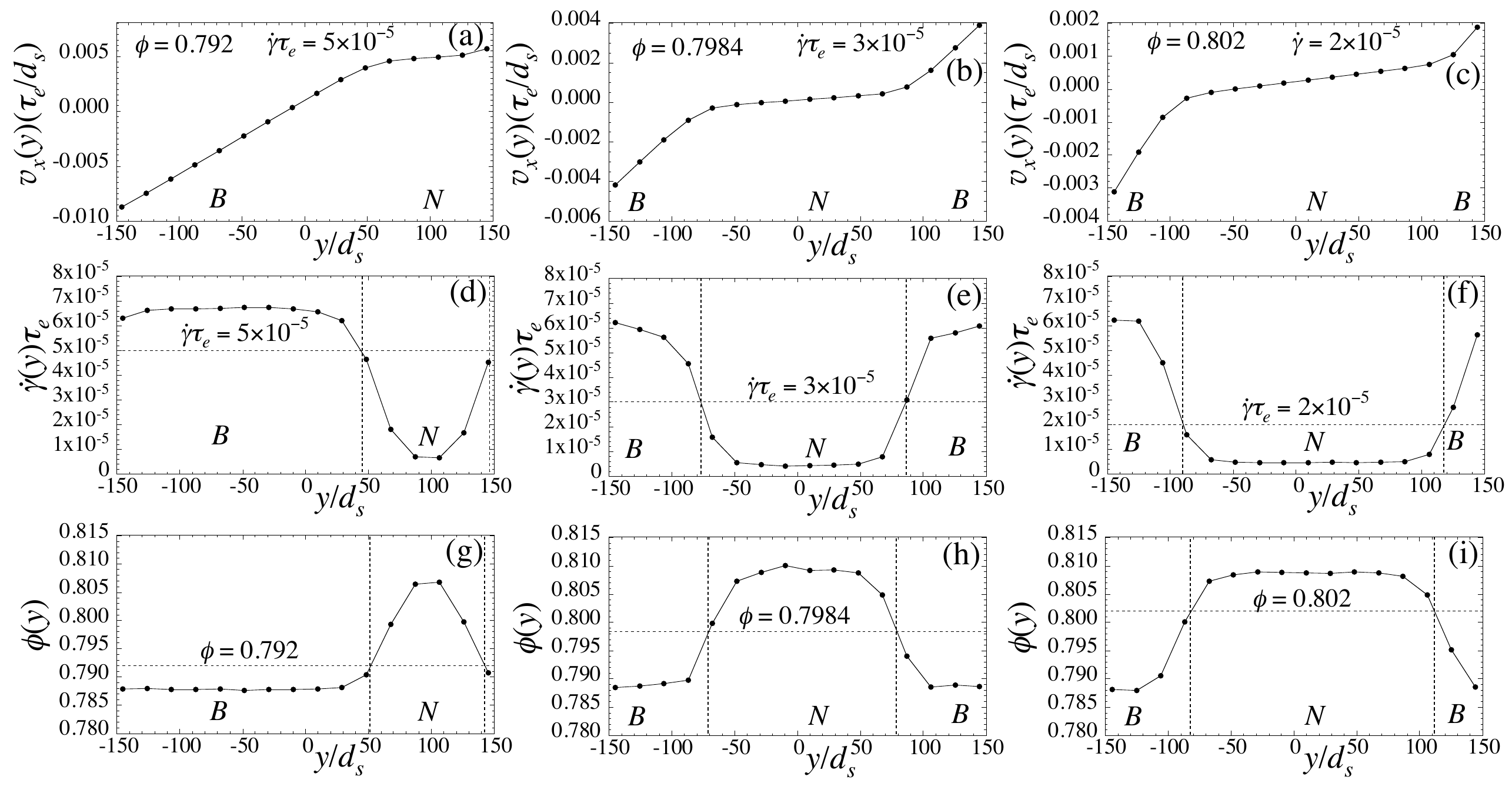}
\caption{Top row: Dimensionless average local velocity $v_x(y)(\tau_e/d_s)$ vs position $y/d_s$ transverse to the flow direction.  Middle row: Average local shear strain rate $\dot\gamma(y)\tau_e$ vs position $y/d_s$.  Bottom row: Average local packing fraction $\phi(y)$ vs $y/d_s$.  Results are shown for three different values of the global parameters, $(\phi= 0.792, \dot\gamma\tau_e=5\times 10^{-5})$ (left column), $(\phi=0.7984, \dot\gamma\tau_e=3\times 10^{-5})$ (middle column) and $(\phi=0.802, \dot\gamma\tau_e=2\times 10^{-5})$ (right column) as one passes through the coexistence region along the contour of constant pressure $P=2\times 10^{-5}$ shown in Fig.~\ref{coexist}.  As $\phi$ increases along this contour, we see that the fraction of the system that is the Bagnoldian phase decreases, while the fraction in the Newtonian phase increases.
Dashed horizontal lines in (d) -- (f) denote the value of the global $\dot\gamma\tau_e$.  Dashed horizontal lines in (g) -- (i) denote the value of the global $\phi$.
Dashed vertical lines in (d) -- (i) separate bands of Bagnoldian (labeled ``B") and Newtonian (labeled ``N") rheology.
Results are for a system with $N=65536$ particles and $Q=1$.
}
\label{vx-gdot-phi-vs-y}
\end{figure*}

It is also interesting to construct the corresponding phase diagram in the $(P,\dot\gamma\tau_e)$ plane, which we show in Fig.~\ref{P-gdot-pd}.  It is interesting to note that, as one increases $\dot\gamma$ at fixed $\phi$, one crosses from a region of Bagnoldian rheology to a region of Newtonian rheology.  In contrast, as one increases $\dot\gamma$ at fixed $P$, it is the reverse; i.e., one crosses from a region of Newtonian rheology to a region of Bagnoldian rheology.  The vertical dashed line cutting through the coexistence region at constant $P=2\times 10^{-5}$ in the $(P,\dot\gamma\tau_e)$ plane of Fig.~\ref{P-gdot-pd}, maps onto the dashed line in the $(\phi,\dot\gamma\tau_e)$ plane of Fig.~\ref{coexist}.

For a direct demonstration that the shaded region in Figs.~\ref{coexist} and \ref{P-gdot-pd} is indeed a region of coexisting Bagnoldian and Newtonian shear bands, we can look explicitly at the spatial variation of packing fraction and strain rate within configurations, as we cross this region.  If the interpretation of a coexistence region is correct, we expect that the system will phase separate into shear bands of Bagnoldian and Newtonian rheology in mechanical equilibrium with each other, i.e. the two bands will be at equal values of pressure $p$ and shear stress $\sigma$.  

We  consider crossing the coexistence region along a contour of constant pressure.  Such a contour at fixed dimensionless pressure $P=2\times 10^{-5}$ is shown as the dashed line in Figs.~\ref{coexist} and \ref{P-gdot-pd}.  In Fig.~\ref{snapshot} we show a gray-scale encoded snapshot of a sample configuration roughly mid way through the coexistence region on this contour, at the global values $(\phi,\dot\gamma\tau_e)=(0.7984,3\times 10^{-5})$.  By ``global values" we mean the values averaged over the entire system, or equivalently, the values of $\phi$ and $\dot\gamma$ that enter as the control parameters of our simulations.  We compute the local value of packing fraction $\phi$ and strain rate $\dot\gamma$ of each particle by averaging over a circle of radius $4d_s$ centered on each particle.  Figure~\ref{snapshot}(a) shows the resulting local $\phi(x,y)$ throughout the configuration, while Fig.~\ref{snapshot}(b) shows the local $\dot\gamma(x,y)\tau_e$.  Because our system is so large, with $N=65536$, it is difficult to distinguish individual particles in Fig.~\ref{snapshot}.  However the variation in the gray-scale allows one to visualize both the local average $\phi$ and $\dot\gamma\tau_e$, as well as fluctuations about the local average, as one varies position within the configuration.  We see clearly the Newtonian band in the center of the system, with a larger local $\phi$ and smaller local $\dot\gamma\tau_e$ as compared to the Bagnoldian band at the edges of the system.  For $\phi$, the fluctuations about the local average appear to be roughly uniform throughout both bands.  For $\dot\gamma\tau_e$, the fluctuations seem more pronounced in the Bagnoldian band.  Presumbaly this is because in the Newtonian band particles  stay in contact for long periods of time, thus reducing fluctuations in local velocity and hence the local strain rate. 

To quantify the results illustrated in Fig.~\ref{snapshot} we  average the behavior of the system along the flow direction $\mathbf{\hat x}$, and over many different configurations in our shearing ensemble, to plot quantities as a function of the position $y$ transverse to the flow direction.
In Fig.~\ref{vx-gdot-phi-vs-y} we show results at three different points along the contour of constant $P=2\times 10^{-5}$ within the coexistence region, corresponding to the global values of  $(\phi,\dot\gamma\tau_e)=(0.792,5\times 10^{-5})$ (left column), $(0.7984,3\times 10^{-5})$ (middle column), and $(0.802,2\times 10^{-5})$ (right column).  
The top row shows the average local velocity $v_x(y)$.  We clearly see that the velocity profile is not linear across the entire width of the system, but is rather piecewise linear with regions of two different slopes (recall, the system obeys Lees-Edwards boundary conditions, in the $y$ direction).
The middle row shows the average local strain rate $\dot\gamma(y)=dv_x/dy$.  The bottom row shows the average local packing fraction $\phi(y)$.  We see that in each case the system has clearly separated into two bands, one with larger local $\dot\gamma\tau_e$ and smaller local $\phi$, which we identify as the Bagnoldian band, and one with smaller local $\dot\gamma\tau_e$ and larger local $\phi$, which we identify as the Newtonian band.  As the global $\phi$ increases, the Newtonian band increases in width while the Bagnoldian band shrinks.  We have explicitly checked that these shear bands are not a transient effect, but rather represent the steady state; computing the profiles as in Fig.~\ref{vx-gdot-phi-vs-y} gives essentially the same results whether we average over a finite strain interval near the beginning of our run, or over a non-overlapping strain interval near the end of our run.

Our results thus show the behavior commonly expected at a first order phase transition.  Within the coexistence region the system separates into two phases in mutual equilibrium, with one phase growing and the other shrinking as one crosses from one side of the coexistence region to the other.  However there is one feature of our results that is not quite the same as for a first order transition.
If we denote as $(\phi^*_B, \dot\gamma^*_B)$ and $(\phi^*_N,\dot\gamma^*_N)$ the local values of the packing fraction and strain rate in the center of the Bagnoldian and Newtonian bands (i.e. away from the interface between the bands) then we see that these values are essentially constant everywhere along the contour of constant $P=2\times 10^{-5}$, with $(\phi^*_B, \dot\gamma^*_B)=(0.7880, 6.2\times 10^{-5})$ and $(\phi^*_N,\dot\gamma^*_N)=(0.8090,4.2\times 10^{-6})$.  This is expected because the condition of mechanical stability of the coexisting bands requires that $P=P(\phi^*_B,\dot\gamma^*_B\tau_e)=P(\phi^*_N,\dot\gamma^*_N\tau_e)$ and $\Sigma=\Sigma(\phi^*_B,\dot\gamma^*_B\tau_e)=\Sigma(\phi^*_N,\dot\gamma^*_N\tau_e)$ must be constant along this contour.  
For a standard first order phase transition, we would expect that these two values $(\phi^*_B,\dot\gamma^*_B\tau_e)$ and $(\phi^*_N,\dot\gamma^*_N\tau_e)$ would correspond to the two points $(\phi_B,\dot\gamma_B\tau_e)$ and $(\phi_N,\dot\gamma_N\tau_e)$ where the contour of constant $P=2\times 10^{-5}$ first enters the coexistence region from the Bagnoldian side and  where it exits the coexistence region on the Newtonian side.  However we find that this is not quite so;  we find that $(\phi_B,\dot\gamma_B\tau_e)=(0.790, 5.06\times 10^{-5})$ and $(\phi_N,\dot\gamma_N\tau_e)=(0.8062, 5.2\times 10^{-6})$, noticeably different from $(\phi^*_B,\dot\gamma^*_B\tau_e)$ and $(\phi^*_N,\dot\gamma^*_N\tau_e)$.  In Appendix B we consider this difference further and argue that it is a finite-size-effect resulting from the finite length of our system $L$ as compared to the finite width of the interface between the Bagnoldian and Newtonian shear bands seen in Fig.~\ref{vx-gdot-phi-vs-y}.

Note, the fact that the shear bands in Fig.~\ref{vx-gdot-phi-vs-y} have constant values of $(\phi^*_B,\dot\gamma^*_B\tau_e)$ and $(\phi^*_N,\dot\gamma^*_N\tau_e)$,  is only because we are crossing the coexistence region on a contour of constant pressure.   If we crossed the coexistence region along any other trajectory, for example at constant $\dot\gamma$ or constant $\phi$, the values of $\sigma$ and $p$ within the two coexisting shear bands at any global value of $(\phi,\dot\gamma\tau_e)$ would always be equal; but these values would be varying as we moved through the coexistence region along the trajectory, and hence the values of $(\phi^*_B,\dot\gamma^*_B\tau_e)$ and $(\phi^*_N,\dot\gamma^*_N\tau_e)$ in the bands would be varying.

%To illustrate this we consider crossing the coexistence region by increasing the global $\dot\gamma\tau_e$ at a fixed global value of $\phi_0=0.802$.  In Fig.~\ref{sig-vs-gdot-phi-802} we plot the resulting steady state dimensionless shear stress $\Sigma$ vs the global strain rate $\dot\gamma\tau_e$.  Defining $\dot\gamma_B(\phi)$ and $\dot\gamma_N(\phi)$ as the inverse functions of $\phi_B(\dot\gamma\tau_e)$ and $\phi_N(\dot\gamma\tau_e)$,
%we see kinks at the $\dot\gamma_B(\phi_0)$ and $\dot\gamma_N(\phi_0)$  where the system enters and leaves the coexistence region, analogous to the kinks seen in Fig.~\ref{pe-vs-phi-65536} for $P^\mathrm{el}$ vs $\phi$ at constant $\dot\gamma$.  For $\dot\gamma\tau_e < \dot\gamma_B(\phi_0)\tau_e$ we see $\Sigma\sim(\dot\gamma\tau_e)^2$, as expected for the Bagnoldian phase.  For $\dot\gamma\tau_e>\dot\gamma_N(\phi_0)\tau_e$ we see $\Sigma\sim\dot\gamma\tau_e$, as expected for the Newtonian phase.  Between these limits we have coexisting shear bands.

%By considering the spatial profiles of the packing fraction $\phi(y)$ and the strain rate $\dot\gamma(y)$, as in Fig.~\ref{vx-gdot-phi-vs-y}, we can determine the local packing fraction and strain rate in the coexisting shear bands as a function of the global $\dot\gamma$ at $\phi_0$.  We will define these, for the Bagnoldian and Newtonian bands respectively, as,
%\begin{align}
%\phi^{(B)}(\dot\gamma;\phi_0),\quad & \dot\gamma^{(B)}(\dot\gamma;\phi_0), \\
%\phi^{(N)}(\dot\gamma;\phi_0), \quad & \dot\gamma^{(N)}(\dot\gamma;\phi_0). 
%\end{align}

Finally, as is the case for an ordinary first order equilibrium phase transition, we expect that the coexistence region of finite width in the $(\phi,\dot\gamma\tau_e)$ phase diagram of Fig.~\ref{coexist} will collapse to a sharp line when plotted in a phase diagram expressed in terms of the appropriate variables that define the stability condition between the coexisting phases.  In the present case this is equality of stress.  We therefore consider the phase diagram expressed in the $(P,\Sigma)$ plane, where we may regard the dimensionless shear stress $\Sigma$ as the conjugate variable to $\dot\gamma\tau_e$, and the dimensionless pressure $P$ as the conjugate variable to $\phi$.

\begin{figure}[h!] 
\includegraphics[width=3.2in]{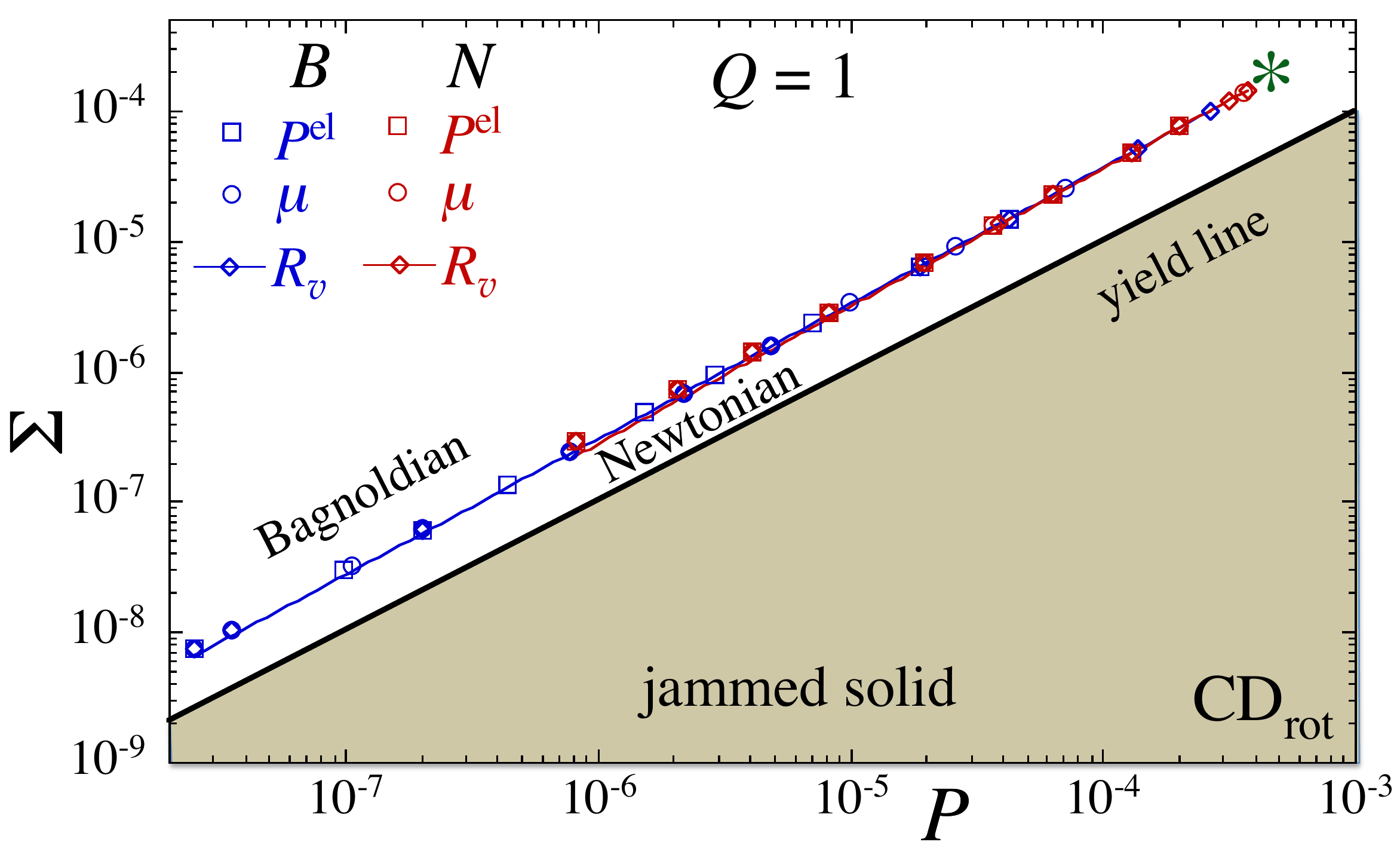} 
\caption{(Color online) Phase diagram in the $(P,\Sigma)$ plane.  Regions of Bagnoldian and Newtonian rheology are separated by a sharp first-order transition line that terminates at a critical end-point ``$*$".  The location of the transition line is determined from the values of dimensionless pressure $P$ and dimensionless shear stress $\Sigma$ along the coexistence region boundaries, $\phi_B(\dot\gamma\tau_e)$ and $\phi_N(\dot\gamma\tau_e)$, of Fig.~\ref{coexist}.  Results are shown from the three different determinations of these phase boundaries, i.e. from $P^\mathrm{el}$ (open squares), $\mu$ (open circles), and $R_v$ (open diamonds).  Also shown is the yield stress line $\Sigma=\mu_J P$, with $\mu_J\approx 0.1$.  Below the yield stress line lies the region of mechanically stable, static, jammed solid states.  Results are for a system with $N=65536$ particles and $Q=1$.
}
\label{stot-vs-ptot-coexist}
\end{figure}

In Fig.~\ref{stot-vs-ptot-coexist} we plot the values of $(P,\Sigma)$ along the Bagnoldian and Newtonian boundaries of the coexistence region of Fig.~\ref{coexist}, specifically the pairs of points $P(\phi_B(\dot\gamma\tau_e), \dot\gamma\tau_e)$ and $\Sigma(\phi_B(\dot\gamma\tau_e),\dot\gamma\tau_e)$, and the pairs of points   $P(\phi_N(\dot\gamma\tau_e), \dot\gamma\tau_e)$ and $\Sigma(\phi_N(\dot\gamma\tau_e),\dot\gamma\tau_e)$.  We plot such points for the three different determinations of $\phi_B(\dot\gamma\tau_e)$ and $\phi_N(\dot\gamma\tau_e)$ obtained from $P^\mathrm{el}$, $\mu$, and $R_v$.  We see that all points indeed collapse to a single sharp line, separating a region of Newtonian rheology from a region of Bagnoldian rheology.  Upon continuously crossing this line under conditions of controlled $P$ and $\Sigma$, the system will show a discontinuous jump in shear strain rate $\dot\gamma\tau_e$ and packing fraction $\phi$.
This first-order transition line  terminates at a critical end-point, indicated schematically by the ``$*$" in the figure.  For reference, we also indicate in the figure the yield stress line, defined by $\Sigma=\mu_J P$, with $\mu_J\approx 0.1$ the macroscopic friction at the jamming transition \cite{VOT_RDCD}.  Below the yield stress line lies the region of mechanically stable, static, jammed solid states. As one approaches the yield stress line from above, the Newtonian rheology just below the first-order transition line will merge continuously with the Herschel-Bulkley rheology at finite $P$ just above the yield line.

\subsection{Model CD$_\mathrm{rot}$: Discontinuous shear thickening}
\label{s3DST}

Discontinuous shear thickening (DST) is the observed phenomenon in which there is a discontinuous jump in the shear stress upon continuously increasing the shear strain rate.  In this section we show how the first order rheological transition that gives rise to the shear banding also  can provide a possible mechanism for DST.

\begin{figure}[h!] 
\includegraphics[width=3.2in]{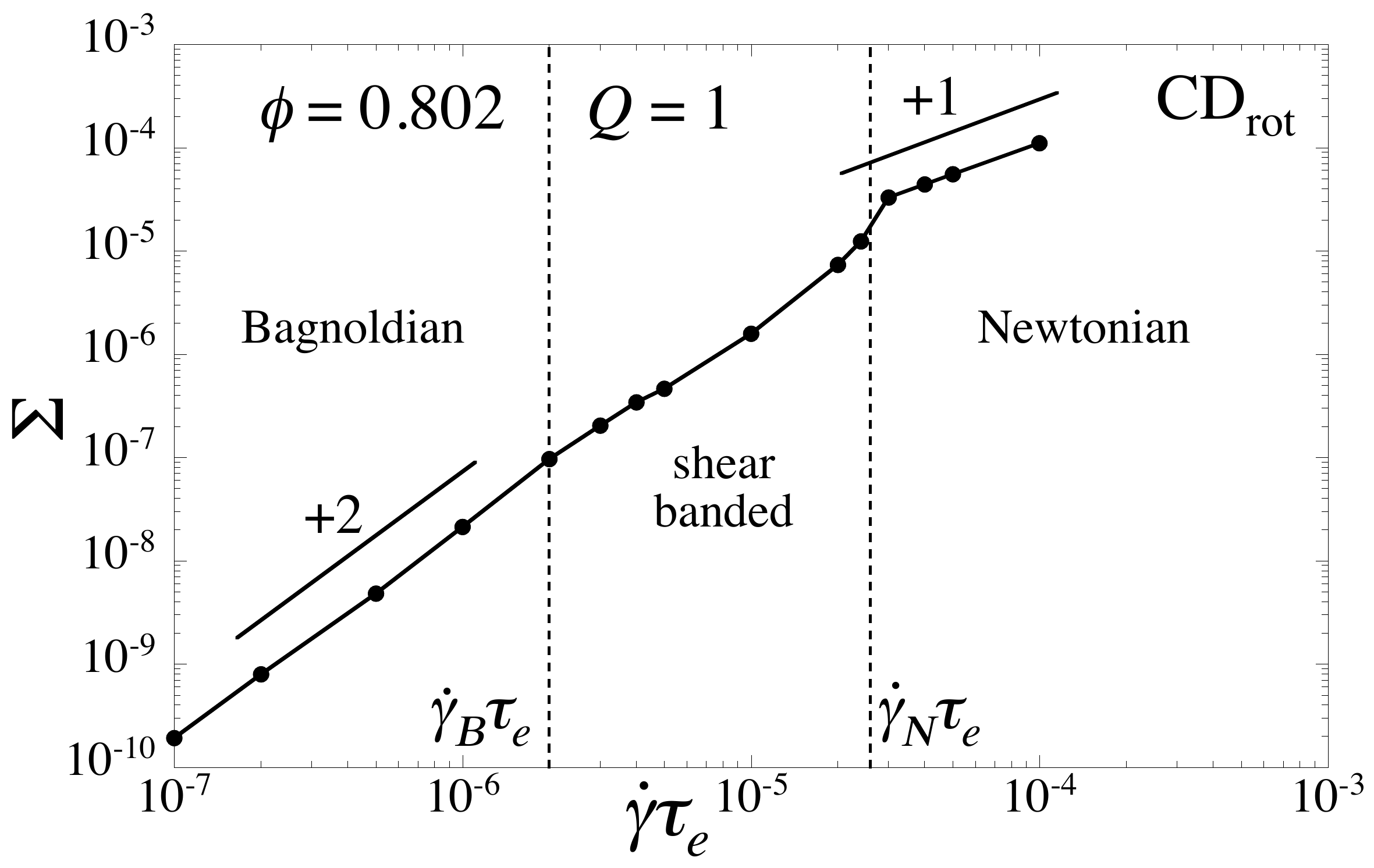} 
\caption{Dimensionless steady-state shear stress $\Sigma$ vs $\dot\gamma\tau_e$ at fixed $\phi=0.802$, as one passes through the shear banded coexistence region bounded by $\dot\gamma_B\tau_e$ and $\dot\gamma_N\tau_e$.  Results are for a system with $N=65536$ particles and $Q=1$.
}
\label{sig-vs-gdot-phi-802}
\end{figure}

Consider sitting at a fixed packing fraction  (i.e. fixed volume) $\phi<\phi_c$ and crossing the coexistence region by continuously increasing the strain rate $\dot\gamma$.  If $\dot\gamma$ is varied sufficiently slowly, so that the system has time to reach the true shear banded steady state at each instantaneous value of $\dot\gamma$, then the resulting $\Sigma(\dot\gamma\tau_e)$ will be continuous, with kinks where $\dot\gamma$ enters and leaves the coexistence region at $\dot\gamma_B(\phi)$ and $\dot\gamma_N(\phi)$ (these are the inverse functions of $\phi_B(\dot\gamma\tau_e)$ and $\phi_N(\dot\gamma\tau_e))$.  
We show such a plot of the steady state dimensionless $\Sigma$ vs $\dot\gamma\tau_e$ at fixed $\phi=0.802$ in Fig.~\ref{sig-vs-gdot-phi-802}.

However if $\dot\gamma$ is varied too rapidly for the system to achieve steady state, then we expect to find the phenomenon generally characteristic of first-order transitions.  Upon increasing $\dot\gamma$, the system will fall out of the true steady state, overshooting the coexistence region phase boundary to persist in the Bagnoldian phase for some range of $\dot\gamma>\dot\gamma_B(\phi)$.  Increasing $\dot\gamma$ further, the system will at some point become too far out of the proper steady state and will switch to the Newtonian phase with a resulting jump in $\Sigma$.  Decreasing $\dot\gamma$, one will see a similar phenomenon, but with hysteresis in the location of the stress jump.

To test this expectation we perform simulations at fixed $\phi$ in which the strain rate $\dot\gamma$ is ramped up and down at a fixed rate $\kappa$,
\begin{equation}
\dfrac{d\dot\gamma}{dt}=\pm\kappa\dot\gamma,
\end{equation}
where $(+)$ is for ramping up, and $(-)$ is for ramping down.  We implement this in our simulations by changing $\dot\gamma$ at time intervals $\Delta t=\tau_e$ according to, $\dot\gamma(t+\tau_e)=\dot\gamma(t){\rm e}^{\pm\kappa \tau_e}$.  For convenience we define the dimensionless rate,
\begin{equation}
%\tilde\kappa\equiv (\log_{10} \mathrm{e})\kappa\tau_e,
\tilde\kappa\equiv \kappa\tau_e/\ln 10,
\end{equation}
so that $\dot\gamma(t+\tau_e)=\dot\gamma(t)10^{\pm\tilde\kappa}$.  Each time $\dot\gamma$ is updated we also rescale particle velocities, 
\begin{equation}
\mathbf{v}_i(t+\tau_e)=\mathbf{v}_i(t) \dfrac{\dot\gamma(t+\tau_e)}{\dot\gamma(t)},
\end{equation}
and angular velocities,
\begin{equation}
\dot\theta_i(t+\tau_e)=\dot\theta_i(t) \dfrac{\dot\gamma(t+\tau_e)}{\dot\gamma(t)},
\end{equation}
so that the change in strain rate is imposed throughout the bulk of the system, rather than only at the boundary.
We then average $P$, $\Sigma$ and $\dot\gamma\tau_e$ over a time interval $10^3\tau_e$, so as to reduce statistical fluctuations, and thus obtain our plotted results.

\begin{figure}[h!] 
\includegraphics[width=3.2in]{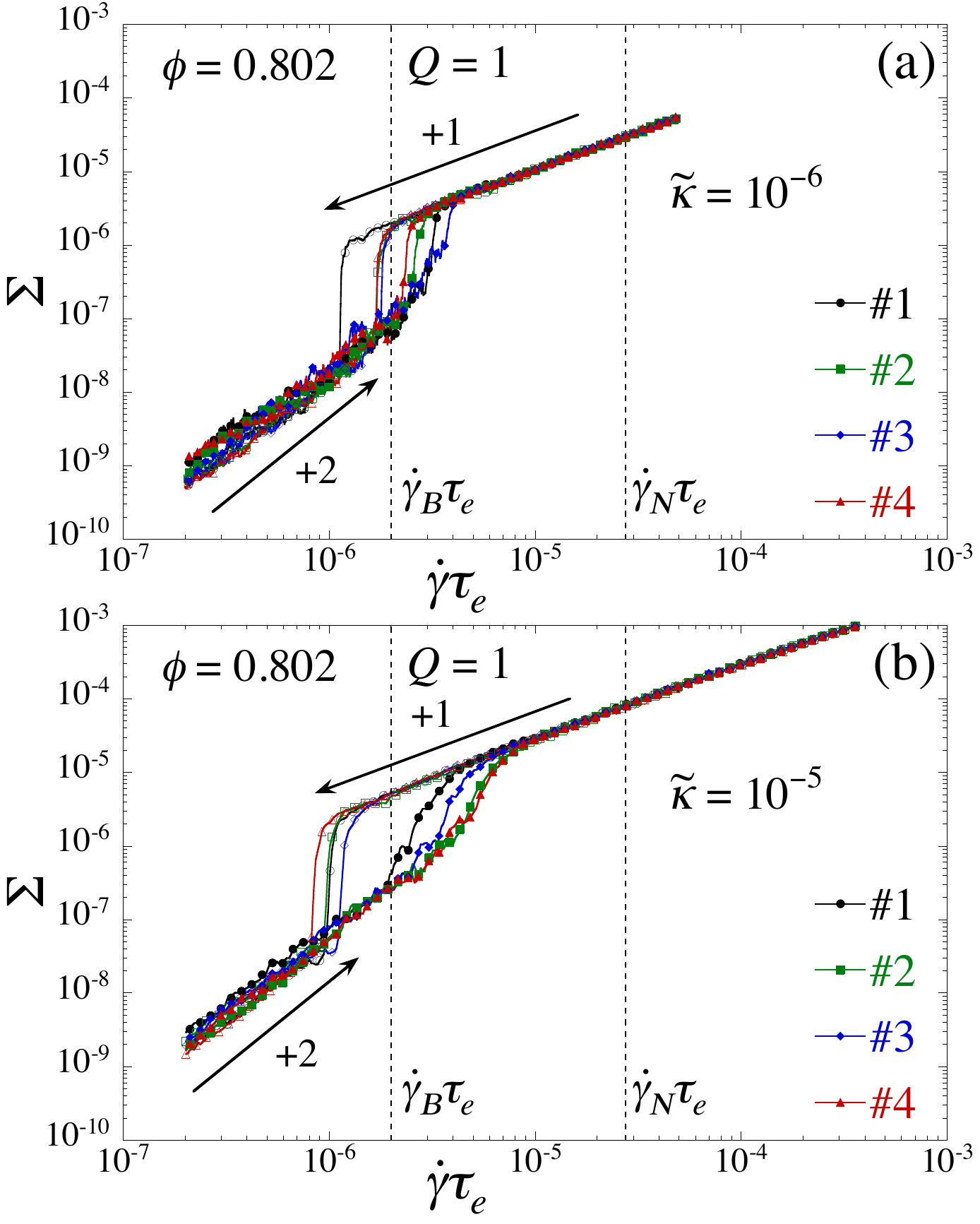} 
\caption{(Color online)  Dimensionless shear stress $\Sigma$ vs shear strain rate $\dot\gamma\tau_e$, as the strain rate is cyclically ramped up and down at a constant rate of (a) $\tilde\kappa=10^{-6}$ and (b) $\tilde\kappa=10^{-5}$.  In each case results are shown for four different initial configurations, with different symbol shapes used to distinguish among the different runs.  In each case a solid symbol is used for the ramping up part of the cycle, while the corresponding open symbol is used for the ramping down part of the cycle.  In (a) we show only every 20th data point, and in (b) only every 5th data point, so as to improve clarity.  The upward pointing arrow with slope 2 denotes the Bagnoldian branch, while the downward pointing arrow with slope of unity denotes the Newtonian branch.  Vertical dashed lines denote the boundaries of the coexistence region at $\dot\gamma_B\tau_e$ and $\dot\gamma_N\tau_e$.  The system is at $\phi=0.802$ with $Q=1$, and has $N=65536$ particles.  
}
\label{Sig-vs-gdot}
\end{figure}

We start at an initial low strain rate well below $\dot\gamma_B$, and increase to a largest strain rate well above $\dot\gamma_N$.  We then ramp back down to the initial strain rate.
In Fig.~\ref{Sig-vs-gdot} we show results for the dimensionless shear stress $\Sigma$ vs $\dot\gamma$ as obtained from this process at the fixed $\phi=0.802$.  In Fig.~\ref{Sig-vs-gdot}(a) we show results for the ramping rate $\tilde\kappa=10^{-6}$, while in \ref{Sig-vs-gdot}(b) we show results for $\tilde\kappa=10^{-5}$.  In each case we show results for four independent initial configurations.  

We see behavior as described above. Upon increasing $\dot\gamma\tau_e$ the system stays on the Bagnoldian branch, with $\Sigma\sim\dot\gamma^2\tau_e^2$, as $\dot\gamma\tau_e$ increases above $\dot\gamma_B\tau_e$, before transitioning  to the Newtonian branch, with $\Sigma\sim\dot\gamma\tau_e$, inside the coexistence region.  Upon decreasing $\dot\gamma\tau_e$ the system stays on the Newtonian branch throughout the coexistence region, transitioning to the Bagnoldian branch only below $\dot\gamma_B\tau_e$.  The jump upwards in $\Sigma$ upon increasing $\dot\gamma\tau_e$ is smaller and more gradual than the jump downwards upon decreasing $\dot\gamma\tau_e$.  The location of the jumps are found to vary, without any obvious trend, as the initial configuration is varied, while the width of the hysteresis region shrinks as the ramping rate $\tilde\kappa$ decreases.  We note that we have observed similar stress jumps in model CD when ramping the strain rate near the Bagnoldian-to-Newtonian transition line $Q_c(\phi)$. 
We thus find that the region of coexisting shear bands in the $(\phi,\dot\gamma\tau)$ phase diagram provides a mechanism for the system to manifest DST, which is seen to result when the strain rate varies too rapidly for the system to relax to the shear-banded steady state corresponding to the instantaneous value of $\dot\gamma$.  Our results in Fig.~\ref{Sig-vs-gdot} appear qualitatively similar to numerical results obtained previously for particles with intergranular elastic friction \cite{OH_friction,Grob}, except in those works the large $\dot\gamma$ branch displayed behavior characteristic of the jammed solid while in our case it is a Newtonian liquid.

\section{Conclusions}
\label{s4}

We have demonstrated that in simple two dimensional models of athermal, soft-core, bidisperse, frictionless  massive disks, we have a sharp first-order rheological transition from Bagnoldian rheology to Newtonian rheology as one varies the packing fraction $\phi$ and inelasticity of the collisions $Q$.  Key to the existence of this transition is the presence of a viscous dissipation acting in the tangential direction at the point of contact of two particles in collision.  
Within this framework, we find that the system always becomes Newtonian sufficiently close to the jamming transition $\phi_J$, however the behavior of the system at small $Q$ (strongly inelastic) depends dramatically on the precise form of the tangential dissipation.  When the dissipation couples only to the translational motion (our model CD) then the system at small $Q$ is always Newtonian for any packing fraction $\phi$.  But when the rotational motion is coupled to the translational motion (our model CD$_\mathrm{rot}$), then rotational motion can compensate for relative translational motion, and act to effectively eliminate the tangential dissipation.  In this case we find that our system has Bagnoldian rheology at small $\phi<\phi_c$, and Newtonian rheology at large $\phi>\phi_c$.

We find that the main distinguishing feature between systems with Newtonian vs Bagnoldian rheology is that in Newtonian rheology the average number of contacts per particle $Z$ stays finite as the strain rate $\dot\gamma\to 0$, while in Bagnoldian rheology $Z\to 0$ as $\dot\gamma\to 0$.  Thus Newtonian rheology is associated with the formation of large connected clusters.
The transition from Bagnoldian to Newtonian rheology upon decreasing $Q$ below $Q_c(\phi)$ in either model CD or CD$_\mathrm{rot}$ results from the increasing inelasticity of the collisions, which act to damp out relative motion; below the critical $Q_c(\phi)$, the duration of a collision $\tau_\mathrm{dur}$ grows dramatically and particles remain in contact. The nature of the transition from Bagnoldian to Newtonian rheology upon increasing $\phi$ at small $Q$ in model CD$_\mathrm{rot}$ is somewhat more subtle.  At low $\phi$, the particles are able to adjust their rotational motion so that the tangential component of the relative {\em contact} velocity at a collision $v_{ij,T}^C$ is very small or vanishing, greatly reducing or eliminating the tangential dissipation.  As a consequence, the tangential relative motion of the particles' centers of mass $v_{ij,T}$ is not damped, particles may separate after a collision, so no large persistent clusters form and the rheology stays Bagnoldian.  As $\phi$ increases however, we find that the average number of particles participating in any given collision increases, and so we speculate that above a critical $\phi_c(Q)$, such multi-particle collisions serve to constrain the rotational motion of the particles in contact, tangential dissipation rises,  relative  motion is damped, and large persistent clusters form.

We have further studied this Bagnoldian-to-Newtonian rheological transition at small $Q$ in model CD$_\mathrm{rot}$
to explore the behavior of the system as a function of increasing shear strain rate $\dot\gamma$.  We find that as $\dot\gamma$ increases, the discontinuous first-order rheological transition opens up into a region of finite width in the $(\phi,\dot\gamma)$ plane consisting of coexisting shear bands of Bagnoldian and Newtonian rheology in mechanical equilibrium with each other.  Thus the interplay of inertia and tangential dissipation gives rise to a mechanism for shear banding, even in our simple homogeneous system of frictionless, repulsive, spherical particles.

We have also considered the behavior of our model CD$_\mathrm{rot}$ as one crosses the above coexistence region by ramping the shear strain rate $\dot\gamma$ up and down, holding the system at constant volume.  When $\dot\gamma$ is ramped at a rate too fast for the system to settle into the true steady-state shear-banded configurations, then one can have hysteretic, discontinuous, jumps in stress as the system transitions between the Bagnoldian and Newtonian phases, similar to such phenomena in first-order equilibrium phase transitions.  Thus the shear banding coexistence region can provide a possible mechanism for discontinuous shear thickening, even for particles with no microscopic inter-granular friction  \cite{Hayakawa2}.

It will be interesting to investigate how robust is the shear-banding phase diagram we have found to the inclusion of other physical effects commonly found in experimental systems of interest.  In our models the energy dissipation is taken to occur only via the collisions between particles.  In non-Brownian suspensions, emulsions, and colloids, however, the particles are embedded within some fluid host matrix, and so one expects there to be a viscous damping, via Stokes drag, between the particles and the host fluid.  There should thus be an additional viscous force $-\zeta [\mathbf{v}_i-\mathbf{u}(\mathbf{r}_i)]$, where $\mathbf{v}_i$ is the velocity of particle $i$, and $\mathbf{u}(\mathbf{r}_i)$ is the local fluid velocity at particle $i$; a similar damping of rotational motion should also be included.
%, and a similar damping of rotational motion.  
Such a viscous drag will result in a rheology that is always Newtonian at sufficiently small $\dot\gamma$.  However, depending on the value of the viscous drag coefficient $\zeta$, upon increasing $\dot\gamma$ inertial effects can become important again and there can be a crossover $\dot\gamma_\times$ where the system crosses smoothly from Newtonian to Bagnoldian-like behavior \cite{Bagnold}.  Indeed, it has been argued that in granular suspensions $\dot\gamma_\times$ vanishes linearly as the packing fraction increases towards the jamming $\phi_J$ \cite{Fall}.  It thus may be that $\dot\gamma_\times$ is sufficiently small near our Bagnoldian-to-Newtonian transition at $\phi_c$, and so the coexistence region may survive for some finite range of $\dot\gamma$.

Another interesting modification of our model would be to include random fluctuating forces to model finite temperature effects in colloids.  Such thermal forces should also result in a Newtonian rheology at sufficiently small $\dot\gamma$.  However, if $\tau_\mathrm{th}$ is the thermal relaxation time, one might expect such thermal effects to become unimportant when the shear strain rate is sufficiently large, $\dot\gamma > \tau_\mathrm{th}^{-1}$ \cite{OT-thermal,Ikeda2,Kawasaki}.  Indeed, the shear bands observed in the colloids of Ref.~\cite{Chikkadi} were exactly in this region of $\dot\gamma>\tau_\mathrm{th}^{-1}$. Thus, if $\tau_\mathrm{th}^{-1}$ near $\phi_c$ is sufficiently small, again our coexistence region may survive.  

Finally it will be interesting to add inter-particle microscopic friction to our model, and study the interplay between the shear banding we observe in this work, and the DST of the frictional models studied in Refs.~\cite{OH_friction, Heussinger, Grob, Grob2}.  Aside from the possible DST in our present model, as discussed in Sec.~\ref{s3DST}, it is worth noting that our shear banding transition shares some of the physical phenomena believed to be at work in the DST of the frictional models.  In particular, our shear banding transition marks the transition from a Bagnoldian phase, where there are few instantaneous particle contacts and such contacts are short lived, to a Newtonian phase, where there are $O(1)$ contacts per particle and the contacts are persistent.  The DST of frictional models is similarly believed to result from a transition between a region where there are few direct particle contacts, to a region where there are many.  It thus may be that our shear banding transition will serve as a trigger for the onset of strong frictional contacts that leads to enhanced DST.  We must leave these further investigations, outlined above, to future work.

\section*{Acknowledgements}
This work was supported in part by the European Research Council under the European UnionÕs Seventh Framework Programme (FP7/2007-2013), ERC Grant Agreement No. 306845.
Early stages of this work were supported by National Science Foundation Grant No. DMR-1205800 and the Swedish Research Council Grant No. 2010-3725. Simulations were performed on resources provided by the Swedish National Infrastructure for Computing (SNIC) at PDC and HPC2N.  

\section*{Appendix A}

\begin{figure}[h!] 
\includegraphics[width=3.2in]{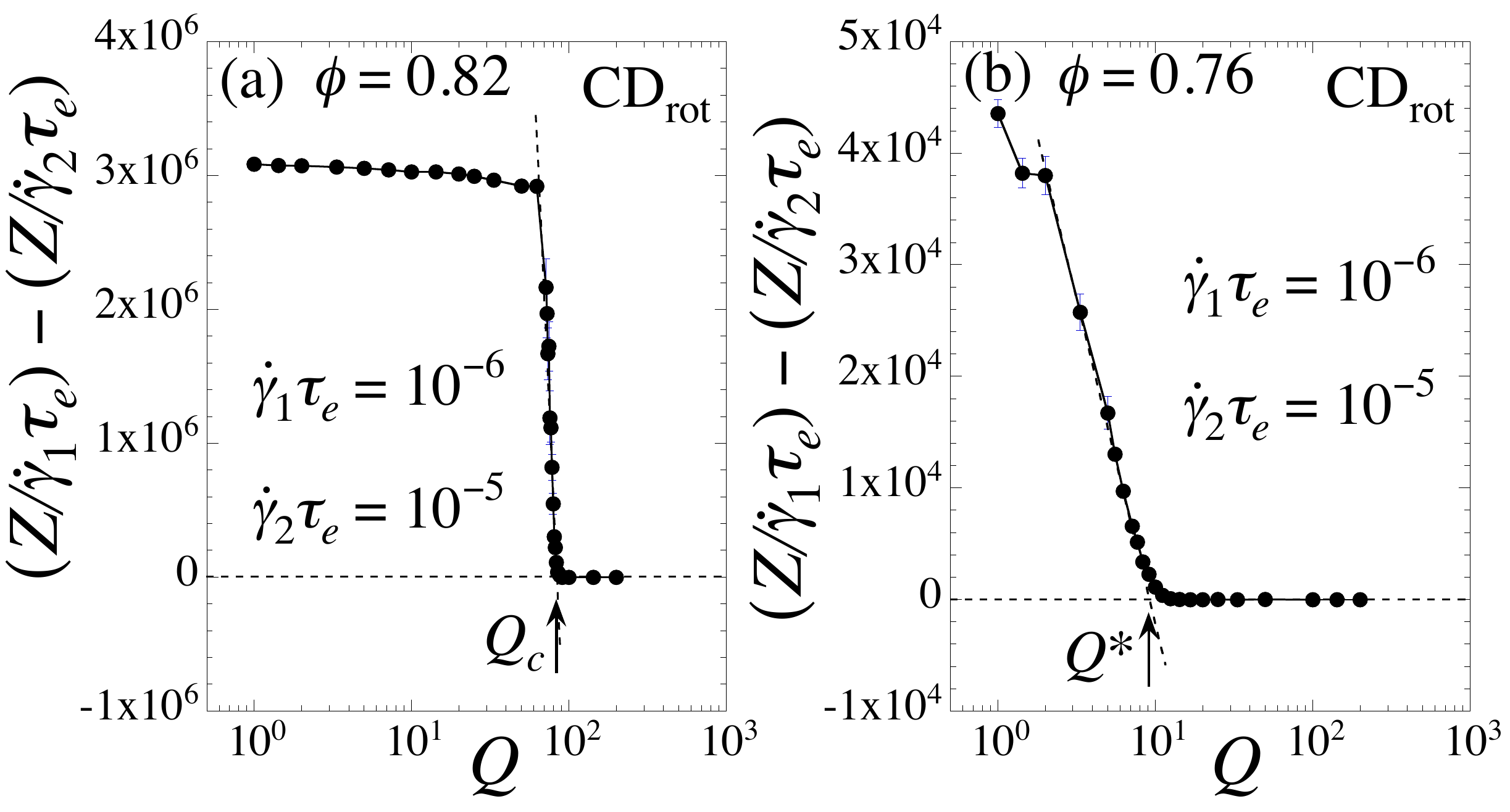} 
\caption{$(Z/\dot\gamma_1\tau_e)-(Z/\dot\gamma_2\tau_e)$, with $\dot\gamma_1\tau_e=10^{-6}$ and $\dot\gamma_2\tau_e=10^{-5}$, vs $Q$ for (a) $\phi=0.82$ and (b) $\phi=0.76$ in model CD$_\mathrm{rot}$.  The intersection of the dashed lines estimates the location of the transition $Q_c(\phi)$ in (a) and the crossover $Q^*(\phi)$ in (b).  System has $N=1024$ particles.
}
\label{Dz_gdot-rot}
\end{figure}

In this Appendix we describe the specific methods we use to map out the phase diagram of model CD$_\mathrm{rot}$, shown in Fig.~\ref{Qc-Qstar-phic}.  To determine the transition line $Q_c(\phi)$ 
%that separates the Newtonian region $Q<Q_c$ from the Bagnoldian region $Q>Q_c$ 
at large $\phi \gtrsim 0.8$, we use our result, as discussed in connection with Figs.~\ref{tau-nu-vs-Q-rot}(a) and (b), that for the Bagnoldian region $Q>Q_c$ we have $Z\sim\dot\gamma\tau_e$ as $\dot\gamma\tau_e\to 0$, while for the Newtonian region $Q<Q_c$ we have $Z\to $ constant as $\dot\gamma\tau_e\to 0$.  Thus, for sufficiently small $\dot\gamma\tau_e$ at  fixed $Q$ and $\phi$, we expect $Z/\dot\gamma\tau_e$ will be independent of $\dot\gamma\tau_e$ for $Q>Q_c$, while $Z/\dot\gamma\tau_e$ will increase with decreasing $\dot\gamma\tau_e$ for $Q<Q_c$.  We thus consider two small strain rates $\dot\gamma_1\tau_e=10^{-6}$ and $\dot\gamma_2\tau_e=10^{-5}$, and compute the difference $(Z/\dot\gamma_1\tau_e) - (Z/\dot\gamma_2\tau_e)$ as a function of $Q$ at fixed $\phi$.  By the preceding observations, this quantity should vanish for $Q>Q_c$ and be non-zero for $Q<Q_c$.  In Fig.~\ref{Dz_gdot-rot}(a) we plot this quantity vs $Q$ for the case with packing fraction $\phi=0.82$.  The point where $(Z/\dot\gamma_1\tau_e) - (Z/\dot\gamma_2\tau_e)$ increases from zero upon decreasing $Q$ thus locates the transition $Q_c$.
 
To locate the crossover line $Q^*(\phi)$ at smaller $\phi\lesssim 0.8$, that separates the weakly inelastic Bagnoldian region $Q>Q^*$ from the strongly inelastic Bagnoldian region $Q<Q^*$, we use the same approach.  As discussed in connection with Figs.~\ref{tau-nu-vs-Q-rot}(c) and (d), for $Q>Q^*$ we again have $Z\sim \dot\gamma\tau_e$.  For $Q<Q^*$ we found that $Z\to 0$ as $\dot\gamma\tau_e\to 0$, but that this dependence is slower than linear in $\dot\gamma\tau_e$.  Thus $(Z/\dot\gamma_1\tau_e)-(Z/\dot\gamma_2\tau_e)$ will again vanish for $Q>Q^*$ and be non zero for $Q<Q^*$.  In Fig.~\ref{Dz_gdot-rot}(b) we plot this quantity vs $Q$ for the case with $\phi=0.76$.  The point where $(Z/\dot\gamma_1\tau_e) - (Z/\dot\gamma_2\tau_e)$ increases from zero upon decreasing $Q$ thus locates the crossover $Q^*$.  Proceeding in this manner at other values of fixed $\phi$ we thus map out the curves $Q_c(\phi)$ and $Q^*(\phi)$ in the phase diagram of Fig.~\ref{Qc-Qstar-phic}.

\begin{figure}[h!] 
\includegraphics[width=3.2in]{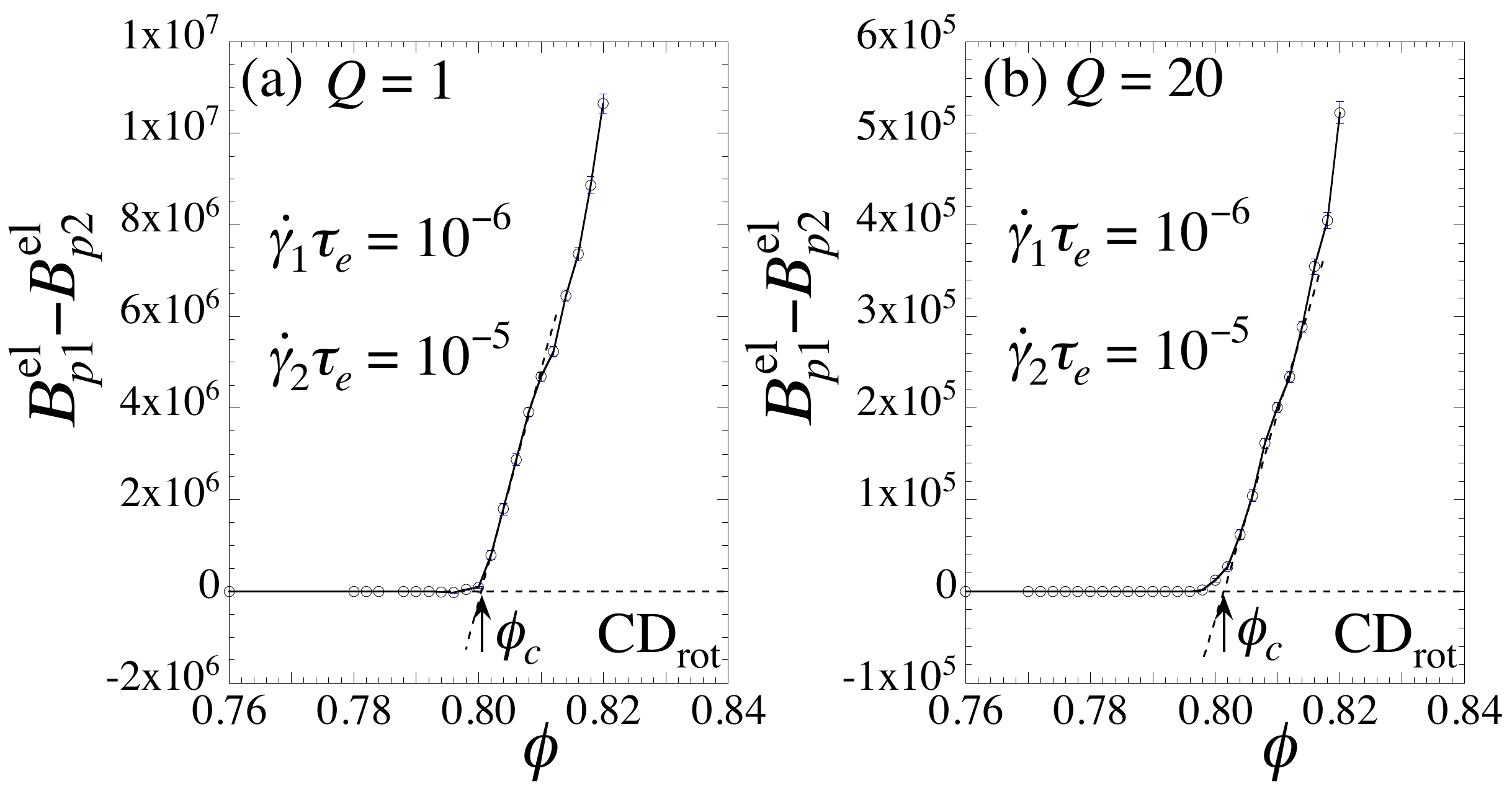} 
\caption{$B_{p1}^\mathrm{el}-B_{p2}^\mathrm{el}$, with $\dot\gamma_1\tau_e=10^{-6}$ and $\dot\gamma_2\tau_e=10^{-5}$, vs $\phi$ for (a) $Q=1$ and (b) $Q=20$  in model CD$_\mathrm{rot}$.  The intersection of the dashed lines estimates the location of the transition $\phi_c(Q)$.  System has $N=1024$ particles.
}
\label{Dbp-vs-phi-rot}
\end{figure}

To determine the transition $\phi_c(Q)$ at small $Q$, that separates the Bagnoldian region $\phi<\phi_c$  from the Newtonian region at $\phi>\phi_c$, we consider the Bagnold coefficient $B_p^\mathrm{el}$.  In the Bagnoldian region at sufficiently small $\dot\gamma\tau_e$ we have $B_p^\mathrm{el}$ is independent of $\dot\gamma\tau_e$. In the Newtonian region, where $\eta_p^\mathrm{el}$ is independent of $\dot\gamma\tau_e$,  we have from Eqs.~(\ref{eB}) and (\ref{eeta}) that $B_p^\mathrm{el}\equiv P^\mathrm{el}/(\dot\gamma\tau_e)^2 =\eta_p^\mathrm{el}/(Q\dot\gamma\tau_e)\sim 1/\dot\gamma\tau_e$. Thus if we compute the difference $B_{p1}^\mathrm{el}-B_{p2}^\mathrm{el}$, where the two Bagnold coefficients are computed at $\dot\gamma_1\tau_e=10^{-6}$ and $\dot\gamma_2\tau_e=10^{-5}$ respectively, then this quantity will vanish in the Bagnoldian region $\phi<\phi_c$, and  become non-zero in the Newtonian region $\phi>\phi_c$.  In Fig.~\ref{Dbp-vs-phi-rot} we show examples of this analysis for the cases $Q=1$ and $Q=20$.  The point where $B_{p1}^\mathrm{el}-B_{p2}^\mathrm{el}$ increases from zero upon increasing $\phi$ thus locates the transition $\phi_c$.  Proceeding in this manner at other values of fixed $Q$ we thus map out the curve $\phi_c(Q)$ to complete the phase diagram of Fig.~\ref{Qc-Qstar-phic}.  We caution that the precise location of the transition lines in this phase diagram may be expected to shift slightly if one considers smaller strain rates $\dot\gamma\tau_e$ and larger system sizes.

\section*{Appendix B}

In this Appendix we consider the contour of constant pressure $P=2\times 10^{-5}$  that passes through the coexistence region of model CD$_\mathrm{rot}$ (the dashed line in Fig.~\ref{coexist}), and 
address the issue of why the values of $(\phi^*_B,\dot\gamma^*_B\tau_e)$ and $(\phi^*_N,\dot\gamma^*_N\tau_e)$, that describe the local packing fraction and strain rate of the two shear bands in Fig.~\ref{vx-gdot-phi-vs-y}, are found to be slightly different from the values $(\phi_B,\dot\gamma_B\tau_e)$ and $(\phi_N,\dot\gamma_N\tau_e)$ at which the contour enters and exits the coexistence region.  We will argue that this difference is due to a finite-size-effect.

\begin{figure}
\includegraphics[width=3.4in]{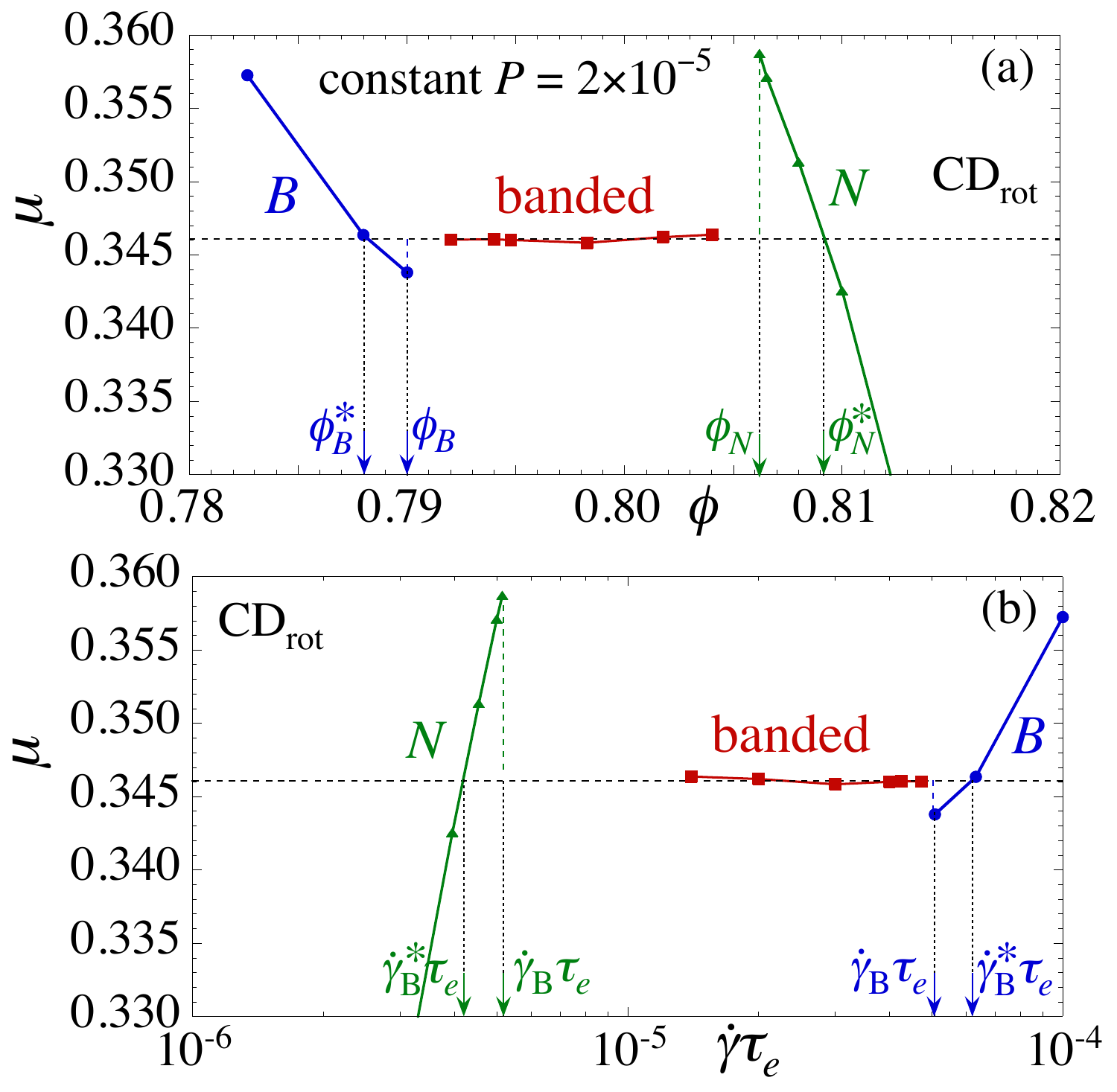}
\caption{(Color online) $\mu=\sigma/p$ vs  (a) the global $\phi$ and (b) the global $\dot\gamma\tau_e$, as one passes through the shear banded coexistence region on a contour of constant pressure $P=2\times 10^{-5}$.  Circles denote points in the homogeneous Bagnoldian phase,  triangles denote points in the homogeneous Newtonian phase, and squares denote points in the shear banded coexistence region.  The values of $\phi_B$ and $\phi_N$, that mark the ends of the homogeneous phases, and the values of $\phi^*_B$ and $\phi^*_N$, that denote the values found in the coexisting phases in the shear banded region, are indicated in (a).  The corresponding values of $\dot\gamma_B\tau_e$ and $\dot\gamma_N\tau_e$ and of $\dot\gamma^*_B\tau_e$ and $\dot\gamma^*_N\tau_e$ are shown in (b).  The system is at $Q=1$ and has $N=65536$ particles.
}
\label{mu-constP}
\end{figure}

To highlight the difference between $(\phi^*_B,\dot\gamma^*_B\tau_e)$ and $(\phi_B,\dot\gamma_B\tau_e)$, and between $(\phi^*_N,\dot\gamma^*_N\tau_e)$ and $(\phi_N,\dot\gamma_N\tau_e)$, we consider the behavior of the macroscopic friction $\mu=\sigma/p=\Sigma/P$ as we pass through the coexistence region on the contour of constant pressure. 
In Fig.~\ref{mu-constP}(a)  we plot  $\mu$ vs the global $\phi$ along this contour of constant  $P=2\times 10^{-5}$.  In this figure circles represent points in the homogeneous Bagnoldian phase, triangles are points in the homogeneous Newtonian phase, while squares are points in the shear banded coexistence region.
We define $\phi_B$ as the highest packing fraction in the homogeneous  Bagnoldian phase, and $\phi_N$ as the lowest packing fraction in the homogeneous Newtonian phase, as indicated in the figure.

Since we are varying $\phi$ at constant $P$, and since in the coexistence region the shear bands must be in mechanical equilibrium with equal $P$ and $\Sigma$, we thus expect that $\mu=P/\Sigma$ must also remain constant as we pass through the coexistence region at constant $P$.  We indeed see in Fig.~\ref{mu-constP}(a) that this is so.
However this constant $\mu=0.346$ is not the value of $\mu$ at either $\phi_B$ or $\phi_N$, but rather we have to enter deeper into the homogeneous phase regions to $\phi^*_B$ and $\phi^*_N$, as indicated in the figure, to find homogeneous states with this value of $\mu=0.346$.
In Fig.~\ref{mu-constP}(b) we make a similar plot of $\mu$ vs the global $\dot\gamma\tau_e$, along this same contour of constant $P$.  
Fig.~\ref{mu-constP} thus allows us to read off the above defined values of $(\phi^*_B,\dot\gamma^*_B\tau_e)$ and $(\phi^*_N,\dot\gamma^*_N\tau_e)$, and we find to these to agree well with the values found in Fig.~\ref{vx-gdot-phi-vs-y} for the local packing fractions and strain rates within the two coexisting bands.
We can also read off from Fig.~\ref{mu-constP} the values of $(\phi_B,\dot\gamma_B\tau_e)$ and $(\phi_N,\dot\gamma_N\tau_e)$ which give the coexistence region boundaries.  We thus see that $ \mu(\phi_B,\dot\gamma_B\tau_e)<\mu(\phi^*_B,\dot\gamma^*_B\tau_e) =\mu(\phi^*_N,\dot\gamma^*_N\tau_e)<\mu(\phi_N,\dot\gamma_N\tau_e)$, and there appears to be small but sharp jumps in $\mu$ (and hence $\Sigma$) as we cross from the homogeneous regions into the coexistence region.

\begin{figure}
\includegraphics[width=3.4in]{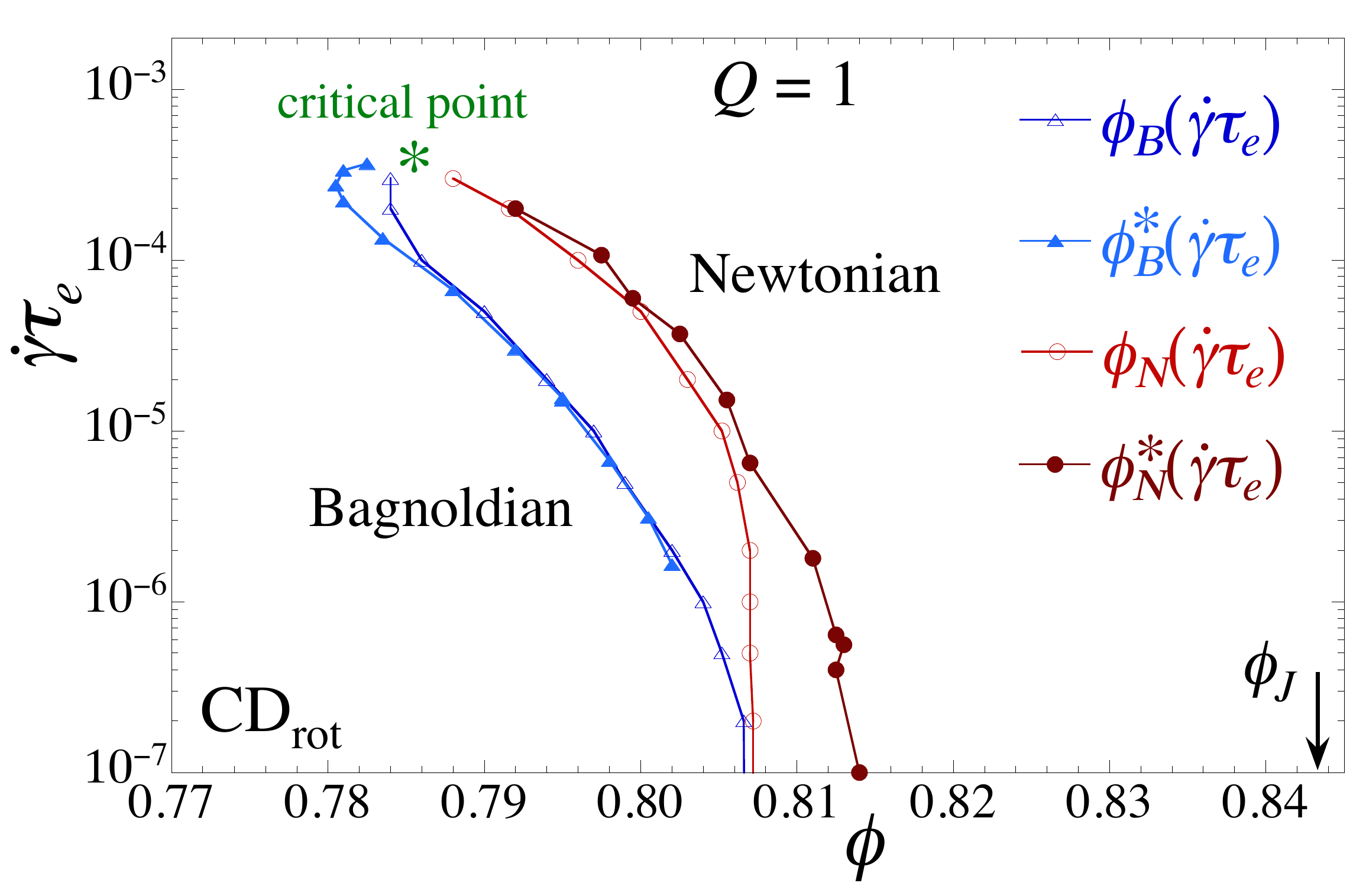}
\caption{(Color online) Phase diagram in the $(\phi,\dot\gamma\tau_e)$ plane.  $\phi_B(\dot\gamma\tau_e)$ (open triangles) denotes the boundary between the region of homogeneous Bagnoldian rheology and the shear banding coexistence region;  $\phi_N(\dot\gamma\tau_e)$ (open circles) denotes the boundary of the coexistence region with the region of homogeneous Newtonian rheology.  These points have been determined by the shear banding parameter $R_v$.
$\phi_B^*(\dot\gamma\tau_e)$ (solid triangles) gives the values of packing fraction and strain rate found within the Bagnoldian shear band of the coexistence region;  $\phi_N^*(\dot\gamma\tau_e)$ (solid circles) gives the values found within the Newtonian shear band of the coexistence region.  We attribute the difference between $\phi_N(\dot\gamma\tau_e)$ and $\phi_N^*(\dot\gamma\tau_e)$, and between $\phi_B(\dot\gamma\tau_e)$ and $\phi_B^*(\dot\gamma\tau_e)$, to finite size effects.
The system is at $Q=1$ and has $N=65536$ particles.
}
\label{coexist2}
\end{figure}

By sampling other points throughout the shear banded coexistence region, and measuring the packing fraction and strain rate in the two coexisting phases, we can determine the values of $(\phi^*_B,\dot\gamma^*_B\tau_e)$ and $(\phi^*_N, \dot\gamma^*_N\tau_e)$ corresponding to other fixed pressures.  We find that these values all lie on common curves $\phi^*_B(\dot\gamma\tau_e)$ and $\phi^*_N(\dot\gamma\tau_e)$ within the $(\phi,\dot\gamma\tau_e)$ plane, as we show in Fig.~\ref{coexist2}.  We see that the difference between $\phi_N(\dot\gamma\tau_e)$ and $\phi_N^*(\dot\gamma\tau_e)$ is greatest at low $\dot\gamma\tau_e$, while the difference between $\phi_B(\dot\gamma\tau_e)$ and $\phi_B^*(\dot\gamma\tau_e)$ is greatest near the critical end point.

Returning to Fig.~\ref{mu-constP}, we note that the constant $\mu$ within the shear banding coexistence region is analogous to the Maxwell construction of a mean-field first-order equilibrium phase transition \cite{Max}.  However for a first-order equilibrium transition, we would expect this curve to be continuous and monotonic.  Defining $f$ as the fraction of the system in the Bagnoldian phase, we would have expected that $f$ varies smoothly from $f=1$ to $f=0$ as the system crosses from the Bagnoldian side to the Newtonian side.  However the difference between $\phi_N$ and $\phi_N^*$, and between $\phi_B$ and $\phi_B^*$ shows that this does not happen.  Since we must have for the global average $\phi=f\phi_B^* + (1-f)\phi_N^*$, and since $\phi_B^*<\phi_B$, it must be the case that $f<1$ as the system enters the coexistence region at $\phi_B$, and similarly since $\phi_N^*>\phi_N$, we must have $f>0$ when the system exists the coexistence region at $\phi_N$.

%One way to interpret this result in analogy with the equilibrium first-order transition, is to imagine that $\phi_B^*(\dot\gamma\tau_e)$ and $\phi_N^*(\dot\gamma\tau_e)$ are the true boundaries of the coexistence region, and $\phi_B(\dot\gamma\tau_e)$ and $\phi_N(\dot\gamma\tau_e)$ are analogous to the spinodal line.  Inside the spinodal line a homogeneous phase is unstable to local fluctuations and a phase separated state necessarily results.  Between the coexistence and spinodal lines, the homogeneous phase is not the free energy minimum, but is metastable with respect to local fluctuations.  In our case, however, the homogenous phase appears to be the stable steady state in this region.  If, for example, we initiate the system in a shear banded state within the region between $\phi_N(\dot\gamma\tau)$ and $\phi_N^*(\dot\gamma\tau_e)$, we find that it decays to a homogeneous Newtonian state.

To interpret this unexpected result, we speculate that $\phi^*_B(\dot\gamma\tau_e)$ and $\phi^*_N(\dot\gamma\tau_e)$ may be the true boundaries of the coexistence region, and that the 
stability of the homogeneous phase within the region between $\phi^*_B(\dot\gamma\tau_e)$ and $\phi_B(\dot\gamma\tau_e)$ and between $\phi_N(\dot\gamma\tau_e)$ and $\phi_N^*(\dot\gamma\tau_e)$ is  a finite size effect, explained as follows.  In the coexistence region, the width of the interface over which the system transitions from the Bagnoldian band to the Newtonian band should have some characteristic size  set by a correlation length $\xi$ that remains finite as long as one is not at the critical end point.  The widths of the bands themselves are set by the system width $L_y$.  As $L_y\to\infty$, the width of the interface always becomes negligible compared to the widths of the bands.  However, for finite $L_y$, as one approaches close to the coexistence line from within the shear banded region, the width of one of the bands is decreasing and will ultimately become of order $\xi$.  At that point the decrease in free energy (assuming our athermal, non-equilibrium, system has something analogous to a free energy) obtained by phase separating may be smaller than the increase in free energy needed to create the interface between the bands.  The homogeneous phase may thus become stabilized.
Looking at the profiles of $\dot\gamma(y)$ and  $\phi(y)$ in Fig.~\ref{vx-gdot-phi-vs-y}, we see that the width of the interface is generally non-negligible compared to the width $L_y$ of our system, and so such a finite size effect in our system may be large and noticeable.  

\begin{figure}
\includegraphics[width=3.4in]{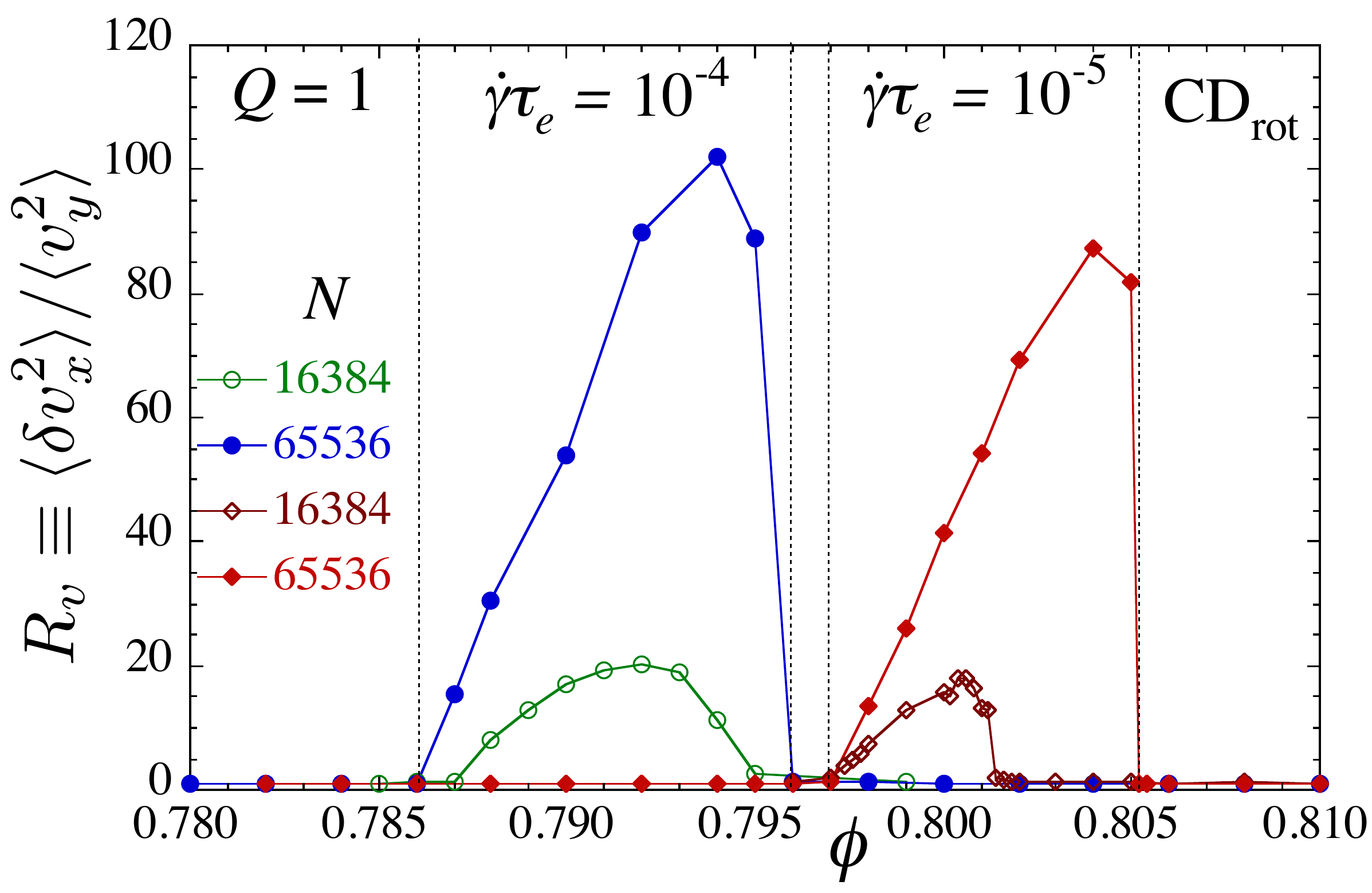}
\caption{(Color online) Shear banding parameter $R_v$ vs packing fraction $\phi$, for two different size systems with $N=16384$ (open symbols) and $N=65536$ (solid symbols) particles.  Results are shown for strain rates $\dot\gamma\tau_e=10^{-5}$ (diamonds) and $\dot\gamma\tau_e=10^{-4}$ (circles), as indicated. Vertical dotted lines denote the boundaries of the shear banded coexistence region for the $N=65535$ size system.
 We see that the width of the coexistence region shrinks as the system size gets smaller, thus indicating a finite size effect in the location of the coexistence region boundaries $\phi_B(\dot\gamma\tau_e)$ and $\phi_N(\dot\gamma\tau_e)$.
}
\label{FFStest}
\end{figure}

We now test this notion by checking whether the boundaries $\phi_B(\dot\gamma\tau_e)$ and $\phi_N(\dot\gamma\tau_e)$ show any noticeable dependence on the system size. In Fig.~\ref{FFStest} we plot the shear banding parameter $R_v$ of Eq.~(\ref{eRv}) vs $\phi$, comparing results for our system with $N=65536$ particles against a system of half the length, with $N=16384$ particles.  We show results for two different strain rates, $\dot\gamma\tau_e=10^{-5}$ and $10^{-4}$.  For $\dot\gamma\tau_e=10^{-5}$, for which Fig.~\ref{coexist2} shows essentially no difference between $\phi_B$ and $\phi^*_B$ but a noticeable difference between $\phi_N$ and $\phi^*_N$, we see that there is little finite size effect for $\phi_B$, but a strong finite size effect for $\phi_N$.
For $\dot\gamma\tau_e=10^{-4}$, for which Fig.~\ref{coexist2} shows a noticeable difference between both $\phi_B$ and $\phi^*_B$ and between $\phi_N$ and $\phi^*_N$, we see a noticeable finite size effect in the location of both $\phi_B$ and $\phi_N$.  For both strain rates, we find that the width of the coexistence region  shrinks as the system size gets smaller.
We thus find, in support of our above speculation, that finite size effects appear in the locations of $\phi_B$ and $\phi_N$ wherever $\phi_B>\phi^*_B$ and $\phi_N<\phi^*_N$.
We thus suppose that  we would find $\phi_N(\dot\gamma\tau_e)\to\phi^*_N(\dot\gamma\tau_e)$, and $\phi_B(\dot\gamma\tau_e)\to\phi^*_B(\dot\gamma\tau_e)$, if we could look at systems with much larger $L_y$.

We note that another manifestation of such a finite size effect may be directly seen in Fig.~\ref{vx-gdot-phi-vs-y}. We  can determine the width $w_B$ of the Bagnoldian band by the intersection of the packing fraction profile $\phi(y)$ with the global average $\phi$, so that the fraction of the system in the Bagnoldian band is $f=w_B/L_y$.   Such a definition gives the correct relation $\phi=f\phi^*_B + (1-f)\phi^*_N$.  We can then proceed similarly using the strain rate profile $\dot\gamma(y)$, to define an $f^\prime$ so that $\dot\gamma=f^\prime\dot\gamma^*_B+(1-f^\prime)\dot\gamma^*_N$.  We then find that, although $f$ and $f^\prime$ are close, they are not equal.  This can be verified by comparing the location of the vertical dashed lines in Figs.~\ref{vx-gdot-phi-vs-y} (d) -- (f) with those in (g) -- (i).  The difference between $f$ and $f^\prime$ is a consequence of the fact that the width of the interface is a non-negligible fraction of the system width $L_y$.

\end{document}